# Selfish Pups: Weaning Conflict and Milk Theft in Free-Ranging Dogs

**Weaning Conflict and Milk Theft in Dogs**

**Manabi Paul[1] and Anindita Bhadra[1,*]**

[1]Behaviour and Ecology Lab, Department of Biological Sciences, Indian Institute of Science Education and Research Kolkata, India

[*]Behaviour and Ecology Lab, Department of Biological Sciences,

Indian Institute of Science Education and Research Kolkata

Mohanpur,

PIN 741246, West Bengal, INDIA

*tel.* 91-33-66340000-1223

*fax* +91-33-25873020

*e-mail:* abhadra@iiserkol.ac.in

**Abstract**

Parent-offspring conflict theory predicts the emergence of weaning conflict between a mother and her offspring arising from skewed relatedness benefits. Empirical observations of weaning conflict has not been carried out in canids. In a field-based study on free-ranging dogs we observed that suckling bout durations reduce, proportion of mother-initiated suckling bouts reduce and mother-initiated suckling terminations increase, with pup age. We identified the 7$^{th}$ - 13$^{th}$ week period of pup age as the zone of conflict between the mother and her pups, beyond which suckling solicitations cease, and before which suckling refusals are few. We also report for the first time milk theft by pups who take advantage of the presence of multiple lactating females, due to the promiscuous mating system of the dogs. This behaviour, though apparently disadvantageous for the mothers, is perhaps adaptive for the dogs in the face of high mortality and competition for resources.

**Introduction**

Maternal care in the form of suckling is an obligatory behaviour in mammals, at least in the early stages of development of young. While in some species suckling occurs for a few days, in others it continues for months, and well beyond the natural weaning period, in humans [1]. Suckling makes heavy demands on the maternal resources, requiring mothers to strike a balance between their investments in current offspring and lifetime reproductive success. Parent-offspring conflict (POC) theory suggests the emergence of a zone of conflict over weaning between the mother and her offspring, due to skewed cost-benefit ratios from relatedness estimates [2]. Various models and empirical tests have validated this theory in contexts as diverse as brood size optimization, reproduction, resource sharing, mate choice, etc.  [3,4]. Most studies of POC focus on maternal investment and benefits, while overlooking the offspring's perspective to a large extent. However, there is evidence that offspring can manipulate mothers to their own advantage, for example, in preeclampsia (pregnancy induced hypertension) in humans induced by the high energy demands of the fetus [5]. Hence it is interesting to investigate the dynamics between the mother's voluntary investment in parental care and the offspring's tendency to exploit her as a resource.

POC over extended parental care has been observed in free-ranging dogs (*Canis familiaris*) [6,7]. Mothers begin competing over food with them during weaning; the conflict levels increasing with pup age [6,7]. This conflict is resource-dependent, mothers choosing to be more aggressive when richer resources are present, thereby showing preference towards lifetime reproductive success (LRS) over immediate benefits from the current litter [6]. Indian free-ranging dogs are scavengers dependent on human-generated wastes for their sustenance [8], and have existed in similar conditions for centuries, living in close proximity with humans, but not under their supervision [9]. They defend territories as groups, but typically prefer to forage singly, though capable of forming large packs to hunt [9,10]. They have a promiscuous mating system, and no reproductive hierarchies within the groups, though allocare by adults is possible [11,12]. This suggests a highly flexible social structure, as opposed to the more structured and stringent social organization in related canids like wolves (*Canis lupus*) and coyotes (*Canis latrans*) [13]. The free-ranging dog mothers face a challenging environment, with resource limitation, competition and high pup mortality [14], leading to low assured fitness per litter [15]. Striking a balance between investment per litter and LRS can lead to higher stability for the population; thus the free-ranging dogs provide an ideal model system for testing the implications of POC theory. We conducted a study on suckling behaviour in free-ranging dogs to understand if a zone of conflict over suckling exists between mothers and pups in dogs.

## Results

### Suckling behaviour

All mothers were observed to suckle their pups (1378 bouts in total), and in some cases allomothers were also observed to suckle non-filial pups, which was termed as allo-nursing. At the 3rd week, mothers invested (18 ± 9.28)% of their total time in suckling, thereby investing (41 ± 27.7)% of their total active time in this behaviour. This time reduced with pup age (Linear regression: $R^2 = 0.876$, $\beta = -0.936$, $P < 0.0001$), until suckling stopped completely at the 13th week. Pups at their 3rd week of age spent (14.6 ± 9.14)% of their time in suckling from their mothers, which comprised of (16.5 ± 9.5)% of their total activity. The mismatch in the mother and offspring's time investment in suckling is due to competition between siblings during suckling, due to which all pups do not get to suckle from the mother at equal rates. Not only the time spent in suckling but also the duration of suckling decreased as pups grew older (duration ~ age: $p < 0.0001$, Model 1 in S1 Text, Fig 1).

### Suckling refusals

Of the total number of suckling events in a week, the proportion of mother-initiated suckling (mo) and pup-initiated suckling (pup) changed with pup age (Fig 2a) and the current litter size (mo/ pup ~ age: $p < 0.0001$; mo/ pup ~ litter size: $p = 0.00029$. Model 2 in S1 Text). Mothers increasingly refused suckling solicitations (rfmo) as the pups grew older, and mother-mediated termination of suckling (through refusals) increased with pup age (rfmo ~ age: $p < 0.0001$.

Model 3 in S1 Text) (Fig 2b). The duration of mother-initiated suckling reduced drastically beyond the $7^{th}$ week (Fig 2c), when mother-mediated terminations reached 100%, and the mothers stopped initiating suckling completely by the $11^{th}$ week (Fig 2b).

Suckling duration seems to be controlled by a combination of the initiator, pup age and the current litter size (Table 1, Model 4 in S1 text). A three-way interaction in the GLMM suggests a significant role of both the pup age and the current litter size for mother and pup initiated suckling duration. A two-way significant interaction suggests that the duration of mother and pup initiated suckling vary for different litter sizes (Table 1, Model 4 in S1 Text).

Allonursing or milk theft?

In nine of the 11 groups, we observed high incidence of allonursing (594 bouts), and in eight of these nine groups, the allonursing female was related to the pups. 100% of the allonursing bouts were initiated by pups, irrespective of their age, and the allomothers never volunteered to allonurse (Fig 2a). Most of the allonursing bouts were terminated by refusals from the allomothers, irrespective of the age of the pups (Fig 3a). The duration of suckling from mothers over the entire period of observations was significantly longer than that from allomothers (2 tailed t test; $t = 13.519$, $df = 1961.714$, $p < 0.0001$; Fig 3b). This difference was consistent even over the zone of conflict, from the $7^{th}$ to the $13^{th}$ week of pup age (2 tailed t test; $t = 4.279$, $df = 1050.915$, $p < 0.0001$). Thus while mothers suckled their pups voluntarily till the onset of weaning and on the whole showed a large degree of care, allomothers never volunteered to nurse non-filial pups but were victims of milk theft.

## Discussion

According to the predictions of POC theory, a mother is expected to invest in LRS, such that her chances of increasing fitness are high [2]. Kin selection theory suggests that individuals can increase their inclusive fitness benefits by investing in the survival of close kin [16]. Thus, while mothers are expected to show conflict towards their offspring over weaning in order to maximize direct fitness, they can benefit by additionally providing altruistic care to the offspring of related females to ensure indirect fitness benefits. The duration of time allocated to a behaviour is a good estimate of investment in the behaviour in terms of time activity budgets of individuals, and we considered the time spent in suckling as a measure of investment by the mothers and allomothers. Though the use of suckling duration as a measure of investment by mothers has been debated, it continues to be used as an indirect measurement of maternal investment [17,18]. Moreover, since we compared suckling durations for the same set of individuals over time, this could be considered as a reliable marker for changing interest of the mothers and pups in suckling.

Our observations of the reducing proportion of mother-initiated and increasing proportion of pup-initiated suckling bouts with pup age, and longer mother-initiated suckling bouts in the early weeks of the pups' development as compared to the pup-initiated bouts suggest the mother's reduced interest in suckling as the pups grew older. Refusal of suckling by the mothers reaching 100% in the 7$^{th}$ week of pup age and complete seizure of solicitations from the pups by the 13$^{th}$ week is a good representation of the zone of conflict over weaning in the free-ranging dogs. Before the 7$^{th}$ week the mother encourages suckling solicitations and also initiates suckling, and beyond the 13th week, neither the mother nor the pups are interested in suckling, leading to the resolution of conflict.

It has been suggested that allonursing can evolve when costs associated with the behaviour are low [19]. Allonursing is more likely to evolve in polytocous species, and in carnivorous cooperative breeders rather than omnivorous or herbivorous ones [19]. Dogs are promiscuous breeders with a scavenging lifestyle [11]. Multiple females often give birth to pups simultaneously, in close proximity of each other, increasing the chances of communal breeding. Allonursing started as early as the 4th week of pup age, all bouts being pup-initiated and allomother-terminated, and much shorter than maternal suckling bouts. Pups appear to adopt a snatch and run strategy while suckling from allomothers, as allomothers terminate the suckling whenever they identify free-riders and do not altruistically offer milk to non-filial pups. Thus pups benefit my using a selfish strategy of maximizing milk intake from any available lactating female. Hence allonursing in free-ranging dogs is an example of milk theft by pups, a behaviour more common in primates and ungulates [20] than in carnivores, and less likely in cooperative and communal breeders[19]. In a competitive and disturbed environment with limited resources and high mortality, this ability to snatch milk would be highly adaptive for the pups. Though apparently maladaptive for the mothers, this might indeed be an evolutionarily stable strategy if related females tend to den close to each other. Our observations provide support for this idea, as most of the observed allomothers were related to the pups. Coupled long term behavioural and genetic studies in the future will help to understand whether milk theft by pups could lead to increased inclusive fitness benefits and LRS for the mothers.

## Materials and Methods

We observed 15 dog groups with 22 mothers and their pups (78) from the 3rd to 17th weeks of pup age, in West Bengal, India (Table in S2 Table). Pups are immobile and restricted to the dens closely guarded by mothers during the first two weeks after birth [21], and so observations were impossible before the third week. Here we use "weeks" to refer to pup age throughout this paper. Each group was observed for two morning (0900-1200h) and two evening (1400-1700h) sessions over blocks of two weeks. Each session consisted of 18 scans and 18 all occurrences sessions (AOS), amounting to a total of 8712 scans and AOS each, of one and five minutes respectively. For 11 of the groups we collected data on each suckling bout (suckling). A combination of the initiator's identity and the duration of suckling bouts was used to estimate the interest of mothers to offer (hence cooperation), and pups to receive, suckling. Both termination of suckling bouts by the mother and failed solicitations by pups was considered as refusal to offer suckling (hence conflict), while termination of suckling by the pups was considered voluntary. The free-ranging dogs experience quite high rates of mortality in the early stages of their lives, and thus the litter size often changes as the pups grow, reducing from the litter size at birth. So we considered the current litter size for every week of observations in the analyses.

### Statistics

Here we used the data from instantaneous scans to estimate the proportion of time spent by the mother in maternal care behaviours. All occurrences were used to check the rate of suckling (frequency per hour) at different pup ages. From the detailed records of suckling bouts of 11 mother-litter units, we measured the duration of each bout and calculated the proportion of mother and pup initiated suckling bouts over pup ages. For statistical analysis, we used

StatistiXL 1.10, Statistica 12 and R ("lme4" package) [22]. To check whether the identity of the suckling initiator (initiator), pup age in weeks (age) and the current litter size (LS) had any effect on the rate of care received by the individual pup in terms of suckling, we ran a generalized linear mixed effect model (GLMM) incorporating the predictor variables (initiator, age and LS) as the fixed effects and the group identity as the random effect. We started with the full model, i.e., with all possible three and two-way interactions among the fixed effects. If the three or two-way interaction was non-significant then we reduced the model using the standard protocol of backward selection method and ended up with the optimal model.

**Acknowledgements**

We thank Sreejani Sen Majumder for her help in some of the fieldwork. This work was funded by projects from CSIR and SERB, India and supported by IISER Kolkata.

**Ethical statement:** No dogs were harmed during this work. All work reported was observation based. The methods were approved by the IISER-Kolkata animal ethics committee of (1385/ac/10/CPCSEA), and followed approved guidelines of animal rights regulations of the Government of India.

**Supporting Information**

**S1 Text. Text file presenting the detailed description of the generalized linear mixed effect models (GLMM).** Four models have been described in details along with the fixed and random effects that have been incorporated in the models.

(DOC)

**S2 Table. A table showing the details of the observed dog groups.** The table represents the group identity of each observed mother-litter units along with their litter size at birth, year of data collection, location and habitat type of the observed units, etc. Presence or absence of allonursing has also been tabulated here.

| **Fixed Effects** | | | | |
|---|---|---|---|---|
| | **Estimate** | **Std. Error** | **t value** | **Pr(>\|z\|)** |
| **Intercept** | 10.20989 | 1.32828 | 7.687 | 3.46e-14*** |
| **Initiator** | -5.17190 | 1.34665 | -3.841 | 0.000126*** |
| **Age** | -0.76597 | 0.27380 | -2.798 | 0.005189** |
| **LS** | -0.51550 | 0.30588 | -1.685 | 0.092067**.** |
| **Initiator *Age** | 0.54558 | 0.28016 | 1.947 | 0.051599**.** |
| **Initiator *LS** | 0.85418 | 0.33336 | 2.562 | 0.010456* |
| **Age*LS** | 0.07249 | 0.06490 | 1.117 | 0.264121 |
| **Initiator *Age*LS** | -0.17554 | 0.06731 | -2.608 | 0.009160** |

.

| **Random Effects** | | | |
|---|---|---|---|
| **Groups** | **Name** | **Variance** | **Std.Dev.** |
| Group Identity | Intercept | 1.156 | 1.075 |
| Residual | | 6.222 | 2.494 |

**Table 1. Tabulated results of the generalized linear mixed effect model (GLMM) showing the effects of the initiator (Initiator), pup age (Age) and current litter size for a particular mother-litter group (LS) on the duration of suckling bouts.**

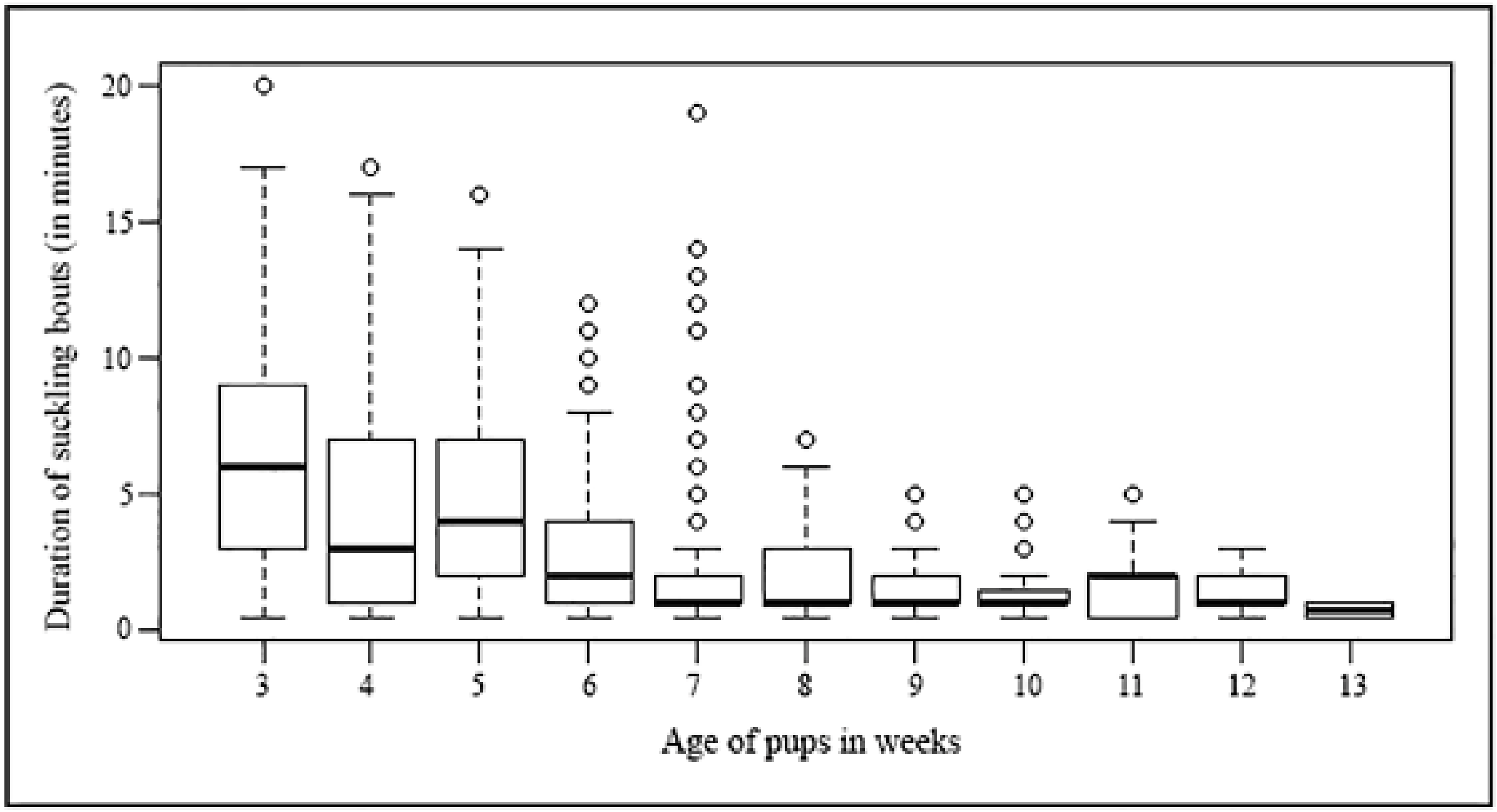

**Figure 1. The duration of suckling bouts decreased as pups grew older.** Box-whisker plots showing the decreasing pattern in the duration of suckling bouts (irrespective of the initiator) over increasing pup age.

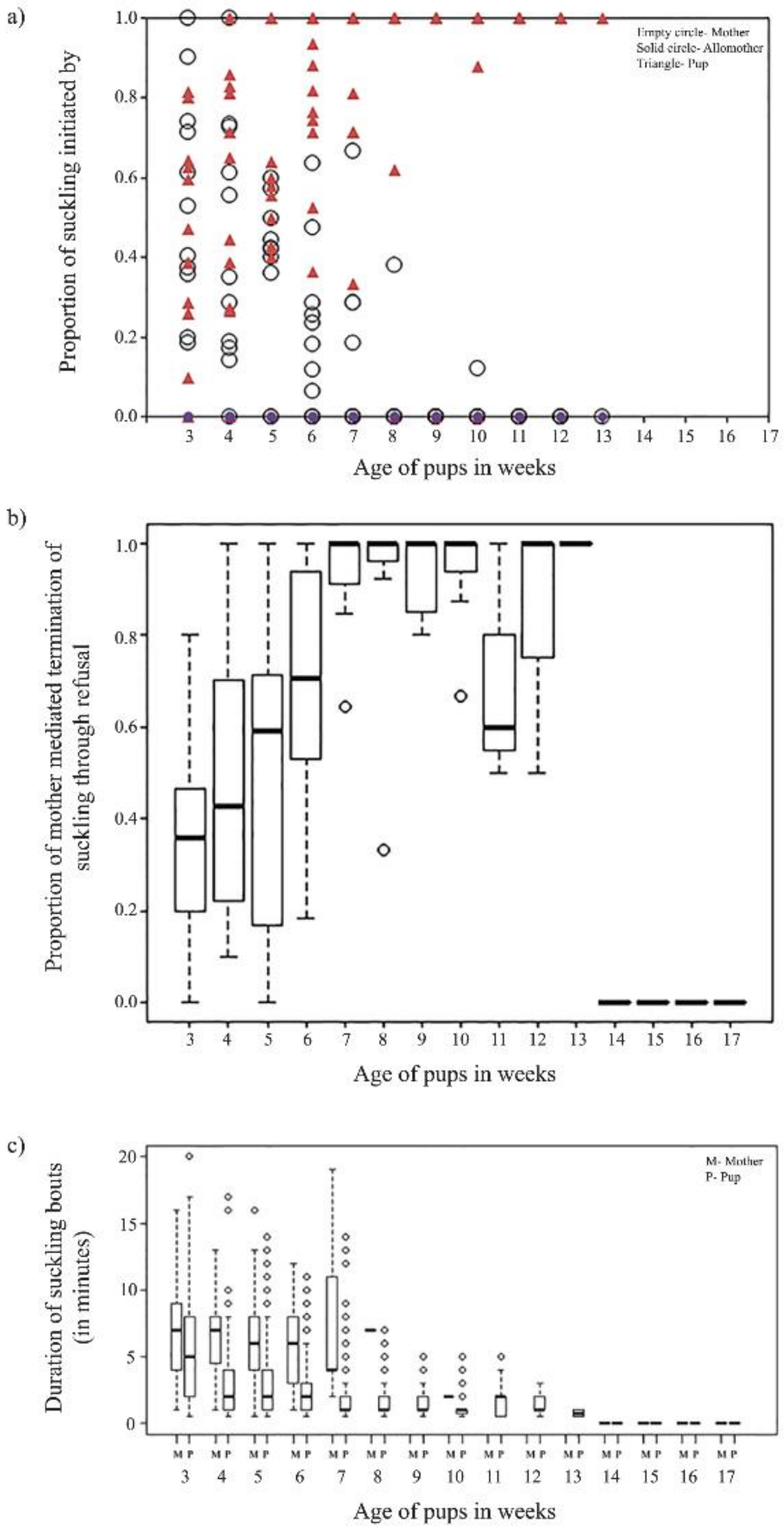
a)
Proportion of suckling initiated by
Age of pups in weeks
Empty circle- Mother
Solid circle- Allomother
Triangle- Pup
b)
Proportion of mother mediated termination of suckling through refusal
Age of pups in weeks
c)
Duration of suckling bouts (in minutes)
Age of pups in weeks
M- Mother
P- Pup

**Figure 2. Graphical representation of mother and allomother mediated suckling bouts. (a)** Scatter plot showing the proportion of total suckling initiated by the mother, pups and allomothers over pup age. **(b)** Box-whisker plot showing the increased trend of mother-mediated termination of suckling through refusal over pup age. **(c)** Box- whisker plot showing the duration of mother and pup initiated suckling bouts separately over pup age (N = 11 litters).

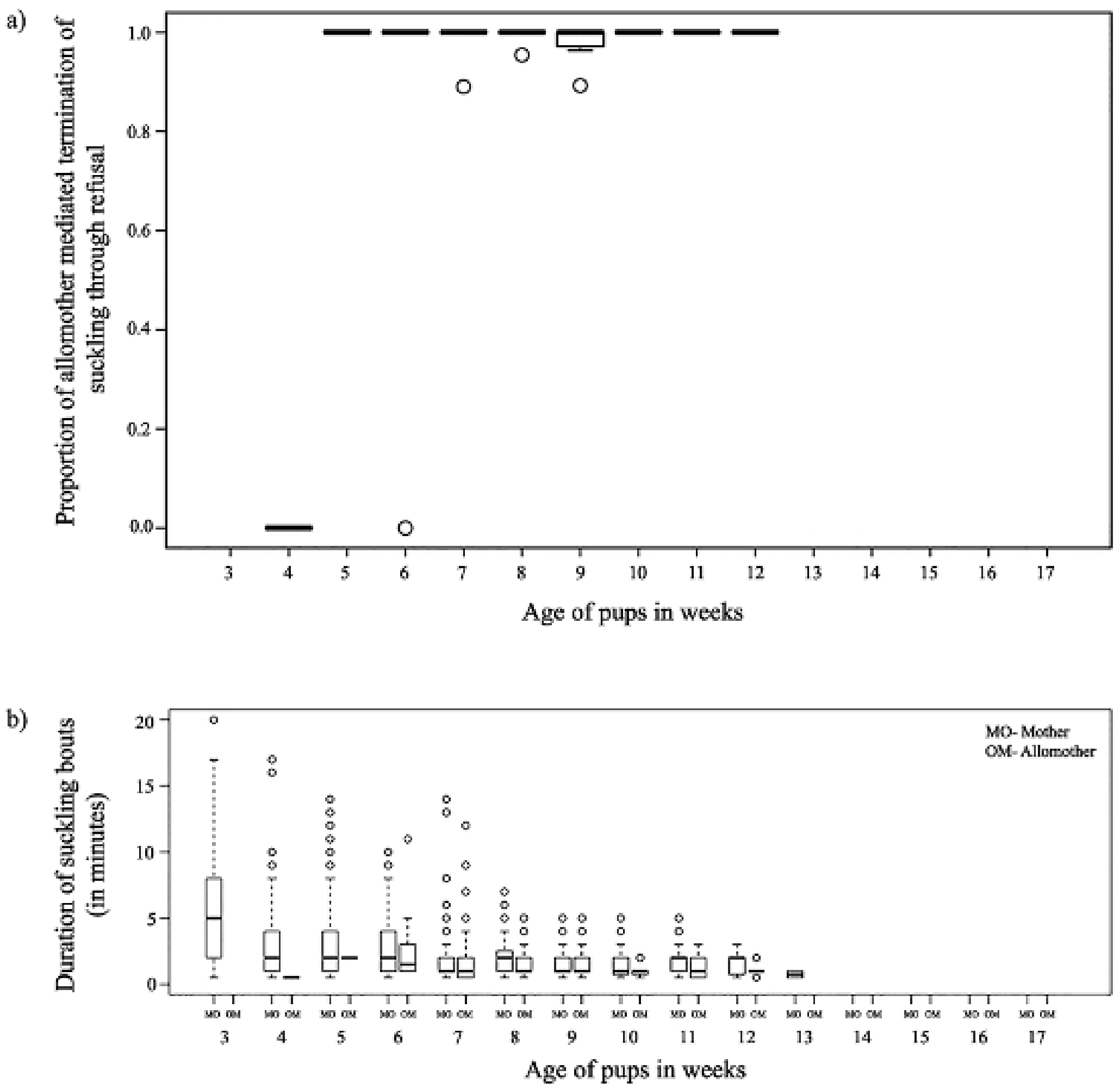

**Figure 3. (a)** Box- whisker plot showing allomother-mediated termination of suckling through refusal over pup age. **(b)** Box- whisker plot showing the duration of suckling bouts for pups suckling from mothers and allomothers, for different ages of pups.

**Supplementary Information**

**Model 1.**

We recorded the start and end time of each suckling bouts observed along with the details of the initiators. Suckling bouts' duration was calculated from the observed start and end time.

In order to check if the pup age had any impact on the duration of suckling bouts, we ran a generalized linear mixed effect model (GLMM) considering the duration of suckling bouts as the response variable, pup age as the fixed effects and the group identity as the random effects. A Ga ussian distribution was considered for the response variable in the model.

**Variables used in the model:**

Response variable:

Duration of suckling bouts (in minutes)- **dur**

Fixed effects:

Age of pups in weeks- **age**

Random effects:

Group identity- **fgr**

**Model: glmer (dur ~ +age + (1 | fgr), family = gaussian)**

**Results**

Random effects:

| | | |
|---|---|---|
| Groups | Name | Variance |
| | Std.Deffgr | (Intercept) |
| 1.101 | 1.049 | |

Fixed effects:

| | Estimate | Std. Error | t value | Pr(>\|t\|) |
|---|---|---|---|---|
| (Intercept) | 7.58515 | 0.35466 | 21.39 | 1.94e-12 *** |
| **age** | **-0.67677** | **0.02414** | **-28.03** | **< 2e-16** *** |

Significant codes: 0 ‘***’ 0.001 ‘**’ 0.01 ‘*’ 0.05 ‘.’ 0.1 ‘ ’ 1

**Model 2.**

For generalized linear mixed-effects model (GLMM) analysis we categorized the initiators of each suckling bouts into two categories i.e. initiated by the mother and initiated by the pup and used the Binomial family of distributions in R statistics. Hence suckling was incorporated in the model as a binary variable, either initiated by the mother or the pup. We used the function “cbind” that created a matrix by binding the column vectors containing the numbers of “mother initiated suckling” and “pup initiated suckling” measured for each group per week. In this case R adds the two columns together to produce the correct binomial denominator.

The two parameters, i.e., the current litter size and age of the pups in weeks were considered as the fixed effects while the identity of the groups was taken as the random effect. We started with the full model, i.e., with all possible two-way interactions among the fixed effects. The two-way interaction was non-significant; we then reduced the model using standard protocol of backward selection method and ended up with the optimal model, the result of which is described below.

**Variables used in the model:**

Response variable:

Number of mother initiated suckling bouts (for each week) - **mo**

Number of pup initiated suckling bouts (for each week) - **pup**

Fixed effects:

Age of pups in weeks- **age**

Current litter size- **ls**

Random effects:

Group identity- **fgr**

**Model: glmer (cbind(mo, pup) ~ +age + ls + (1 | fgr), family = binomial)**

**Results**

Random effects:

| Groups | Name | Variance | Std.Dev. |
|---|---|---|---|
| fgr | (Intercept) | 0.5259 | 0.7252 |

Fixed effects:

| | Estimate | Std. Error | z value | Pr(>\|t\|) |
|---|---|---|---|---|
| (Intercept) | 0.7904 | 0.5881 | 1.344 | 0.17893 |
| **age** | **-0.6931** | **0.0414** | **-16.740** | **< 2e-16** *** |
| **ls** | **0.5173** | **0.1428** | **3.624** | **0.00029***** |

Significant codes: 0 '***' 0.001 '**' 0.01 '*' 0.05 '.' 0.1 ' ' 1

**Model 3.**

For each suckling bouts observed, we recorded the time duration of bouts, the initiator (i.e. who i nitiated the suckling bout) and the terminator (i.e. who terminated the suckling bout) along with t he behavior of the terminator to the pup who were suckled. Out of the total termination of the suckling bouts ob served, mother mediated suckling termination was counted to calculate the proportion. We ran a GLMM considering the proportion of mother mediated termination of suckling bouts as the response vari able, pup age and the current litter size as fixed effects and the group identity as the random effects.

We started with the full model, i.e., with all possible two-way interactions among the fixed effects. The two-way interaction was non-significant; we then reduced the model using standard protocol of backward selection method and ended up with the optimal model, the result of which is described below.

**Variables used in the model:**

Response variable:

Proportion of mother mediated termination of suckling bouts through refusal - **rfmo**

Fixed effects:

Age of pups in weeks- **age**

Current litter size- **ls**

Random effects:

Group identity- **fgr**

**Model: glmer (rfmo ~ +age + ls + (1 | fgr), family = gaussian)**

**Results**

Random effects:

| Groups | Name | Variance | Std.Dev. |
|---|---|---|---|
| fgr | (Intercept) | 0.01030 | 0.1015 |

Fixed effects:

| | Estimate | Std. Error | t value | Pr(>\|t\|) |
|---|---|---|---|---|
| (Intercept) | 0.10251 | 0.13779 | 0.744 | 0.467 |
| **age** | **0.07669** | **0.01144** | **6.705** | **2.85e-09** *** |
| ls | 0.03154 | 0.02951 | 1.069 | 0.306 |

Significant codes: 0 '***' 0.001 '**' 0.01 '*' 0.05 '.' 0.1 ' ' 1

**Model 4.**

In order to check the effect of pup age, current litter size and the identity of the initiator (either th e mother or the pup) on the suckling duration, we ran a GLMM considering the Gaussian distribu tion for the response variable i.e. the suckling duration. Initiators' identity, pup age and the current litter size was incorporated in the model as the fixed effects whereas the group identity was considered as the ra ndom effect.

We started with the full model, i.e., with all possible two-way and three-way interactions among the fixed effects. The three-way interaction was significant, hence we kept the model as the final one .

**Variables used in the model:**

Response variable:

Duration of suckling bouts in minutes: **dur**

Fixed effects:

Mother or pup initiated suckling: **Initiator**

Age of pups in weeks- **age**

Current litter size- **ls**

Random effects:

Group identity- **fgr**

**Model: glmer (dur ~ + Initiator * age * ls + (1 | fgr), family = gaussian)**

**Results**

Random effects:

| Groups | Name | Variance | Std.Dev. |
|---|---|---|---|
| fgr | (Intercept) | 1.156 | 1.075 |

Fixed effects:

| | Estimate | Std. Error | t value | Pr(>\|t\|) |
|---|---|---|---|---|
| (Intercept) | 10.20989 | 1.32828 | 7.687 | 3.46e-14*** |
| **Initiator** | **-5.17190** | **1.34665** | **-3.841** | **0.000126*** |
| **age** | **-0.76597** | **0.27380** | **-2.798** | **0.005189** |
| ls | -0.51550 | 0.30588 | -1.685 | 0.092067**.** |
| Initiator*age | 0.54558 | 0.28016 | 1.947 | 0.051599**.** |
| **Initiator*ls** | **0.85418** | **0.33336** | **2.562** | **0.010456***|
| age*ls | 0.07249 | 0.06490 | 1.117 | 0.264121 |
| **Initiator*age*ls** | **-0.17554** | **0.06731** | **-2.608** | **0.009160****|

Significant codes:  0 ‘***’ 0.001 ‘**’ 0.01 ‘*’ 0.05 ‘.’ 0.1 ‘ ’ 1

**Raw data for Model 1 and 4**

| Year | Group | Age | Current litter size | Duration (min) of suckling bouts | Initiator | Suckled by | |
|---|---|---|---|---|---|---|---|
| | | | | | | Mother | Allomother |
| 4 | CAN | 3 | 6 | 11 | MO | 1 | |
| 4 | CAN | 3 | 6 | 8 | MO | 1 | |
| 4 | CAN | 3 | 6 | 11 | MO | 1 | |
| 4 | CAN | 3 | 6 | 8 | MO | 1 | |
| 4 | CAN | 3 | 6 | 11 | MO | 1 | |
| 4 | CAN | 3 | 6 | 11 | MO | 1 | |
| 4 | CAN | 3 | 6 | 7 | MO | 1 | |
| 4 | CAN | 3 | 6 | 7 | MO | 1 | |
| 4 | CAN | 3 | 6 | 7 | MO | 1 | |
| 4 | CAN | 3 | 6 | 7 | MO | 1 | |
| 4 | CAN | 3 | 6 | 7 | MO | 1 | |
| 4 | CAN | 3 | 6 | 7 | MO | 1 | |
| 4 | CAN | 3 | 6 | 4 | MO | 1 | |
| 4 | CAN | 3 | 6 | 8 | MO | 1 | |
| 4 | CAN | 3 | 6 | 4 | MO | 1 | |
| 4 | CAN | 3 | 6 | 4 | MO | 1 | |

| | | | | | | |
|---|---|---|---|---|---|---|
| 4 | CAN | 3 | 6 | 4 | MO | 1 |
| 4 | CAN | 3 | 6 | 4 | MO | 1 |
| 4 | CAN | 3 | 6 | 8 | Pup | 1 |
| 4 | CAN | 3 | 6 | 5 | Pup | 1 |
| 4 | MDB | 3 | 5 | 8 | MO | 1 |
| 4 | MDB | 3 | 5 | 8 | MO | 1 |
| 4 | MDB | 3 | 5 | 8 | MO | 1 |
| 4 | MDB | 3 | 5 | 8 | MO | 1 |
| 4 | MDB | 3 | 5 | 8 | MO | 1 |
| 4 | MDB | 3 | 5 | 7 | MO | 1 |
| 4 | MDB | 3 | 5 | 7 | MO | 1 |
| 4 | MDB | 3 | 5 | 7 | MO | 1 |
| 4 | MDB | 3 | 5 | 7 | MO | 1 |
| 4 | MDB | 3 | 5 | 7 | MO | 1 |
| 4 | MDB | 3 | 5 | 11 | MO | 1 |
| 4 | MDB | 3 | 5 | 11 | MO | 1 |
| 4 | MDB | 3 | 5 | 11 | MO | 1 |
| 4 | MDB | 3 | 5 | 11 | MO | 1 |
| 4 | MDB | 3 | 5 | 11 | MO | 1 |
| 4 | PF1 | 3 | 5 | 1 | MO | 1 |
| 4 | PF1 | 3 | 5 | 1 | MO | 1 |
| 4 | PF1 | 3 | 5 | 1 | MO | 1 |
| 4 | PF1 | 3 | 5 | 1 | MO | 1 |
| 4 | PF1 | 3 | 5 | 1 | MO | 1 |
| 4 | PF1 | 3 | 5 | 11 | Pup | 1 |
| 4 | PF1 | 3 | 5 | 12 | Pup | 1 |
| 4 | PF1 | 3 | 5 | 9 | Pup | 1 |
| 4 | PF1 | 3 | 5 | 9 | Pup | 1 |
| 4 | PF1 | 3 | 5 | 12 | Pup | 1 |
| 4 | PF1 | 3 | 5 | 1 | Pup | 1 |
| 4 | PF1 | 3 | 5 | 2 | MO | 1 |
| 4 | PF1 | 3 | 5 | 2 | MO | 1 |
| 4 | PF1 | 3 | 5 | 2 | MO | 1 |
| 4 | PF1 | 3 | 5 | 2 | MO | 1 |
| 4 | PF1 | 3 | 5 | 2 | MO | 1 |
| 4 | PF1 | 3 | 5 | 12 | MO | 1 |
| 4 | PF1 | 3 | 5 | 12 | MO | 1 |
| 4 | PF1 | 3 | 5 | 12 | MO | 1 |
| 4 | PF1 | 3 | 5 | 8 | MO | 1 |
| 4 | PF1 | 3 | 5 | 12 | MO | 1 |
| 4 | PF1 | 3 | 5 | 16 | MO | 1 |
| 4 | PF1 | 3 | 5 | 11 | MO | 1 |
| 4 | PF1 | 3 | 5 | 15 | MO | 1 |
| 4 | PF1 | 3 | 5 | 11 | MO | 1 |
| 4 | PF1 | 3 | 5 | 16 | MO | 1 |
| 4 | PF1 | 3 | 5 | 6 | Pup | 1 |
| 4 | PF1 | 3 | 5 | 2 | Pup | 1 |
| 4 | PF1 | 3 | 5 | 2 | Pup | 1 |
| 4 | PF1 | 3 | 5 | 2 | Pup | 1 |
| 4 | PF1 | 3 | 5 | 2 | Pup | 1 |
| 4 | PF1 | 3 | 5 | 2 | Pup | 1 |
| 4 | PF1 | 3 | 5 | 3 | MO | 1 |
| 4 | PF1 | 3 | 5 | 3 | MO | 1 |
| 4 | PF1 | 3 | 5 | 3 | MO | 1 |
| 4 | PF1 | 3 | 5 | 3 | MO | 1 |
| 4 | PF1 | 3 | 5 | 3 | MO | 1 |
| 4 | PF1 | 3 | 5 | 8 | MO | 1 |
| 4 | PF1 | 3 | 5 | 10 | MO | 1 |
| 4 | PF1 | 3 | 5 | 7 | MO | 1 |
| 4 | PF1 | 3 | 5 | 8 | MO | 1 |
| 4 | PF1 | 3 | 5 | 7 | MO | 1 |
| 4 | RS1 | 3 | 2 | 4 | MO | 1 |
| 4 | RS1 | 3 | 2 | 4 | MO | 1 |
| 4 | RS1 | 3 | 2 | 7 | MO | 1 |
| 4 | RS1 | 3 | 2 | 7 | MO | 1 |
| 4 | RS1 | 3 | 2 | 1 | Pup | 1 |
| 4 | RS1 | 3 | 2 | 3 | MO | 1 |
| 4 | RS1 | 3 | 2 | 2 | MO | 1 |
| 4 | RS1 | 3 | 2 | 4 | Pup | 1 |
| 4 | RS1 | 3 | 2 | 11 | Pup | 1 |
| 4 | RS1 | 3 | 2 | 10 | Pup | 1 |
| 4 | RS1 | 3 | 2 | 8 | Pup | 1 |
| 4 | RS1 | 3 | 2 | 8 | Pup | 1 |
| 4 | RS1 | 3 | 2 | 4 | Pup | 1 |
| 4 | RS1 | 3 | 2 | 1 | Pup | 1 |
| 4 | RS1 | 3 | 2 | 2 | Pup | 1 |
| 4 | RS1 | 3 | 2 | 9 | Pup | 1 |
| 4 | RS1 | 3 | 2 | 9 | Pup | 1 |
| 4 | RS1 | 3 | 2 | 10 | MO | 1 |
| 4 | RS1 | 3 | 2 | 7 | MO | 1 |
| 4 | RS1 | 3 | 2 | 1 | Pup | 1 |
| 4 | RS1 | 3 | 2 | 5 | Pup | 1 |
| 4 | RS1 | 3 | 2 | 1 | Pup | 1 |
| 4 | RS1 | 3 | 2 | 11 | Pup | 1 |
| 4 | RS1 | 3 | 2 | 10 | Pup | 1 |
| 4 | RS1 | 3 | 2 | 6 | MO | 1 |
| 4 | RS1 | 3 | 2 | 6 | MO | 1 |
| 4 | RS1 | 3 | 2 | 3 | Pup | 1 |
| 4 | RS1 | 3 | 2 | 3 | Pup | 1 |
| 4 | RS2 | 3 | 4 | 15 | MO | 1 |
| 4 | RS2 | 3 | 4 | 15 | MO | 1 |
| 4 | RS2 | 3 | 4 | 15 | MO | 1 |
| 4 | RS2 | 3 | 4 | 15 | MO | 1 |
| 4 | RS2 | 3 | 4 | 4 | MO | 1 |
| 4 | RS2 | 3 | 4 | 4 | MO | 1 |
| 4 | RS2 | 3 | 4 | 4 | MO | 1 |
| 4 | RS2 | 3 | 4 | 4 | MO | 1 |
| 4 | RS2 | 3 | 4 | 1 | MO | 1 |
| 4 | RS2 | 3 | 4 | 1 | MO | 1 |
| 4 | RS2 | 3 | 4 | 1 | MO | 1 |
| 4 | RS2 | 3 | 4 | 1 | MO | 1 |
| 4 | RS2 | 3 | 4 | 1 | MO | 1 |
| 4 | RS2 | 3 | 4 | 1 | MO | 1 |
| 4 | RS2 | 3 | 4 | 1 | MO | 1 |
| 4 | RS2 | 3 | 4 | 1 | MO | 1 |
| 4 | RS2 | 3 | 4 | 7 | Pup | 1 |
| 4 | RS2 | 3 | 4 | 7 | Pup | 1 |
| 4 | RS2 | 3 | 4 | 7 | Pup | 1 |
| 4 | RS2 | 3 | 4 | 7 | Pup | 1 |
| 4 | RS2 | 3 | 4 | 2 | MO | 1 |
| 4 | RS2 | 3 | 4 | 2 | MO | 1 |
| 4 | RS2 | 3 | 4 | 2 | MO | 1 |
| 4 | RS2 | 3 | 4 | 2 | MO | 1 |
| 4 | RS2 | 3 | 4 | 10 | Pup | 1 |
| 4 | RS2 | 3 | 4 | 10 | Pup | 1 |
| 4 | RS2 | 3 | 4 | 10 | Pup | 1 |
| 4 | RS2 | 3 | 4 | 10 | Pup | 1 |
| 4 | RS2 | 3 | 4 | 20 | Pup | 1 |
| 4 | RS2 | 3 | 4 | 20 | Pup | 1 |
| 4 | RS2 | 3 | 4 | 20 | Pup | 1 |
| 4 | RS2 | 3 | 4 | 20 | Pup | 1 |
| 4 | RS2 | 3 | 4 | 15 | MO | 1 |
| 4 | RS2 | 3 | 4 | 15 | MO | 1 |
| 4 | RS2 | 3 | 4 | 15 | MO | 1 |
| 4 | RS2 | 3 | 4 | 15 | MO | 1 |
| 4 | RS2 | 3 | 4 | 9 | MO | 1 |
| 4 | RS2 | 3 | 4 | 9 | MO | 1 |
| 4 | RS2 | 3 | 4 | 9 | MO | 1 |
| 4 | RS2 | 3 | 4 | 9 | MO | 1 |
| 4 | RS2 | 3 | 4 | 11 | MO | 1 |
| 4 | RS2 | 3 | 4 | 11 | MO | 1 |
| 4 | RS2 | 3 | 4 | 11 | MO | 1 |
| 4 | RS2 | 3 | 4 | 11 | MO | 1 |
| 4 | RS2 | 3 | 4 | 6 | MO | 1 |
| 4 | RS2 | 3 | 4 | 6 | MO | 1 |
| 4 | RS2 | 3 | 4 | 6 | MO | 1 |
| 4 | RS2 | 3 | 4 | 6 | MO | 1 |
| 4 | RS2 | 3 | 4 | 7 | Pup | 1 |
| 4 | RS2 | 3 | 4 | 7 | Pup | 1 |
| 4 | RS3 | 3 | 2 | 8 | MO | 1 |
| 4 | RS3 | 3 | 2 | 14 | Pup | 1 |
| 4 | RS3 | 3 | 2 | 6 | Pup | 1 |
| 4 | RS3 | 3 | 2 | 2 | Pup | 1 |
| 4 | RS3 | 3 | 2 | 2 | Pup | 1 |
| 4 | RS3 | 3 | 2 | 3 | Pup | 1 |
| 4 | RS3 | 3 | 2 | 3 | Pup | 1 |
| 4 | RS3 | 3 | 2 | 14 | MO | 1 |
| 4 | RS3 | 3 | 2 | 14 | MO | 1 |
| 4 | RS3 | 3 | 2 | 17 | Pup | 1 |
| 4 | RS3 | 3 | 2 | 16 | Pup | 1 |
| 4 | RS3 | 3 | 2 | 10 | Pup | 1 |
| 4 | RS3 | 3 | 2 | 7 | Pup | 1 |
| 4 | RS3 | 3 | 2 | 8 | Pup | 1 |
| 4 | RS3 | 3 | 2 | 8 | Pup | 1 |
| 4 | RS3 | 3 | 2 | 4 | Pup | 1 |
| 4 | RS3 | 3 | 2 | 4 | Pup | 1 |
| 4 | RS3 | 3 | 2 | 9 | MO | 1 |
| 4 | RS3 | 3 | 2 | 9 | MO | 1 |
| 4 | RS3 | 3 | 2 | 3 | Pup | 1 |
| 4 | RS3 | 3 | 2 | 3 | Pup | 1 |
| 4 | RS3 | 3 | 2 | 14 | Pup | 1 |
| 4 | RS3 | 3 | 2 | 8 | Pup | 1 |
| 4 | RS3 | 3 | 2 | 5 | Pup | 1 |
| 4 | RS3 | 3 | 2 | 7 | Pup | 1 |
| 5 | BRN | 3 | 2 | 6 | MO | 1 |
| 5 | BRN | 3 | 2 | 5 | Pup | 1 |
| 5 | BRN | 3 | 2 | 7 | Pup | 1 |
| 5 | BRN | 3 | 2 | 1 | Pup | 1 |
| 5 | BRN | 3 | 2 | 7 | Pup | 1 |
| 5 | BRN | 3 | 2 | 0.5 | Pup | 1 |
| 5 | BRN | 3 | 2 | 2 | Pup | 1 |
| 5 | BRN | 3 | 2 | 0.5 | Pup | 1 |
| 5 | BRN | 3 | 2 | 0.5 | Pup | 1 |
| 5 | BRN | 3 | 2 | 1 | Pup | 1 |
| 5 | BRN | 3 | 2 | 0.5 | Pup | 1 |
| 5 | BRN | 3 | 2 | 0.5 | Pup | 1 |
| 5 | BRN | 3 | 2 | 1 | Pup | 1 |
| 5 | BRN | 3 | 2 | 3 | Pup | 1 |
| 5 | BRN | 3 | 2 | 2 | Pup | 1 |
| 5 | BRN | 3 | 2 | 5 | Pup | 1 |
| 5 | BRN | 3 | 2 | 9 | Pup | 1 |
| 5 | BRN | 3 | 2 | 1 | Pup | 1 |
| 5 | BRN | 3 | 2 | 1 | Pup | 1 |
| 5 | BRN | 3 | 2 | 7 | Pup | 1 |
| 5 | BRN | 3 | 2 | 5 | Pup | 1 |
| 5 | BRN | 3 | 2 | 6 | MO | 1 |
| 5 | BRN | 3 | 2 | 10 | MO | 1 |
| 5 | BRN | 3 | 2 | 7 | MO | 1 |
| 5 | BRN | 3 | 2 | 7 | MO | 1 |
| 5 | BRN | 3 | 2 | 8 | Pup | 1 |
| 5 | BRN | 3 | 2 | 8 | Pup | 1 |
| 5 | KTI | 3 | 2 | 10 | MO | 1 |
| 5 | KTI | 3 | 2 | 10 | MO | 1 |
| 5 | KTI | 3 | 2 | 8 | MO | 1 |
| 5 | KTI | 3 | 2 | 5 | MO | 1 |
| 5 | KTI | 3 | 2 | 1 | Pup | 1 |
| 5 | KTI | 3 | 2 | 9 | Pup | 1 |
| 5 | KTI | 3 | 2 | 5 | Pup | 1 |
| 5 | KTI | 3 | 2 | 7 | Pup | 1 |
| 5 | KTI | 3 | 2 | 7 | Pup | 1 |
| 5 | KTI | 3 | 2 | 5 | Pup | 1 |
| 5 | KTI | 3 | 2 | 4 | Pup | 1 |
| 5 | KTI | 3 | 2 | 4 | MO | 1 |
| 5 | KTI | 3 | 2 | 4 | MO | 1 |
| 5 | KTI | 3 | 2 | 2 | Pup | 1 |
| 5 | KTI | 3 | 2 | 9 | Pup | 1 |
| 5 | KTI | 3 | 2 | 9 | Pup | 1 |
| 5 | BBR | 3 | 4 | 6 | MO | 1 |
| 5 | BBR | 3 | 4 | 6 | MO | 1 |
| 5 | BBR | 3 | 4 | 4 | Pup | 1 |
| 5 | BBR | 3 | 4 | 8 | Pup | 1 |
| 5 | BBR | 3 | 4 | 12 | Pup | 1 |
| 5 | BBR | 3 | 4 | 14 | Pup | 1 |
| 5 | BBR | 3 | 4 | 8 | Pup | 1 |
| 5 | BBR | 3 | 4 | 0.5 | Pup | 1 |
| 5 | BBR | 3 | 4 | 6 | MO | 1 |
| 5 | BBR | 3 | 4 | 6 | MO | 1 |
| 5 | BBR | 3 | 4 | 3 | MO | 1 |
| 5 | BBR | 3 | 4 | 2 | MO | 1 |
| 5 | BBR | 3 | 4 | 4 | Pup | 1 |
| 5 | BBR | 3 | 4 | 4 | Pup | 1 |
| 5 | BBR | 3 | 4 | 7 | Pup | 1 |
| 5 | BBR | 3 | 4 | 8 | Pup | 1 |
| 5 | BBR | 3 | 4 | 3 | MO | 1 |
| 5 | BBR | 3 | 4 | 9 | MO | 1 |
| 5 | BBR | 3 | 4 | 10 | MO | 1 |
| 5 | BBR | 3 | 4 | 8 | MO | 1 |
| 5 | BBR | 3 | 4 | 9 | Pup | 1 |
| 5 | BBR | 3 | 4 | 3 | Pup | 1 |
| 5 | BBR | 3 | 4 | 2 | Pup | 1 |
| 5 | BBR | 3 | 4 | 4 | Pup | 1 |
| 5 | BBR | 3 | 4 | 3 | Pup | 1 |
| 5 | BBR | 3 | 4 | 5 | MO | 1 |
| 5 | BBR | 3 | 4 | 5 | MO | 1 |
| 5 | BBR | 3 | 4 | 5 | MO | 1 |
| 5 | BBR | 3 | 4 | 5 | MO | 1 |
| 5 | BBR | 3 | 4 | 10 | MO | 1 |
| 5 | BBR | 3 | 4 | 8 | MO | 1 |
| 5 | BBR | 3 | 4 | 7 | MO | 1 |
| 5 | BBR | 3 | 4 | 10 | MO | 1 |
| 5 | BBR | 3 | 4 | 8 | MO | 1 |
| 5 | BBR | 3 | 4 | 5 | MO | 1 |
| 5 | BBR | 3 | 4 | 6 | MO | 1 |
| 5 | BBR | 3 | 4 | 8 | MO | 1 |
| 5 | BBR | 3 | 4 | 1 | Pup | 1 |
| 5 | BBR | 3 | 4 | 0.5 | Pup | 1 |
| 5 | BBR | 3 | 4 | 0.5 | Pup | 1 |
| 5 | BBR | 3 | 4 | 1 | Pup | 1 |
| 5 | BBR | 3 | 4 | 8 | MO | 1 |
| 5 | BBR | 3 | 4 | 8 | MO | 1 |
| 5 | BBR | 3 | 4 | 8 | MO | 1 |
| 5 | BBR | 3 | 4 | 8 | MO | 1 |
| 5 | BBR | 3 | 4 | 2 | MO | 1 |
| 5 | BBR | 3 | 4 | 2 | MO | 1 |
| 5 | BBR | 3 | 4 | 2 | MO | 1 |
| 5 | BBR | 3 | 4 | 2 | MO | 1 |
| 5 | WHI | 3 | 2 | 6 | MO | 1 |
| 5 | WHI | 3 | 2 | 9 | Pup | 1 |
| 5 | WHI | 3 | 2 | 5 | Pup | 1 |
| 5 | WHI | 3 | 2 | 6 | MO | 1 |
| 5 | WHI | 3 | 2 | 6 | MO | 1 |
| 5 | WHI | 3 | 2 | 9 | MO | 1 |
| 5 | WHI | 3 | 2 | 9 | MO | 1 |
| 5 | WHI | 3 | 2 | 2 | Pup | 1 |
| 5 | WHI | 3 | 2 | 9 | Pup | 1 |
| 5 | WHI | 3 | 2 | 8 | Pup | 1 |
| 5 | WHI | 3 | 2 | 2 | MO | 1 |
| 5 | WHI | 3 | 2 | 2 | MO | 1 |
| 5 | WHI | 3 | 2 | 4 | Pup | 1 |
| 5 | WHI | 3 | 2 | 3 | Pup | 1 |
| 5 | WHI | 3 | 2 | 1 | Pup | 1 |
| 5 | WHI | 3 | 2 | 7 | MO | 1 |
| 5 | WHI | 3 | 2 | 8 | MO | 1 |
| 5 | RS5 | 3 | 3 | 7 | MO | 1 |
| 5 | RS5 | 3 | 3 | 4 | MO | 1 |
| 5 | RS5 | 3 | 3 | 10 | MO | 1 |
| 5 | RS5 | 3 | 3 | 5 | Pup | 1 |
| 5 | RS5 | 3 | 3 | 6 | MO | 1 |
| 5 | RS5 | 3 | 3 | 6 | MO | 1 |
| 5 | RS5 | 3 | 3 | 6 | MO | 1 |
| 5 | RS5 | 3 | 3 | 2 | Pup | 1 |
| 5 | RS5 | 3 | 3 | 1 | Pup | 1 |
| 5 | RS5 | 3 | 3 | 2 | Pup | 1 |
| 5 | RS5 | 3 | 3 | 8 | MO | 1 |
| 5 | RS5 | 3 | 3 | 7 | MO | 1 |
| 5 | RS5 | 3 | 3 | 6 | MO | 1 |
| 5 | RS5 | 3 | 3 | 4 | MO | 1 |
| 5 | RS5 | 3 | 3 | 4 | MO | 1 |
| 5 | RS5 | 3 | 3 | 2 | MO | 1 |
| 5 | RS5 | 3 | 3 | 3 | Pup | 1 |
| 5 | RS5 | 3 | 3 | 1 | Pup | 1 |
| 5 | RS5 | 3 | 3 | 1 | Pup | 1 |
| 5 | RS5 | 3 | 3 | 1 | Pup | 1 |
| 5 | RS5 | 3 | 3 | 3 | Pup | 1 |
| 5 | RS5 | 3 | 3 | 16 | Pup | 1 |
| 5 | RS5 | 3 | 3 | 2 | Pup | 1 |
| 5 | RS5 | 3 | 3 | 6 | Pup | 1 |
| 5 | RS5 | 3 | 3 | 6 | Pup | 1 |
| 5 | RS5 | 3 | 3 | 0.5 | Pup | 1 |
| 5 | RS5 | 3 | 3 | 1 | Pup | 1 |
| 5 | RS5 | 3 | 3 | 0.5 | Pup | 1 |
| 5 | RS5 | 3 | 3 | 10 | Pup | 1 |
| 5 | RS5 | 3 | 3 | 3 | Pup | 1 |
| 5 | RS5 | 3 | 3 | 2 | Pup | 1 |
| 5 | RS5 | 3 | 3 | 4 | MO | 1 |
| 5 | RS5 | 3 | 3 | 5 | MO | 1 |
| 5 | RS5 | 3 | 3 | 5 | MO | 1 |
| 5 | RS5 | 3 | 3 | 3 | Pup | 1 |
| 5 | RS5 | 3 | 3 | 2 | Pup | 1 |
| 5 | RS5 | 3 | 3 | 2 | Pup | 1 |
| 4 | CAN | 4 | 6 | 8 | MO | 1 |
| 4 | CAN | 4 | 6 | 8 | MO | 1 |
| 4 | CAN | 4 | 6 | 8 | MO | 1 |
| 4 | CAN | 4 | 6 | 8 | MO | 1 |
| 4 | CAN | 4 | 6 | 8 | MO | 1 |
| 4 | CAN | 4 | 6 | 8 | MO | 1 |
| 4 | CAN | 4 | 6 | 8 | MO | 1 |
| 4 | CAN | 4 | 6 | 8 | MO | 1 |
| 4 | CAN | 4 | 6 | 8 | MO | 1 |
| 4 | CAN | 4 | 6 | 8 | MO | 1 |
| 4 | CAN | 4 | 6 | 8 | MO | 1 |
| 4 | CAN | 4 | 6 | 8 | MO | 1 |
| 4 | CAN | 4 | 6 | 9 | MO | 1 |
| 4 | CAN | 4 | 6 | 9 | MO | 1 |
| 4 | CAN | 4 | 6 | 9 | MO | 1 |
| 4 | CAN | 4 | 6 | 9 | MO | 1 |
| 4 | CAN | 4 | 6 | 9 | MO | 1 |
| 4 | CAN | 4 | 6 | 9 | MO | 1 |
| 4 | MDB | 4 | 5 | 1 | MO | 1 |
| 4 | MDB | 4 | 5 | 1 | MO | 1 |
| 4 | MDB | 4 | 5 | 1 | MO | 1 |
| 4 | MDB | 4 | 5 | 1 | MO | 1 |
| 4 | MDB | 4 | 5 | 1 | MO | 1 |
| 4 | MDB | 4 | 5 | 6 | MO | 1 |
| 4 | MDB | 4 | 5 | 6 | MO | 1 |
| 4 | MDB | 4 | 5 | 6 | MO | 1 |
| 4 | MDB | 4 | 5 | 6 | MO | 1 |
| 4 | MDB | 4 | 5 | 6 | MO | 1 |
| 4 | MDB | 4 | 5 | 8 | Pup | 1 |
| 4 | MDB | 4 | 5 | 6 | Pup | 1 |
| 4 | MDB | 4 | 5 | 9 | MO | 1 |
| 4 | MDB | 4 | 5 | 9 | MO | 1 |
| 4 | MDB | 4 | 5 | 9 | MO | 1 |
| 4 | MDB | 4 | 5 | 9 | MO | 1 |

| | | | | | | | |
|---|---|---|---|---|---|---|---|
| 4 | MDB | 4 | 5 | 9 | MO | 1 | |
| 4 | MDB | 4 | 5 | 4 | MO | 1 | |
| 4 | MDB | 4 | 5 | 4 | MO | 1 | |
| 4 | MDB | 4 | 5 | 4 | MO | 1 | |
| 4 | MDB | 4 | 5 | 4 | MO | 1 | |
| 4 | MDB | 4 | 5 | 4 | MO | 1 | |
| 4 | MDB | 4 | 5 | 2 | MO | 1 | |
| 4 | MDB | 4 | 5 | 2 | MO | 1 | |
| 4 | MDB | 4 | 5 | 1 | Pup | 1 | |
| 4 | MDB | 4 | 5 | 1 | Pup | 1 | |
| 4 | MDB | 4 | 5 | 1 | Pup | 1 | |
| 4 | MDB | 4 | 5 | 2 | Pup | 1 | |
| 4 | MDB | 4 | 5 | 2 | Pup | 1 | |
| 4 | MDB | 4 | 5 | 4 | Pup | 1 | |
| 4 | PF1 | 4 | 5 | 2 | Pup | 1 | |
| 4 | PF1 | 4 | 5 | 2 | Pup | 1 | |
| 4 | PF1 | 4 | 5 | 2 | Pup | 1 | |
| 4 | PF1 | 4 | 5 | 2 | Pup | 1 | |
| 4 | PF1 | 4 | 5 | 2 | Pup | 1 | |
| 4 | PF1 | 4 | 5 | 1 | Pup | 1 | |
| 4 | PF1 | 4 | 5 | 1 | Pup | 1 | |
| 4 | PF1 | 4 | 5 | 1 | Pup | 1 | |
| 4 | PF1 | 4 | 5 | 3 | Pup | 1 | |
| 4 | PF1 | 4 | 5 | 3 | Pup | 1 | |
| 4 | PF1 | 4 | 5 | 0.5 | Pup | | 1 |
| 4 | PF1 | 4 | 5 | 0.5 | Pup | | 1 |
| 4 | PF1 | 4 | 5 | 2 | Pup | 1 | |
| 4 | PF1 | 4 | 5 | 2 | Pup | 1 | |
| 4 | PF1 | 4 | 5 | 2 | Pup | 1 | |
| 4 | PF1 | 4 | 5 | 3 | Pup | 1 | |
| 4 | PF1 | 4 | 5 | 3 | Pup | 1 | |
| 4 | PF1 | 4 | 5 | 3 | Pup | 1 | |
| 4 | PF1 | 4 | 5 | 3 | Pup | 1 | |
| 4 | PF1 | 4 | 5 | 3 | Pup | 1 | |
| 4 | PF1 | 4 | 5 | 3 | Pup | 1 | |
| 4 | PF1 | 4 | 5 | 3 | Pup | 1 | |
| 4 | PF1 | 4 | 5 | 6 | Pup | 1 | |
| 4 | PF1 | 4 | 5 | 6 | Pup | 1 | |
| 4 | PF1 | 4 | 5 | 6 | Pup | 1 | |
| 4 | PF1 | 4 | 5 | 6 | Pup | 1 | |
| 4 | PF1 | 4 | 5 | 6 | Pup | 1 | |
| 4 | PF1 | 4 | 5 | 0.5 | Pup | 1 | |

| | | | | | | |
|---|---|---|---|---|---|---|
| 4 | PF1 | 4 | 5 | 1 | Pup | 1 |
| 4 | PF1 | 4 | 5 | 1 | Pup | 1 |
| 4 | PF1 | 4 | 5 | 3 | Pup | 1 |
| 4 | PF1 | 4 | 5 | 3 | Pup | 1 |
| 4 | PF1 | 4 | 5 | 3 | Pup | 1 |
| 4 | PF1 | 4 | 5 | 2 | Pup | 1 |
| 4 | PF1 | 4 | 5 | 2 | Pup | 1 |
| 4 | PF1 | 4 | 5 | 2 | Pup | 1 |
| 4 | PF1 | 4 | 5 | 3 | Pup | 1 |
| 4 | PF1 | 4 | 5 | 3 | Pup | 1 |
| 4 | PF1 | 4 | 5 | 3 | Pup | 1 |
| 4 | RS1 | 4 | 2 | 8 | MO | 1 |
| 4 | RS1 | 4 | 2 | 8 | MO | 1 |
| 4 | RS1 | 4 | 2 | 1 | Pup | 1 |
| 4 | RS1 | 4 | 2 | 7 | MO | 1 |
| 4 | RS1 | 4 | 2 | 7 | MO | 1 |
| 4 | RS1 | 4 | 2 | 1 | Pup | 1 |
| 4 | RS1 | 4 | 2 | 4 | MO | 1 |
| 4 | RS1 | 4 | 2 | 3 | MO | 1 |
| 4 | RS1 | 4 | 2 | 9 | MO | 1 |
| 4 | RS1 | 4 | 2 | 7 | MO | 1 |
| 4 | RS1 | 4 | 2 | 3 | Pup | 1 |
| 4 | RS1 | 4 | 2 | 2 | Pup | 1 |
| 4 | RS1 | 4 | 2 | 3 | MO | 1 |
| 4 | RS1 | 4 | 2 | 3 | MO | 1 |
| 4 | RS1 | 4 | 2 | 7 | Pup | 1 |
| 4 | RS1 | 4 | 2 | 7 | Pup | 1 |
| 4 | RS1 | 4 | 2 | 7 | Pup | 1 |
| 4 | RS1 | 4 | 2 | 8 | Pup | 1 |
| 4 | RS2 | 4 | 4 | 8 | MO | 1 |
| 4 | RS2 | 4 | 4 | 8 | MO | 1 |
| 4 | RS2 | 4 | 4 | 8 | MO | 1 |
| 4 | RS2 | 4 | 4 | 8 | MO | 1 |
| 4 | RS2 | 4 | 4 | 9 | Pup | 1 |
| 4 | RS2 | 4 | 4 | 9 | Pup | 1 |
| 4 | RS2 | 4 | 4 | 9 | Pup | 1 |
| 4 | RS2 | 4 | 4 | 9 | Pup | 1 |
| 4 | RS2 | 4 | 4 | 9 | Pup | 1 |
| 4 | RS2 | 4 | 4 | 9 | Pup | 1 |
| 4 | RS2 | 4 | 4 | 1 | Pup | 1 |
| 4 | RS2 | 4 | 4 | 1 | Pup | 1 |
| 4 | RS2 | 4 | 4 | 1 | Pup | 1 |
| 4 | RS2 | 4 | 4 | 1 | Pup | 1 |
| 4 | RS2 | 4 | 4 | 1 | Pup | 1 |
| 4 | RS2 | 4 | 4 | 1 | Pup | 1 |
| 4 | RS2 | 4 | 4 | 9 | Pup | 1 |
| 4 | RS2 | 4 | 4 | 9 | Pup | 1 |
| 4 | RS2 | 4 | 4 | 9 | Pup | 1 |
| 4 | RS2 | 4 | 4 | 9 | Pup | 1 |
| 4 | RS2 | 4 | 4 | 9 | Pup | 1 |
| 4 | RS2 | 4 | 4 | 9 | Pup | 1 |
| 4 | RS2 | 4 | 4 | 2 | Pup | 1 |
| 4 | RS2 | 4 | 4 | 2 | Pup | 1 |
| 4 | RS2 | 4 | 4 | 2 | Pup | 1 |
| 4 | RS2 | 4 | 4 | 2 | Pup | 1 |
| 4 | RS2 | 4 | 4 | 7 | Pup | 1 |
| 4 | RS2 | 4 | 4 | 7 | Pup | 1 |
| 4 | RS2 | 4 | 4 | 1 | Pup | 1 |
| 4 | RS2 | 4 | 4 | 1 | Pup | 1 |
| 4 | RS2 | 4 | 4 | 1 | Pup | 1 |
| 4 | RS2 | 4 | 4 | 1 | Pup | 1 |
| 4 | RS2 | 4 | 4 | 1 | Pup | 1 |
| 4 | RS2 | 4 | 4 | 1 | Pup | 1 |
| 4 | RS2 | 4 | 4 | 1 | Pup | 1 |
| 4 | RS2 | 4 | 4 | 1 | Pup | 1 |
| 4 | RS2 | 4 | 4 | 1 | Pup | 1 |
| 4 | RS2 | 4 | 4 | 1 | Pup | 1 |
| 4 | RS2 | 4 | 4 | 0.5 | Pup | 1 |
| 4 | RS2 | 4 | 4 | 0.5 | Pup | 1 |
| 4 | RS2 | 4 | 4 | 1 | Pup | 1 |
| 4 | RS2 | 4 | 4 | 1 | Pup | 1 |
| 4 | RS2 | 4 | 4 | 2 | Pup | 1 |
| 4 | RS2 | 4 | 4 | 2 | Pup | 1 |
| 4 | RS2 | 4 | 4 | 2 | Pup | 1 |
| 4 | RS2 | 4 | 4 | 1 | Pup | 1 |
| 4 | RS2 | 4 | 4 | 1 | Pup | 1 |
| 4 | RS2 | 4 | 4 | 1 | Pup | 1 |
| 4 | RS2 | 4 | 4 | 0.5 | Pup | 1 |
| 4 | RS2 | 4 | 4 | 0.5 | Pup | 1 |
| 4 | RS2 | 4 | 4 | 0.5 | Pup | 1 |
| 4 | RS2 | 4 | 4 | 0.5 | Pup | 1 |
| 4 | RS2 | 4 | 4 | 0.5 | Pup | 1 |
| 4 | RS2 | 4 | 4 | 0.5 | Pup | 1 |
| 4 | RS2 | 4 | 4 | 2 | Pup | 1 |
| 4 | RS2 | 4 | 4 | 2 | Pup | 1 |
| 4 | RS2 | 4 | 4 | 0.5 | Pup | 1 |
| 4 | RS2 | 4 | 4 | 0.5 | Pup | 1 |
| 4 | RS2 | 4 | 4 | 0.5 | Pup | 1 |
| 4 | RS2 | 4 | 4 | 1 | Pup | 1 |
| 4 | RS2 | 4 | 4 | 5 | Pup | 1 |
| 4 | RS2 | 4 | 4 | 1 | Pup | 1 |
| 4 | RS2 | 4 | 4 | 8 | MO | 1 |
| 4 | RS2 | 4 | 4 | 8 | MO | 1 |
| 4 | RS2 | 4 | 4 | 8 | MO | 1 |
| 4 | RS2 | 4 | 4 | 8 | MO | 1 |
| 4 | RS2 | 4 | 4 | 6 | MO | 1 |
| 4 | RS2 | 4 | 4 | 6 | MO | 1 |
| 4 | RS2 | 4 | 4 | 6 | MO | 1 |
| 4 | RS2 | 4 | 4 | 6 | MO | 1 |
| 4 | RS3 | 4 | 2 | 6 | Pup | 1 |
| 4 | RS3 | 4 | 2 | 2 | Pup | 1 |
| 4 | RS3 | 4 | 2 | 9 | Pup | 1 |
| 4 | RS3 | 4 | 2 | 5 | MO | 1 |
| 4 | RS3 | 4 | 2 | 5 | MO | 1 |
| 4 | RS3 | 4 | 2 | 7 | Pup | 1 |
| 4 | RS3 | 4 | 2 | 4 | Pup | 1 |
| 4 | RS3 | 4 | 2 | 1 | MO | 1 |
| 4 | RS3 | 4 | 2 | 1 | MO | 1 |
| 4 | RS3 | 4 | 2 | 16 | Pup | 1 |
| 4 | RS3 | 4 | 2 | 17 | Pup | 1 |
| 4 | RS3 | 4 | 2 | 2 | Pup | 1 |
| 4 | RS3 | 4 | 2 | 1 | Pup | 1 |
| 4 | RS3 | 4 | 2 | 1 | Pup | 1 |
| 5 | BRN | 4 | 2 | 5 | MO | 1 |
| 5 | BRN | 4 | 2 | 5 | MO | 1 |
| 5 | BRN | 4 | 2 | 3 | Pup | 1 |
| 5 | BRN | 4 | 2 | 2 | Pup | 1 |
| 5 | BRN | 4 | 2 | 2 | Pup | 1 |
| 5 | BRN | 4 | 2 | 2 | Pup | 1 |
| 5 | BRN | 4 | 2 | 4 | Pup | 1 |
| 5 | BRN | 4 | 2 | 3 | Pup | 1 |
| 5 | BRN | 4 | 2 | 2 | Pup | 1 |
| 5 | BRN | 4 | 2 | 1 | Pup | 1 |
| 5 | BRN | 4 | 2 | 0.5 | Pup | 1 |
| 5 | BRN | 4 | 2 | 1 | Pup | 1 |
| 5 | BRN | 4 | 2 | 1 | Pup | 1 |
| 5 | BRN | 4 | 2 | 0.5 | Pup | 1 |
| 5 | KTI | 4 | 2 | 4 | MO | 1 |
| 5 | KTI | 4 | 2 | 4 | MO | 1 |
| 5 | KTI | 4 | 2 | 6 | MO | 1 |
| 5 | KTI | 4 | 2 | 7 | MO | 1 |
| 5 | KTI | 4 | 2 | 7 | MO | 1 |
| 5 | KTI | 4 | 2 | 7 | MO | 1 |
| 5 | KTI | 4 | 2 | 7 | MO | 1 |
| 5 | KTI | 4 | 2 | 8 | MO | 1 |
| 5 | KTI | 4 | 2 | 3 | Pup | 1 |
| 5 | KTI | 4 | 2 | 2 | Pup | 1 |
| 5 | KTI | 4 | 2 | 1 | Pup | 1 |
| 5 | KTI | 4 | 2 | 2 | Pup | 1 |
| 5 | KTI | 4 | 2 | 2 | Pup | 1 |
| 5 | KTI | 4 | 2 | 2 | Pup | 1 |
| 5 | KTI | 4 | 2 | 1 | Pup | 1 |
| 5 | KTI | 4 | 2 | 2 | Pup | 1 |
| 5 | KTI | 4 | 2 | 6 | Pup | 1 |
| 5 | KTI | 4 | 2 | 5 | Pup | 1 |
| 5 | KTI | 4 | 2 | 6 | Pup | 1 |
| 5 | KTI | 4 | 2 | 7 | Pup | 1 |
| 5 | KTI | 4 | 2 | 6 | Pup | 1 |
| 5 | KTI | 4 | 2 | 2 | Pup | 1 |
| 5 | KTI | 4 | 2 | 8 | Pup | 1 |
| 5 | KTI | 4 | 2 | 8 | Pup | 1 |
| 5 | KTI | 4 | 2 | 5 | MO | 1 |
| 5 | KTI | 4 | 2 | 5 | MO | 1 |
| 5 | KTI | 4 | 2 | 3 | Pup | 1 |
| 5 | KTI | 4 | 2 | 2 | Pup | 1 |
| 5 | KTI | 4 | 2 | 1 | Pup | 1 |
| 5 | KTI | 4 | 2 | 3 | Pup | 1 |
| 5 | KTI | 4 | 2 | 3 | Pup | 1 |
| 5 | KTI | 4 | 2 | 5 | MO | 1 |
| 5 | KTI | 4 | 2 | 5 | MO | 1 |
| 5 | KTI | 4 | 2 | 2 | Pup | 1 |
| 5 | KTI | 4 | 2 | 9 | Pup | 1 |
| 5 | KTI | 4 | 2 | 8 | Pup | 1 |
| 5 | KTI | 4 | 2 | 4 | MO | 1 |
| 5 | KTI | 4 | 2 | 4 | MO | 1 |
| 5 | KTI | 4 | 2 | 8 | Pup | 1 |
| 5 | KTI | 4 | 2 | 8 | Pup | 1 |
| 5 | BBR | 4 | 4 | 6 | MO | 1 |
| 5 | BBR | 4 | 4 | 6 | MO | 1 |
| 5 | BBR | 4 | 4 | 6 | MO | 1 |
| 5 | BBR | 4 | 4 | 6 | MO | 1 |
| 5 | BBR | 4 | 4 | 4 | MO | 1 |
| 5 | BBR | 4 | 4 | 6 | MO | 1 |
| 5 | BBR | 4 | 4 | 6 | MO | 1 |
| 5 | BBR | 4 | 4 | 6 | MO | 1 |
| 5 | BBR | 4 | 4 | 3 | Pup | 1 |
| 5 | BBR | 4 | 4 | 4 | Pup | 1 |
| 5 | BBR | 4 | 4 | 3 | Pup | 1 |
| 5 | BBR | 4 | 4 | 1 | Pup | 1 |
| 5 | BBR | 4 | 4 | 4 | MO | 1 |
| 5 | BBR | 4 | 4 | 4 | MO | 1 |
| 5 | BBR | 4 | 4 | 4 | MO | 1 |
| 5 | BBR | 4 | 4 | 4 | MO | 1 |
| 5 | BBR | 4 | 4 | 3 | MO | 1 |
| 5 | BBR | 4 | 4 | 3 | MO | 1 |
| 5 | BBR | 4 | 4 | 3 | MO | 1 |
| 5 | BBR | 4 | 4 | 3 | MO | 1 |
| 5 | BBR | 4 | 4 | 4 | Pup | 1 |
| 5 | BBR | 4 | 4 | 1 | Pup | 1 |
| 5 | BBR | 4 | 4 | 3 | Pup | 1 |
| 5 | BBR | 4 | 4 | 0.5 | Pup | 1 |
| 5 | BBR | 4 | 4 | 0.5 | Pup | 1 |
| 5 | BBR | 4 | 4 | 7 | MO | 1 |
| 5 | BBR | 4 | 4 | 7 | MO | 1 |
| 5 | BBR | 4 | 4 | 7 | MO | 1 |
| 5 | BBR | 4 | 4 | 7 | MO | 1 |
| 5 | BBR | 4 | 4 | 7 | MO | 1 |
| 5 | BBR | 4 | 4 | 7 | MO | 1 |
| 5 | BBR | 4 | 4 | 7 | MO | 1 |
| 5 | BBR | 4 | 4 | 7 | MO | 1 |
| 5 | WHI | 4 | 2 | 5 | MO | 1 |
| 5 | WHI | 4 | 2 | 10 | MO | 1 |
| 5 | WHI | 4 | 2 | 4 | Pup | 1 |
| 5 | WHI | 4 | 2 | 1 | Pup | 1 |
| 5 | WHI | 4 | 2 | 4 | Pup | 1 |
| 5 | WHI | 4 | 2 | 1 | Pup | 1 |
| 5 | WHI | 4 | 2 | 0.5 | Pup | 1 |
| 5 | WHI | 4 | 2 | 4 | MO | 1 |
| 5 | WHI | 4 | 2 | 6 | MO | 1 |
| 5 | WHI | 4 | 2 | 4 | Pup | 1 |
| 5 | WHI | 4 | 2 | 3 | Pup | 1 |
| 5 | WHI | 4 | 2 | 2 | Pup | 1 |
| 5 | WHI | 4 | 2 | 2 | Pup | 1 |
| 5 | WHI | 4 | 2 | 7 | MO | 1 |
| 5 | WHI | 4 | 2 | 8 | MO | 1 |
| 5 | WHI | 4 | 2 | 1 | Pup | 1 |
| 5 | WHI | 4 | 2 | 0.5 | Pup | 1 |
| 5 | WHI | 4 | 2 | 4 | Pup | 1 |
| 5 | WHI | 4 | 2 | 7 | Pup | 1 |
| 5 | WHI | 4 | 2 | 6 | MO | 1 |
| 5 | WHI | 4 | 2 | 7 | MO | 1 |
| 5 | WHI | 4 | 2 | 7 | MO | 1 |
| 5 | WHI | 4 | 2 | 8 | MO | 1 |
| 5 | WHI | 4 | 2 | 1 | Pup | 1 |
| 5 | WHI | 4 | 2 | 1 | Pup | 1 |
| 5 | WHI | 4 | 2 | 1 | Pup | 1 |
| 5 | WHI | 4 | 2 | 1 | Pup | 1 |
| 5 | WHI | 4 | 2 | 2 | Pup | 1 |
| 5 | WHI | 4 | 2 | 0.5 | Pup | 1 |
| 5 | WHI | 4 | 2 | 0.5 | Pup | 1 |
| 5 | WHI | 4 | 2 | 1 | Pup | 1 |
| 5 | WHI | 4 | 2 | 2 | Pup | 1 |
| 5 | WHI | 4 | 2 | 0.5 | Pup | 1 |
| 5 | WHI | 4 | 2 | 2 | Pup | 1 |
| 5 | WHI | 4 | 2 | 3 | Pup | 1 |
| 5 | WHI | 4 | 2 | 0.5 | Pup | 1 |
| 5 | WHI | 4 | 2 | 0.5 | Pup | 1 |
| 5 | WHI | 4 | 2 | 0.5 | Pup | 1 |
| 5 | WHI | 4 | 2 | 3 | Pup | 1 |
| 5 | WHI | 4 | 2 | 3 | Pup | 1 |
| 5 | WHI | 4 | 2 | 1 | Pup | 1 |
| 5 | WHI | 4 | 2 | 0.5 | Pup | 1 |
| 5 | WHI | 4 | 2 | 0.5 | Pup | 1 |
| 5 | WHI | 4 | 2 | 1 | Pup | 1 |
| 5 | WHI | 4 | 2 | 1 | Pup | 1 |
| 5 | WHI | 4 | 2 | 1 | Pup | 1 |
| 5 | WHI | 4 | 2 | 1 | Pup | 1 |
| 5 | WHI | 4 | 2 | 1 | Pup | 1 |
| 5 | WHI | 4 | 2 | 3 | Pup | 1 |
| 5 | WHI | 4 | 2 | 7 | Pup | 1 |
| 5 | WHI | 4 | 2 | 3 | Pup | 1 |
| 5 | WHI | 4 | 2 | 4 | Pup | 1 |
| 5 | WHI | 4 | 2 | 2 | Pup | 1 |
| 5 | RS5 | 4 | 3 | 12 | MO | 1 |
| 5 | RS5 | 4 | 3 | 9 | MO | 1 |
| 5 | RS5 | 4 | 3 | 13 | MO | 1 |
| 5 | RS5 | 4 | 3 | 10 | MO | 1 |
| 5 | RS5 | 4 | 3 | 9 | MO | 1 |
| 5 | RS5 | 4 | 3 | 11 | MO | 1 |
| 5 | RS5 | 4 | 3 | 2 | Pup | 1 |
| 5 | RS5 | 4 | 3 | 7 | MO | 1 |
| 5 | RS5 | 4 | 3 | 8 | MO | 1 |
| 5 | RS5 | 4 | 3 | 8 | Pup | 1 |
| 5 | RS5 | 4 | 3 | 2 | Pup | 1 |
| 5 | RS5 | 4 | 3 | 2 | Pup | 1 |
| 5 | RS5 | 4 | 3 | 10 | Pup | 1 |
| 5 | RS5 | 4 | 3 | 9 | Pup | 1 |
| 5 | RS5 | 4 | 3 | 10 | Pup | 1 |
| 5 | RS5 | 4 | 3 | 8 | MO | 1 |
| 5 | RS5 | 4 | 3 | 8 | MO | 1 |
| 5 | RS5 | 4 | 3 | 8 | MO | 1 |
| 4 | CAN | 5 | 6 | 7 | MO | 1 |
| 4 | CAN | 5 | 6 | 7 | MO | 1 |
| 4 | CAN | 5 | 6 | 7 | MO | 1 |
| 4 | CAN | 5 | 6 | 7 | MO | 1 |
| 4 | CAN | 5 | 6 | 7 | MO | 1 |
| 4 | CAN | 5 | 6 | 7 | MO | 1 |
| 4 | CAN | 5 | 6 | 9 | MO | 1 |
| 4 | CAN | 5 | 6 | 9 | MO | 1 |
| 4 | CAN | 5 | 6 | 9 | MO | 1 |
| 4 | CAN | 5 | 6 | 9 | MO | 1 |
| 4 | CAN | 5 | 6 | 6 | MO | 1 |
| 4 | CAN | 5 | 6 | 9 | MO | 1 |
| 4 | CAN | 5 | 6 | 14 | Pup | 1 |
| 4 | CAN | 5 | 6 | 14 | Pup | 1 |
| 4 | CAN | 5 | 6 | 12 | Pup | 1 |
| 4 | CAN | 5 | 6 | 12 | Pup | 1 |
| 4 | CAN | 5 | 6 | 12 | Pup | 1 |
| 4 | CAN | 5 | 6 | 12 | Pup | 1 |
| 4 | CAN | 5 | 6 | 8 | MO | 1 |
| 4 | CAN | 5 | 6 | 8 | MO | 1 |
| 4 | CAN | 5 | 6 | 7 | MO | 1 |
| 4 | CAN | 5 | 6 | 7 | MO | 1 |
| 4 | CAN | 5 | 6 | 6 | MO | 1 |
| 4 | CAN | 5 | 6 | 6 | MO | 1 |
| 4 | CAN | 5 | 6 | 2 | Pup | 1 |
| 4 | CAN | 5 | 6 | 2 | Pup | 1 |
| 4 | CAN | 5 | 6 | 2 | Pup | 1 |
| 4 | CAN | 5 | 6 | 2 | Pup | 1 |
| 4 | CAN | 5 | 6 | 2 | Pup | 1 |
| 4 | CAN | 5 | 6 | 2 | Pup | 1 |
| 4 | CAN | 5 | 6 | 2 | Pup | 1 |
| 4 | CAN | 5 | 6 | 1 | Pup | 1 |
| 4 | CAN | 5 | 6 | 8 | MO | 1 |
| 4 | CAN | 5 | 6 | 4 | MO | 1 |
| 4 | CAN | 5 | 6 | 10 | MO | 1 |
| 4 | CAN | 5 | 6 | 9 | MO | 1 |
| 4 | CAN | 5 | 6 | 9 | MO | 1 |
| 4 | CAN | 5 | 6 | 8 | MO | 1 |
| 4 | CAN | 5 | 6 | 3 | Pup | 1 |
| 4 | CAN | 5 | 6 | 2 | Pup | 1 |
| 4 | PF1 | 5 | 4 | 1 | Pup | 1 |
| 4 | PF1 | 5 | 4 | 1 | Pup | 1 |
| 4 | PF1 | 5 | 4 | 1 | Pup | 1 |
| 4 | PF1 | 5 | 4 | 1 | Pup | 1 |
| 4 | PF1 | 5 | 4 | 1 | Pup | 1 |
| 4 | PF1 | 5 | 4 | 3 | Pup | 1 |
| 4 | PF1 | 5 | 4 | 3 | Pup | 1 |
| 4 | PF1 | 5 | 4 | 3 | Pup | 1 |
| 4 | PF1 | 5 | 4 | 2 | Pup | 1 |
| 4 | PF1 | 5 | 4 | 2 | Pup | 1 |
| 4 | PF1 | 5 | 4 | 2 | Pup | 1 |
| 4 | PF1 | 5 | 4 | 2 | Pup | 1 |
| 4 | PF1 | 5 | 4 | 0.5 | Pup | 1 |
| 4 | PF1 | 5 | 4 | 0.5 | Pup | 1 |
| 4 | PF1 | 5 | 4 | 0.5 | Pup | 1 |
| 4 | PF1 | 5 | 4 | 0.5 | Pup | 1 |
| 4 | PF1 | 5 | 4 | 2 | Pup | 1 |
| 4 | PF1 | 5 | 4 | 2 | Pup | 1 |
| 4 | PF1 | 5 | 4 | 2 | Pup | 1 |
| 4 | PF1 | 5 | 4 | 2 | Pup | 1 |
| 4 | PF1 | 5 | 4 | 0.5 | Pup | 1 |
| 4 | PF1 | 5 | 4 | 1 | Pup | 1 |
| 4 | PF1 | 5 | 4 | 1 | Pup | 1 |
| 4 | PF1 | 5 | 4 | 1 | Pup | 1 |
| 4 | PF1 | 5 | 4 | 1 | Pup | 1 |
| 4 | PF1 | 5 | 4 | 1 | Pup | 1 |
| 4 | PF1 | 5 | 4 | 3 | Pup | 1 |
| 4 | PF1 | 5 | 4 | 3 | Pup | 1 |
| 4 | PF1 | 5 | 4 | 0.5 | Pup | 1 |
| 4 | RS1 | 5 | 2 | 4 | Pup | 1 |
| 4 | RS1 | 5 | 2 | 4 | Pup | 1 |
| 4 | RS1 | 5 | 2 | 8 | MO | 1 |
| 4 | RS1 | 5 | 2 | 6 | MO | 1 |
| 4 | RS1 | 5 | 2 | 2 | Pup | 1 |
| 4 | RS2 | 5 | 4 | 4 | MO | 1 |
| 4 | RS2 | 5 | 4 | 4 | MO | 1 |
| 4 | RS2 | 5 | 4 | 4 | MO | 1 |
| 4 | RS2 | 5 | 4 | 4 | MO | 1 |
| 4 | RS2 | 5 | 4 | 3 | Pup | 1 |
| 4 | RS2 | 5 | 4 | 1 | Pup | 1 |
| 4 | RS2 | 5 | 4 | 6 | MO | 1 |
| 4 | RS2 | 5 | 4 | 8 | MO | 1 |
| 4 | RS2 | 5 | 4 | 5 | MO | 1 |
| 4 | RS2 | 5 | 4 | 5 | MO | 1 |
| 4 | RS2 | 5 | 4 | 3 | Pup | 1 |
| 4 | RS2 | 5 | 4 | 3 | Pup | 1 |
| 4 | RS2 | 5 | 4 | 1 | Pup | 1 |
| 4 | RS2 | 5 | 4 | 1 | Pup | 1 |
| 4 | RS2 | 5 | 4 | 1 | Pup | 1 |
| 4 | RS2 | 5 | 4 | 1 | Pup | 1 |
| 4 | RS2 | 5 | 4 | 3 | MO | 1 |
| 4 | RS2 | 5 | 4 | 3 | MO | 1 |
| 4 | RS2 | 5 | 4 | 3 | MO | 1 |
| 4 | RS2 | 5 | 4 | 3 | MO | 1 |
| 4 | RS2 | 5 | 4 | 1 | Pup | 1 |
| 4 | RS2 | 5 | 4 | 1 | Pup | 1 |
| 4 | RS2 | 5 | 4 | 1 | Pup | 1 |
| 4 | RS2 | 5 | 4 | 1 | Pup | 1 |
| 4 | RS2 | 5 | 4 | 3 | Pup | 1 |
| 4 | RS2 | 5 | 4 | 6 | MO | 1 |
| 4 | RS2 | 5 | 4 | 6 | MO | 1 |
| 4 | RS2 | 5 | 4 | 6 | MO | 1 |
| 4 | RS2 | 5 | 4 | 6 | MO | 1 |
| 4 | RS2 | 5 | 4 | 0.5 | Pup | 1 |
| 4 | RS2 | 5 | 4 | 2 | Pup | 1 |
| 4 | RS2 | 5 | 4 | 2 | Pup | 1 |
| 4 | RS2 | 5 | 4 | 0.5 | Pup | 1 |
| 4 | RS2 | 5 | 4 | 0.5 | Pup | 1 |
| 4 | RS2 | 5 | 4 | 1 | Pup | 1 |
| 4 | RS2 | 5 | 4 | 1 | Pup | 1 |
| 4 | RS2 | 5 | 4 | 4 | MO | 1 |
| 4 | RS2 | 5 | 4 | 4 | MO | 1 |
| 4 | RS2 | 5 | 4 | 4 | MO | 1 |
| 4 | RS2 | 5 | 4 | 4 | MO | 1 |
| 4 | RS2 | 5 | 4 | 3 | MO | 1 |
| 4 | RS2 | 5 | 4 | 3 | MO | 1 |

| | | | | | | | |
|---|---|---|---|---|---|---|---|
| 4 | RS2 | 5 | 4 | 3 | MO | 1 | |
| 4 | RS2 | 5 | 4 | 3 | MO | 1 | |
| 4 | RS2 | 5 | 4 | 2 | Pup | 1 | |
| 4 | RS2 | 5 | 4 | 1 | Pup | 1 | |
| 4 | RS2 | 5 | 4 | 1 | Pup | 1 | |
| 4 | RS2 | 5 | 4 | 1 | Pup | 1 | |
| 4 | RS2 | 5 | 4 | 10 | MO | 1 | |
| 4 | RS2 | 5 | 4 | 10 | MO | 1 | |
| 4 | RS2 | 5 | 4 | 10 | MO | 1 | |
| 4 | RS2 | 5 | 4 | 10 | MO | 1 | |
| 4 | RS2 | 5 | 4 | 2 | Pup | 1 | |
| 4 | RS2 | 5 | 4 | 1 | Pup | 1 | |
| 4 | RS2 | 5 | 4 | 0.5 | Pup | 1 | |
| 4 | RS2 | 5 | 4 | 1 | Pup | 1 | |
| 4 | RS3 | 5 | 2 | 6 | Pup | 1 | |
| 4 | RS3 | 5 | 2 | 16 | MO | 1 | |
| 4 | RS3 | 5 | 2 | 12 | MO | 1 | |
| 4 | RS3 | 5 | 2 | 1 | Pup | 1 | |
| 4 | RS3 | 5 | 2 | 7 | MO | 1 | |
| 4 | RS3 | 5 | 2 | 9 | MO | 1 | |
| 4 | RS3 | 5 | 2 | 4 | Pup | 1 | |
| 4 | RS3 | 5 | 2 | 4 | Pup | 1 | |
| 4 | RS3 | 5 | 2 | 2 | Pup | 1 | |
| 4 | RS3 | 5 | 2 | 2 | Pup | 1 | |
| 4 | RS3 | 5 | 2 | 11 | Pup | 1 | |
| 4 | RS3 | 5 | 2 | 13 | Pup | 1 | |
| 4 | RS3 | 5 | 2 | 3 | Pup | 1 | |
| 4 | RS3 | 5 | 2 | 3 | MO | 1 | |
| 4 | RS3 | 5 | 2 | 3 | MO | 1 | |
| 4 | RS3 | 5 | 2 | 2 | Pup | 1 | |
| 4 | RS3 | 5 | 2 | 2 | Pup | 1 | |
| 4 | RS3 | 5 | 2 | 3 | MO | 1 | |
| 4 | RS3 | 5 | 2 | 3 | MO | 1 | |
| 5 | BRN | 5 | 2 | 2 | Pup | | 1 |
| 5 | BRN | 5 | 2 | 1 | Pup | 1 | |
| 5 | BRN | 5 | 2 | 6 | Pup | 1 | |
| 5 | BRN | 5 | 2 | 4 | Pup | 1 | |
| 5 | BRN | 5 | 2 | 4 | Pup | 1 | |
| 5 | BRN | 5 | 2 | 5 | Pup | 1 | |
| 5 | BRN | 5 | 2 | 5 | Pup | 1 | |
| 5 | BRN | 5 | 2 | 2 | Pup | 1 | |
| 5 | BRN | 5 | 2 | 0.5 | Pup | 1 | |

| | | | | | | |
|---|---|---|---|---|---|---|
| 5 | BRN | 5 | 2 | 4 | Pup | 1 |
| 5 | BRN | 5 | 2 | 2 | Pup | 1 |
| 5 | BRN | 5 | 2 | 2 | Pup | 1 |
| 5 | KTI | 5 | 2 | 13 | MO | 1 |
| 5 | KTI | 5 | 2 | 8 | MO | 1 |
| 5 | KTI | 5 | 2 | 1 | Pup | 1 |
| 5 | KTI | 5 | 2 | 1 | Pup | 1 |
| 5 | KTI | 5 | 2 | 5 | MO | 1 |
| 5 | KTI | 5 | 2 | 7 | MO | 1 |
| 5 | KTI | 5 | 2 | 1 | Pup | 1 |
| 5 | KTI | 5 | 2 | 2 | Pup | 1 |
| 5 | KTI | 5 | 2 | 1 | Pup | 1 |
| 5 | KTI | 5 | 2 | 4 | Pup | 1 |
| 5 | KTI | 5 | 2 | 6 | Pup | 1 |
| 5 | KTI | 5 | 2 | 2 | Pup | 1 |
| 5 | KTI | 5 | 2 | 2 | Pup | 1 |
| 5 | KTI | 5 | 2 | 5 | MO | 1 |
| 5 | KTI | 5 | 2 | 5 | MO | 1 |
| 5 | KTI | 5 | 2 | 7 | Pup | 1 |
| 5 | KTI | 5 | 2 | 1 | Pup | 1 |
| 5 | KTI | 5 | 2 | 9 | MO | 1 |
| 5 | KTI | 5 | 2 | 9 | MO | 1 |
| 5 | KTI | 5 | 2 | 3 | MO | 1 |
| 5 | KTI | 5 | 2 | 3 | MO | 1 |
| 5 | KTI | 5 | 2 | 1 | MO | 1 |
| 5 | KTI | 5 | 2 | 1 | MO | 1 |
| 5 | KTI | 5 | 2 | 5 | Pup | 1 |
| 5 | KTI | 5 | 2 | 5 | Pup | 1 |
| 5 | KTI | 5 | 2 | 3 | Pup | 1 |
| 5 | KTI | 5 | 2 | 4 | Pup | 1 |
| 5 | KTI | 5 | 2 | 4 | Pup | 1 |
| 5 | KTI | 5 | 2 | 6 | Pup | 1 |
| 5 | KTI | 5 | 2 | 0.5 | Pup | 1 |
| 5 | KTI | 5 | 2 | 0.5 | Pup | 1 |
| 5 | KTI | 5 | 2 | 9 | Pup | 1 |
| 5 | KTI | 5 | 2 | 9 | Pup | 1 |
| 5 | KTI | 5 | 2 | 1 | Pup | 1 |
| 5 | KTI | 5 | 2 | 1 | Pup | 1 |
| 5 | KTI | 5 | 2 | 2 | Pup | 1 |
| 5 | KTI | 5 | 2 | 5 | Pup | 1 |
| 5 | KTI | 5 | 2 | 0.5 | MO | 1 |
| 5 | KTI | 5 | 2 | 1 | MO | 1 |
| 5 | BBR | 5 | 4 | 0.5 | MO | 1 |
| 5 | BBR | 5 | 4 | 0.5 | MO | 1 |
| 5 | BBR | 5 | 4 | 0.5 | MO | 1 |
| 5 | BBR | 5 | 4 | 0.5 | MO | 1 |
| 5 | BBR | 5 | 4 | 9 | Pup | 1 |
| 5 | BBR | 5 | 4 | 8 | Pup | 1 |
| 5 | BBR | 5 | 4 | 9 | Pup | 1 |
| 5 | BBR | 5 | 4 | 10 | Pup | 1 |
| 5 | BBR | 5 | 4 | 2 | Pup | 1 |
| 5 | BBR | 5 | 4 | 5 | Pup | 1 |
| 5 | BBR | 5 | 4 | 4 | Pup | 1 |
| 5 | BBR | 5 | 4 | 4 | Pup | 1 |
| 5 | BBR | 5 | 4 | 4 | Pup | 1 |
| 5 | BBR | 5 | 4 | 2 | Pup | 1 |
| 5 | BBR | 5 | 4 | 1 | Pup | 1 |
| 5 | BBR | 5 | 4 | 8 | MO | 1 |
| 5 | BBR | 5 | 4 | 8 | MO | 1 |
| 5 | BBR | 5 | 4 | 8 | MO | 1 |
| 5 | BBR | 5 | 4 | 8 | MO | 1 |
| 5 | BBR | 5 | 4 | 1 | Pup | 1 |
| 5 | BBR | 5 | 4 | 5 | Pup | 1 |
| 5 | BBR | 5 | 4 | 9 | Pup | 1 |
| 5 | BBR | 5 | 4 | 4 | Pup | 1 |
| 5 | BBR | 5 | 4 | 4 | Pup | 1 |
| 5 | BBR | 5 | 4 | 4 | Pup | 1 |
| 5 | BBR | 5 | 4 | 5 | MO | 1 |
| 5 | BBR | 5 | 4 | 5 | MO | 1 |
| 5 | BBR | 5 | 4 | 5 | MO | 1 |
| 5 | BBR | 5 | 4 | 5 | MO | 1 |
| 5 | BBR | 5 | 4 | 8 | MO | 1 |
| 5 | BBR | 5 | 4 | 8 | MO | 1 |
| 5 | BBR | 5 | 4 | 5 | MO | 1 |
| 5 | BBR | 5 | 4 | 8 | MO | 1 |
| 5 | BBR | 5 | 4 | 8 | MO | 1 |
| 5 | BBR | 5 | 4 | 8 | MO | 1 |
| 5 | BBR | 5 | 4 | 8 | MO | 1 |
| 5 | BBR | 5 | 4 | 8 | MO | 1 |
| 5 | BBR | 5 | 4 | 2 | Pup | 1 |
| 5 | BBR | 5 | 4 | 6 | MO | 1 |
| 5 | BBR | 5 | 4 | 6 | MO | 1 |
| 5 | BBR | 5 | 4 | 6 | MO | 1 |
| 5 | BBR | 5 | 4 | 6 | MO | 1 |
| 5 | WHI | 5 | 2 | 7 | MO | 1 |
| 5 | WHI | 5 | 2 | 9 | MO | 1 |
| 5 | WHI | 5 | 2 | 2 | Pup | 1 |
| 5 | WHI | 5 | 2 | 1 | Pup | 1 |
| 5 | WHI | 5 | 2 | 1 | Pup | 1 |
| 5 | WHI | 5 | 2 | 2 | Pup | 1 |
| 5 | WHI | 5 | 2 | 6 | Pup | 1 |
| 5 | WHI | 5 | 2 | 4 | MO | 1 |
| 5 | WHI | 5 | 2 | 4 | MO | 1 |
| 5 | WHI | 5 | 2 | 8 | MO | 1 |
| 5 | WHI | 5 | 2 | 10 | MO | 1 |
| 5 | WHI | 5 | 2 | 1 | Pup | 1 |
| 5 | WHI | 5 | 2 | 8 | Pup | 1 |
| 5 | WHI | 5 | 2 | 6 | Pup | 1 |
| 5 | WHI | 5 | 2 | 3 | MO | 1 |
| 5 | WHI | 5 | 2 | 5 | MO | 1 |
| 5 | WHI | 5 | 2 | 6 | Pup | 1 |
| 5 | WHI | 5 | 2 | 7 | Pup | 1 |
| 5 | WHI | 5 | 2 | 3 | Pup | 1 |
| 5 | WHI | 5 | 2 | 7 | MO | 1 |
| 5 | WHI | 5 | 2 | 9 | MO | 1 |
| 5 | WHI | 5 | 2 | 9 | MO | 1 |
| 5 | WHI | 5 | 2 | 9 | MO | 1 |
| 5 | WHI | 5 | 2 | 2 | Pup | 1 |
| 5 | RS5 | 5 | 3 | 8 | MO | 1 |
| 5 | RS5 | 5 | 3 | 8 | MO | 1 |
| 5 | RS5 | 5 | 3 | 8 | MO | 1 |
| 5 | RS5 | 5 | 3 | 9 | MO | 1 |
| 5 | RS5 | 5 | 3 | 9 | MO | 1 |
| 5 | RS5 | 5 | 3 | 9 | MO | 1 |
| 5 | RS5 | 5 | 3 | 3 | MO | 1 |
| 5 | RS5 | 5 | 3 | 3 | MO | 1 |
| 5 | RS5 | 5 | 3 | 6 | MO | 1 |
| 5 | RS5 | 5 | 3 | 2 | Pup | 1 |
| 5 | RS5 | 5 | 3 | 2 | Pup | 1 |
| 5 | RS5 | 5 | 3 | 2 | Pup | 1 |
| 5 | RS5 | 5 | 3 | 7 | Pup | 1 |
| 5 | RS5 | 5 | 3 | 6 | Pup | 1 |
| 5 | RS5 | 5 | 3 | 4 | Pup | 1 |
| 5 | RS5 | 5 | 3 | 2 | Pup | 1 |
| 5 | RS5 | 5 | 3 | 9 | Pup | 1 |
| 5 | RS5 | 5 | 3 | 9 | Pup | 1 |
| 5 | RS5 | 5 | 3 | 10 | Pup | 1 |
| 5 | RS5 | 5 | 3 | 3 | Pup | 1 |
| 5 | RS5 | 5 | 3 | 1 | Pup | 1 |
| 5 | RS5 | 5 | 3 | 1 | Pup | 1 |
| 5 | RS5 | 5 | 3 | 3 | Pup | 1 |
| 5 | RS5 | 5 | 3 | 1 | Pup | 1 |
| 5 | RS5 | 5 | 3 | 6 | MO | 1 |
| 5 | RS5 | 5 | 3 | 5 | MO | 1 |
| 5 | RS5 | 5 | 3 | 5 | MO | 1 |
| 4 | CAN | 6 | 6 | 1 | Pup | 1 |
| 4 | CAN | 6 | 6 | 3 | MO | 1 |
| 4 | CAN | 6 | 6 | 8 | MO | 1 |
| 4 | CAN | 6 | 6 | 8 | MO | 1 |
| 4 | CAN | 6 | 6 | 12 | MO | 1 |
| 4 | CAN | 6 | 6 | 8 | MO | 1 |
| 4 | CAN | 6 | 6 | 12 | MO | 1 |
| 4 | CAN | 6 | 6 | 12 | MO | 1 |
| 4 | CAN | 6 | 6 | 4 | Pup | 1 |
| 4 | CAN | 6 | 6 | 4 | Pup | 1 |
| 4 | CAN | 6 | 6 | 3 | Pup | 1 |
| 4 | MDB | 6 | 5 | 0.5 | Pup | 1 |
| 4 | MDB | 6 | 5 | 0.5 | Pup | 1 |
| 4 | MDB | 6 | 5 | 0.5 | Pup | 1 |
| 4 | MDB | 6 | 5 | 1 | Pup | 1 |
| 4 | MDB | 6 | 5 | 1 | Pup | 1 |
| 4 | MDB | 6 | 5 | 1 | Pup | 1 |
| 4 | MDB | 6 | 5 | 1 | Pup | 1 |
| 4 | MDB | 6 | 5 | 0.5 | Pup | 1 |
| 4 | MDB | 6 | 5 | 0.5 | Pup | 1 |
| 4 | MDB | 6 | 5 | 0.5 | Pup | 1 |
| 4 | MDB | 6 | 5 | 0.5 | Pup | 1 |
| 4 | MDB | 6 | 5 | 0.5 | Pup | 1 |
| 4 | MDB | 6 | 5 | 3 | Pup | 1 |
| 4 | MDB | 6 | 5 | 3 | Pup | 1 |
| 4 | MDB | 6 | 5 | 3 | Pup | 1 |
| 4 | MDB | 6 | 5 | 3 | Pup | 1 |
| 4 | MDB | 6 | 5 | 3 | Pup | 1 |
| 4 | MDB | 6 | 5 | 1 | Pup | 1 |
| 4 | MDB | 6 | 5 | 1 | Pup | 1 |
| 4 | MDB | 6 | 5 | 1 | Pup | 1 |
| 4 | MDB | 6 | 5 | 0.5 | Pup | 1 |
| 4 | MDB | 6 | 5 | 0.5 | Pup | 1 |
| 4 | MDB | 6 | 5 | 0.5 | Pup | 1 |
| 4 | MDB | 6 | 5 | 0.5 | Pup | 1 |
| 4 | MDB | 6 | 5 | 0.5 | Pup | 1 |
| 4 | MDB | 6 | 5 | 0.5 | Pup | 1 |

| | | | | | | | |
|---|---|---|---|---|---|---|---|
| 4 | MDB | 6 | 5 | 0.5 | Pup | 1 | |
| 4 | MDB | 6 | 5 | 1 | MO | 1 | |
| 4 | MDB | 6 | 5 | 1 | MO | 1 | |
| 4 | MDB | 6 | 5 | 1 | MO | 1 | |
| 4 | MDB | 6 | 5 | 1 | MO | 1 | |
| 4 | MDB | 6 | 5 | 1 | MO | 1 | |
| 4 | MDB | 6 | 5 | 1 | MO | 1 | |
| 4 | MDB | 6 | 5 | 1 | MO | 1 | |
| 4 | MDB | 6 | 5 | 1 | MO | 1 | |
| 4 | MDB | 6 | 5 | 1 | MO | 1 | |
| 4 | MDB | 6 | 5 | 1 | MO | 1 | |
| 4 | MDB | 6 | 5 | 0.5 | Pup | 1 | |
| 4 | MDB | 6 | 5 | 0.5 | Pup | 1 | |
| 4 | PF1 | 6 | 3 | 5 | Pup | | 1 |
| 4 | PF1 | 6 | 3 | 11 | Pup | | 1 |
| 4 | PF1 | 6 | 3 | 5 | Pup | | 1 |
| 4 | PF1 | 6 | 3 | 1 | Pup | | 1 |
| 4 | PF1 | 6 | 3 | 8 | Pup | 1 | |
| 4 | PF1 | 6 | 3 | 7 | Pup | 1 | |
| 4 | PF1 | 6 | 3 | 7 | Pup | 1 | |
| 4 | PF1 | 6 | 3 | 1 | Pup | 1 | |
| 4 | PF1 | 6 | 3 | 2 | Pup | 1 | |
| 4 | PF1 | 6 | 3 | 1 | Pup | 1 | |
| 4 | PF1 | 6 | 3 | 6 | Pup | 1 | |
| 4 | PF1 | 6 | 3 | 6 | Pup | 1 | |
| 4 | PF1 | 6 | 3 | 5 | Pup | 1 | |
| 4 | PF1 | 6 | 3 | 0.5 | Pup | 1 | |
| 4 | PF1 | 6 | 3 | 0.5 | Pup | 1 | |
| 4 | PF1 | 6 | 3 | 0.5 | Pup | 1 | |
| 4 | RS1 | 6 | 2 | 3 | MO | 1 | |
| 4 | RS1 | 6 | 2 | 3 | MO | 1 | |
| 4 | RS1 | 6 | 2 | 3 | Pup | 1 | |
| 4 | RS1 | 6 | 2 | 3 | Pup | 1 | |
| 4 | RS1 | 6 | 2 | 1 | Pup | 1 | |
| 4 | RS1 | 6 | 2 | 1 | Pup | 1 | |
| 4 | RS1 | 6 | 2 | 1 | Pup | 1 | |
| 4 | RS2 | 6 | 4 | 4 | Pup | 1 | |
| 4 | RS2 | 6 | 4 | 4 | Pup | 1 | |
| 4 | RS2 | 6 | 4 | 4 | Pup | 1 | |
| 4 | RS2 | 6 | 4 | 4 | Pup | 1 | |
| 4 | RS2 | 6 | 4 | 3 | Pup | 1 | |

| | | | | | | |
|---|---|---|---|---|---|---|
| 4 | RS2 | 6 | 4 | 3 | Pup | 1 |
| 4 | RS2 | 6 | 4 | 3 | Pup | 1 |
| 4 | RS2 | 6 | 4 | 3 | Pup | 1 |
| 4 | RS2 | 6 | 4 | 3 | Pup | 1 |
| 4 | RS2 | 6 | 4 | 1 | Pup | 1 |
| 4 | RS2 | 6 | 4 | 1 | Pup | 1 |
| 4 | RS2 | 6 | 4 | 1 | Pup | 1 |
| 4 | RS2 | 6 | 4 | 1 | Pup | 1 |
| 4 | RS2 | 6 | 4 | 1 | Pup | 1 |
| 4 | RS2 | 6 | 4 | 0.5 | Pup | 1 |
| 4 | RS2 | 6 | 4 | 5 | Pup | 1 |
| 4 | RS2 | 6 | 4 | 5 | Pup | 1 |
| 4 | RS2 | 6 | 4 | 5 | Pup | 1 |
| 4 | RS2 | 6 | 4 | 5 | Pup | 1 |
| 4 | RS2 | 6 | 4 | 3 | Pup | 1 |
| 4 | RS2 | 6 | 4 | 0.5 | Pup | 1 |
| 4 | RS2 | 6 | 4 | 0.5 | Pup | 1 |
| 4 | RS2 | 6 | 4 | 3 | Pup | 1 |
| 4 | RS2 | 6 | 4 | 2 | Pup | 1 |
| 4 | RS2 | 6 | 4 | 5 | Pup | 1 |
| 4 | RS2 | 6 | 4 | 3 | Pup | 1 |
| 4 | RS2 | 6 | 4 | 3 | Pup | 1 |
| 4 | RS2 | 6 | 4 | 1 | Pup | 1 |
| 4 | RS2 | 6 | 4 | 1 | Pup | 1 |
| 4 | RS2 | 6 | 4 | 1 | Pup | 1 |
| 4 | RS2 | 6 | 4 | 1 | Pup | 1 |
| 4 | RS2 | 6 | 4 | 3 | Pup | 1 |
| 4 | RS2 | 6 | 4 | 3 | Pup | 1 |
| 4 | RS2 | 6 | 4 | 3 | Pup | 1 |
| 4 | RS2 | 6 | 4 | 0.5 | Pup | 1 |
| 4 | RS2 | 6 | 4 | 0.5 | Pup | 1 |
| 4 | RS2 | 6 | 4 | 1 | Pup | 1 |
| 4 | RS2 | 6 | 4 | 1 | Pup | 1 |
| 4 | RS2 | 6 | 4 | 1 | Pup | 1 |
| 4 | RS2 | 6 | 4 | 1 | Pup | 1 |
| 4 | RS2 | 6 | 4 | 0.5 | Pup | 1 |
| 4 | RS2 | 6 | 4 | 0.5 | Pup | 1 |
| 4 | RS2 | 6 | 4 | 1 | Pup | 1 |
| 4 | RS2 | 6 | 4 | 1 | Pup | 1 |
| 4 | RS2 | 6 | 4 | 3 | Pup | 1 |
| 4 | RS2 | 6 | 4 | 2 | Pup | 1 |
| 4 | RS2 | 6 | 4 | 2 | Pup | 1 |
| 4 | RS2 | 6 | 4 | 1 | Pup | 1 |
| 4 | RS2 | 6 | 4 | 1 | Pup | 1 |
| 4 | RS2 | 6 | 4 | 0.5 | Pup | 1 |
| 4 | RS2 | 6 | 4 | 0.5 | Pup | 1 |
| 4 | RS2 | 6 | 4 | 1 | Pup | 1 |
| 4 | RS2 | 6 | 4 | 1 | Pup | 1 |
| 4 | RS2 | 6 | 4 | 0.5 | Pup | 1 |
| 4 | RS2 | 6 | 4 | 2 | Pup | 1 |
| 4 | RS2 | 6 | 4 | 2 | Pup | 1 |
| 4 | RS2 | 6 | 4 | 3 | Pup | 1 |
| 4 | RS2 | 6 | 4 | 3 | Pup | 1 |
| 4 | RS2 | 6 | 4 | 3 | Pup | 1 |
| 4 | RS2 | 6 | 4 | 3 | Pup | 1 |
| 4 | RS2 | 6 | 4 | 3 | Pup | 1 |
| 4 | RS2 | 6 | 4 | 3 | MO | 1 |
| 4 | RS2 | 6 | 4 | 3 | MO | 1 |
| 4 | RS2 | 6 | 4 | 3 | MO | 1 |
| 4 | RS2 | 6 | 4 | 3 | MO | 1 |
| 4 | RS2 | 6 | 4 | 1 | Pup | 1 |
| 4 | RS2 | 6 | 4 | 1 | Pup | 1 |
| 4 | RS2 | 6 | 4 | 1 | Pup | 1 |
| 4 | RS2 | 6 | 4 | 1 | Pup | 1 |
| 4 | RS2 | 6 | 4 | 6 | MO | 1 |
| 4 | RS2 | 6 | 4 | 6 | MO | 1 |
| 4 | RS2 | 6 | 4 | 6 | MO | 1 |
| 4 | RS2 | 6 | 4 | 6 | MO | 1 |
| 4 | RS2 | 6 | 4 | 6 | Pup | 1 |
| 4 | RS2 | 6 | 4 | 6 | Pup | 1 |
| 4 | RS2 | 6 | 4 | 6 | Pup | 1 |
| 4 | RS2 | 6 | 4 | 6 | Pup | 1 |
| 4 | RS2 | 6 | 4 | 2 | Pup | 1 |
| 4 | RS2 | 6 | 4 | 1 | Pup | 1 |
| 4 | RS2 | 6 | 4 | 1 | Pup | 1 |
| 4 | RS2 | 6 | 4 | 1 | Pup | 1 |
| 4 | RS2 | 6 | 4 | 1 | Pup | 1 |
| 4 | RS2 | 6 | 4 | 1 | Pup | 1 |
| 4 | RS2 | 6 | 4 | 1 | Pup | 1 |
| 4 | RS2 | 6 | 4 | 1 | Pup | 1 |
| 4 | RS2 | 6 | 4 | 1 | Pup | 1 |
| 4 | RS2 | 6 | 4 | 1 | Pup | 1 |
| 4 | RS2 | 6 | 4 | 0.5 | Pup | 1 |
| 4 | RS2 | 6 | 4 | 0.5 | Pup | 1 |
| 4 | RS2 | 6 | 4 | 1 | Pup | 1 |
| 4 | RS2 | 6 | 4 | 1 | Pup | 1 |

| | | | | | | | |
|---|---|---|---|---|---|---|---|
| 4 | RS2 | 6 | 4 | 4 | Pup | 1 | |
| 4 | RS2 | 6 | 4 | 4 | Pup | 1 | |
| 4 | RS2 | 6 | 4 | 4 | Pup | 1 | |
| 4 | RS2 | 6 | 4 | 4 | Pup | 1 | |
| 4 | RS2 | 6 | 4 | 4 | Pup | 1 | |
| 4 | RS2 | 6 | 4 | 2 | Pup | 1 | |
| 4 | RS2 | 6 | 4 | 2 | Pup | 1 | |
| 4 | RS2 | 6 | 4 | 2 | Pup | 1 | |
| 4 | RS2 | 6 | 4 | 2 | Pup | 1 | |
| 4 | RS2 | 6 | 4 | 2 | Pup | 1 | |
| 4 | RS2 | 6 | 4 | 1 | Pup | 1 | |
| 4 | RS2 | 6 | 4 | 2 | Pup | 1 | |
| 4 | RS2 | 6 | 4 | 2 | Pup | 1 | |
| 4 | RS2 | 6 | 4 | 3 | Pup | 1 | |
| 4 | RS2 | 6 | 4 | 1 | Pup | 1 | |
| 4 | RS2 | 6 | 4 | 1 | Pup | 1 | |
| 4 | RS2 | 6 | 4 | 2 | Pup | 1 | |
| 4 | RS2 | 6 | 4 | 2 | Pup | 1 | |
| 4 | RS2 | 6 | 4 | 2 | Pup | 1 | |
| 4 | RS2 | 6 | 4 | 2 | Pup | 1 | |
| 4 | RS2 | 6 | 4 | 2 | Pup | 1 | |
| 4 | RS2 | 6 | 4 | 1 | Pup | 1 | |
| 4 | RS2 | 6 | 4 | 1 | Pup | 1 | |
| 4 | RS2 | 6 | 4 | 1 | Pup | 1 | |
| 4 | RS2 | 6 | 4 | 1 | Pup | 1 | |
| 4 | RS2 | 6 | 4 | 1 | Pup | | 1 |
| 4 | RS2 | 6 | 4 | 1 | Pup | | 1 |
| 4 | RS2 | 6 | 4 | 1 | Pup | | 1 |
| 4 | RS2 | 6 | 4 | 1 | Pup | | 1 |
| 4 | RS2 | 6 | 4 | 1 | Pup | | 1 |
| 4 | RS3 | 6 | 2 | 6 | MO | 1 | |
| 4 | RS3 | 6 | 2 | 5 | MO | 1 | |
| 4 | RS3 | 6 | 2 | 1 | Pup | | 1 |
| 4 | RS3 | 6 | 2 | 1 | Pup | | 1 |
| 4 | RS3 | 6 | 2 | 1 | Pup | | 1 |
| 4 | RS3 | 6 | 2 | 3 | Pup | | 1 |
| 4 | RS3 | 6 | 2 | 1 | Pup | | 1 |
| 4 | RS3 | 6 | 2 | 2 | Pup | 1 | |
| 4 | RS3 | 6 | 2 | 0.5 | Pup | 1 | |
| 4 | RS3 | 6 | 2 | 1 | Pup | 1 | |
| 4 | RS3 | 6 | 2 | 2 | Pup | 1 | |
| 5 | BRN | 6 | 1 | 2 | Pup | 1 | |
| 5 | BRN | 6 | 1 | 1 | Pup | 1 | |
| 5 | BRN | 6 | 1 | 0.5 | Pup | 1 | |
| 5 | BRN | 6 | 1 | 4 | Pup | 1 | |
| 5 | BRN | 6 | 1 | 4 | Pup | 1 | |
| 5 | BRN | 6 | 1 | 2 | Pup | 1 | |
| 5 | BRN | 6 | 1 | 3 | Pup | 1 | |
| 5 | BRN | 6 | 1 | 6 | Pup | 1 | |
| 5 | BRN | 6 | 1 | 5 | Pup | 1 | |
| 5 | BRN | 6 | 1 | 0.5 | Pup | 1 | |
| 5 | BRN | 6 | 1 | 0.5 | Pup | 1 | |
| 5 | BRN | 6 | 1 | 3 | Pup | 1 | |
| 5 | BRN | 6 | 1 | 2 | Pup | 1 | |
| 5 | BRN | 6 | 1 | 1 | Pup | 1 | |
| 5 | BRN | 6 | 1 | 1 | Pup | 1 | |
| 5 | BRN | 6 | 1 | 5 | Pup | 1 | |
| 5 | BRN | 6 | 1 | 5 | Pup | 1 | |
| 5 | KTI | 6 | 2 | 2 | Pup | 1 | |
| 5 | KTI | 6 | 2 | 1 | Pup | 1 | |
| 5 | KTI | 6 | 2 | 1 | Pup | | 1 |
| 5 | KTI | 6 | 2 | 1 | Pup | | 1 |
| 5 | KTI | 6 | 2 | 2 | Pup | | 1 |
| 5 | KTI | 6 | 2 | 2 | Pup | | 1 |
| 5 | KTI | 6 | 2 | 2 | Pup | | 1 |
| 5 | KTI | 6 | 2 | 4 | Pup | 1 | |
| 5 | KTI | 6 | 2 | 4 | Pup | 1 | |
| 5 | KTI | 6 | 2 | 1 | Pup | | 1 |
| 5 | KTI | 6 | 2 | 0.5 | Pup | 1 | |
| 5 | KTI | 6 | 2 | 1 | Pup | 1 | |
| 5 | KTI | 6 | 2 | 2 | Pup | 1 | |
| 5 | KTI | 6 | 2 | 1 | Pup | | 1 |
| 5 | KTI | 6 | 2 | 1 | Pup | 1 | |
| 5 | KTI | 6 | 2 | 1 | Pup | 1 | |
| 5 | KTI | 6 | 2 | 1 | Pup | 1 | |
| 5 | KTI | 6 | 2 | 1 | Pup | 1 | |
| 5 | BBR | 6 | 4 | 5 | MO | 1 | |
| 5 | BBR | 6 | 4 | 6 | MO | 1 | |
| 5 | BBR | 6 | 4 | 8 | MO | 1 | |
| 5 | BBR | 6 | 4 | 8 | MO | 1 | |
| 5 | BBR | 6 | 4 | 4 | Pup | 1 | |
| 5 | BBR | 6 | 4 | 2 | Pup | 1 | |
| 5 | BBR | 6 | 4 | 1 | Pup | 1 | |
| 5 | BBR | 6 | 4 | 6 | MO | 1 | |
| 5 | BBR | 6 | 4 | 6 | MO | 1 | |
| 5 | BBR | 6 | 4 | 6 | MO | 1 | |
| 5 | BBR | 6 | 4 | 6 | MO | 1 | |
| 5 | BBR | 6 | 4 | 8 | MO | 1 | |
| 5 | BBR | 6 | 4 | 8 | MO | 1 | |
| 5 | BBR | 6 | 4 | 6 | MO | 1 | |
| 5 | BBR | 6 | 4 | 8 | MO | 1 | |
| 5 | BBR | 6 | 4 | 1 | Pup | | 1 |
| 5 | BBR | 6 | 4 | 1 | Pup | | 1 |
| 5 | BBR | 6 | 4 | 3 | Pup | | 1 |
| 5 | BBR | 6 | 4 | 3 | Pup | | 1 |
| 5 | BBR | 6 | 4 | 3 | Pup | | 1 |
| 5 | BBR | 6 | 4 | 4 | Pup | 1 | |
| 5 | BBR | 6 | 4 | 4 | Pup | 1 | |
| 5 | BBR | 6 | 4 | 1 | Pup | 1 | |
| 5 | BBR | 6 | 4 | 2 | Pup | | 1 |
| 5 | BBR | 6 | 4 | 2 | Pup | | 1 |
| 5 | BBR | 6 | 4 | 2 | Pup | | 1 |
| 5 | BBR | 6 | 4 | 4 | Pup | 1 | |
| 5 | BBR | 6 | 4 | 4 | Pup | 1 | |
| 5 | BBR | 6 | 4 | 3 | Pup | 1 | |
| 5 | BBR | 6 | 4 | 3 | Pup | 1 | |
| 5 | BBR | 6 | 4 | 3 | Pup | 1 | |
| 5 | BBR | 6 | 4 | 5 | Pup | 1 | |
| 5 | BBR | 6 | 4 | 5 | Pup | 1 | |
| 5 | BBR | 6 | 4 | 5 | Pup | 1 | |
| 5 | BBR | 6 | 4 | 5 | Pup | 1 | |
| 5 | BBR | 6 | 4 | 5 | Pup | 1 | |
| 5 | BBR | 6 | 4 | 1 | Pup | | 1 |
| 5 | BBR | 6 | 4 | 1 | Pup | | 1 |
| 5 | BBR | 6 | 4 | 1 | Pup | | 1 |
| 5 | BBR | 6 | 4 | 3 | Pup | | 1 |
| 5 | BBR | 6 | 4 | 3 | Pup | | 1 |
| 5 | BBR | 6 | 4 | 3 | Pup | | 1 |
| 5 | BBR | 6 | 4 | 5 | Pup | 1 | |
| 5 | BBR | 6 | 4 | 4 | Pup | 1 | |
| 5 | BBR | 6 | 4 | 4 | Pup | 1 | |
| 5 | BBR | 6 | 4 | 3 | Pup | 1 | |
| 5 | BBR | 6 | 4 | 1 | Pup | 1 | |
| 5 | BBR | 6 | 4 | 1 | Pup | 1 | |
| 5 | BBR | 6 | 4 | 3 | Pup | | 1 |
| 5 | BBR | 6 | 4 | 3 | Pup | | 1 |
| 5 | BBR | 6 | 4 | 3 | Pup | | 1 |
| 5 | WHI | 6 | 2 | 4 | MO | 1 | |
| 5 | WHI | 6 | 2 | 6 | MO | 1 | |
| 5 | WHI | 6 | 2 | 9 | Pup | 1 | |
| 5 | WHI | 6 | 2 | 5 | Pup | 1 | |
| 5 | WHI | 6 | 2 | 1 | Pup | 1 | |
| 5 | WHI | 6 | 2 | 8 | MO | 1 | |
| 5 | WHI | 6 | 2 | 9 | MO | 1 | |
| 5 | WHI | 6 | 2 | 2 | Pup | 1 | |
| 5 | WHI | 6 | 2 | 1 | Pup | 1 | |
| 5 | WHI | 6 | 2 | 10 | Pup | 1 | |
| 5 | WHI | 6 | 2 | 9 | Pup | 1 | |
| 5 | WHI | 6 | 2 | 6 | MO | 1 | |
| 5 | WHI | 6 | 2 | 6 | MO | 1 | |
| 5 | WHI | 6 | 2 | 7 | MO | 1 | |
| 5 | WHI | 6 | 2 | 9 | MO | 1 | |
| 5 | WHI | 6 | 2 | 2 | Pup | 1 | |
| 5 | WHI | 6 | 2 | 3 | Pup | 1 | |
| 5 | WHI | 6 | 2 | 7 | Pup | 1 | |
| 5 | WHI | 6 | 2 | 5 | Pup | 1 | |
| 5 | WHI | 6 | 2 | 8 | MO | 1 | |
| 5 | WHI | 6 | 2 | 11 | MO | 1 | |
| 5 | RS5 | 6 | 3 | 3 | Pup | 1 | |
| 5 | RS5 | 6 | 3 | 3 | Pup | 1 | |
| 5 | RS5 | 6 | 3 | 3 | Pup | 1 | |
| 5 | RS5 | 6 | 3 | 5 | Pup | 1 | |
| 5 | RS5 | 6 | 3 | 5 | Pup | 1 | |
| 5 | RS5 | 6 | 3 | 5 | Pup | 1 | |
| 5 | RS5 | 6 | 3 | 3 | MO | 1 | |
| 5 | RS5 | 6 | 3 | 3 | MO | 1 | |
| 5 | RS5 | 6 | 3 | 2 | MO | 1 | |
| 5 | RS5 | 6 | 3 | 2 | Pup | 1 | |
| 5 | RS5 | 6 | 3 | 5 | Pup | 1 | |
| 5 | RS5 | 6 | 3 | 5 | Pup | 1 | |
| 5 | RS5 | 6 | 3 | 4 | Pup | 1 | |
| 5 | RS5 | 6 | 3 | 2 | Pup | 1 | |
| 5 | RS5 | 6 | 3 | 2 | Pup | 1 | |
| 5 | RS5 | 6 | 3 | 2 | Pup | 1 | |
| 5 | RS5 | 6 | 3 | 8 | Pup | 1 | |
| 5 | RS5 | 6 | 3 | 3 | Pup | 1 | |
| 5 | RS5 | 6 | 3 | 7 | Pup | 1 | |
| 5 | RS5 | 6 | 3 | 5 | Pup | 1 | |
| 5 | RS5 | 6 | 3 | 3 | Pup | 1 | |
| 5 | RS5 | 6 | 3 | 2 | Pup | 1 | |
| 5 | RS5 | 6 | 3 | 2 | Pup | 1 | |
| 5 | RS5 | 6 | 3 | 2 | Pup | 1 | |
| 5 | RS5 | 6 | 3 | 2 | Pup | 1 | |
| 4 | CAN | 7 | 5 | 19 | MO | 1 | |
| 4 | CAN | 7 | 5 | 19 | MO | 1 | |
| 4 | CAN | 7 | 5 | 19 | MO | 1 | |
| 4 | CAN | 7 | 5 | 19 | MO | 1 | |
| 4 | CAN | 7 | 5 | 19 | MO | 1 | |
| 4 | CAN | 7 | 5 | 14 | Pup | 1 | |
| 4 | CAN | 7 | 5 | 14 | Pup | 1 | |
| 4 | CAN | 7 | 5 | 13 | Pup | 1 | |
| 4 | CAN | 7 | 5 | 13 | Pup | 1 | |
| 4 | CAN | 7 | 5 | 13 | Pup | 1 | |
| 4 | CAN | 7 | 5 | 11 | MO | 1 | |
| 4 | CAN | 7 | 5 | 11 | MO | 1 | |
| 4 | CAN | 7 | 5 | 11 | MO | 1 | |
| 4 | CAN | 7 | 5 | 11 | MO | 1 | |
| 4 | CAN | 7 | 5 | 11 | MO | 1 | |
| 4 | PF1 | 7 | 3 | 0.5 | Pup | | 1 |
| 4 | PF1 | 7 | 3 | 1 | Pup | | 1 |
| 4 | PF1 | 7 | 3 | 0.5 | Pup | | 1 |
| 4 | PF1 | 7 | 3 | 0.5 | Pup | | 1 |
| 4 | PF1 | 7 | 3 | 1 | Pup | 1 | |
| 4 | PF1 | 7 | 3 | 1 | Pup | 1 | |
| 4 | PF1 | 7 | 3 | 2 | Pup | 1 | |
| 4 | RS1 | 7 | 2 | 1 | Pup | | 1 |
| 4 | RS1 | 7 | 2 | 1 | Pup | | 1 |
| 4 | RS1 | 7 | 2 | 2 | Pup | 1 | |
| 4 | RS1 | 7 | 2 | 2 | Pup | 1 | |
| 4 | RS1 | 7 | 2 | 0.5 | Pup | 1 | |
| 4 | RS1 | 7 | 2 | 1 | Pup | | 1 |
| 4 | RS1 | 7 | 2 | 4 | Pup | 1 | |
| 4 | RS1 | 7 | 2 | 4 | Pup | 1 | |
| 4 | RS1 | 7 | 2 | 0.5 | Pup | | 1 |
| 4 | RS1 | 7 | 2 | 1 | Pup | | 1 |
| 4 | RS1 | 7 | 2 | 0.5 | Pup | | 1 |
| 4 | RS1 | 7 | 2 | 12 | Pup | | 1 |
| 4 | RS1 | 7 | 2 | 9 | Pup | | 1 |
| 4 | RS1 | 7 | 2 | 1 | Pup | | 1 |
| 4 | RS1 | 7 | 2 | 0.5 | Pup | | 1 |
| 4 | RS1 | 7 | 2 | 0.5 | Pup | | 1 |
| 4 | RS1 | 7 | 2 | 1 | Pup | | 1 |
| 4 | RS1 | 7 | 2 | 0.5 | Pup | | 1 |
| 4 | RS1 | 7 | 2 | 0.5 | Pup | | 1 |
| 4 | RS1 | 7 | 2 | 0.5 | Pup | | 1 |
| 4 | RS1 | 7 | 2 | 0.5 | Pup | | 1 |
| 4 | RS1 | 7 | 2 | 1 | Pup | | 1 |
| 4 | RS1 | 7 | 2 | 5 | Pup | | 1 |
| 4 | RS1 | 7 | 2 | 3 | Pup | | 1 |
| 4 | RS1 | 7 | 2 | 2 | Pup | | 1 |
| 4 | RS1 | 7 | 2 | 0.5 | Pup | | 1 |
| 4 | RS1 | 7 | 2 | 1 | Pup | | 1 |
| 4 | RS1 | 7 | 2 | 0.5 | Pup | | 1 |
| 4 | RS1 | 7 | 2 | 1 | Pup | 1 | |
| 4 | RS1 | 7 | 2 | 1 | Pup | 1 | |
| 4 | RS1 | 7 | 2 | 0.5 | Pup | | 1 |
| 4 | RS1 | 7 | 2 | 2 | Pup | | 1 |
| 4 | RS1 | 7 | 2 | 2 | Pup | | 1 |
| 4 | RS1 | 7 | 2 | 0.5 | Pup | | 1 |
| 4 | RS1 | 7 | 2 | 1 | Pup | | 1 |
| 4 | RS1 | 7 | 2 | 1 | Pup | | 1 |
| 4 | RS1 | 7 | 2 | 1 | Pup | | 1 |
| 4 | RS1 | 7 | 2 | 1 | Pup | | 1 |
| 4 | RS1 | 7 | 2 | 1 | Pup | | 1 |
| 4 | RS1 | 7 | 2 | 1 | Pup | | 1 |
| 4 | RS1 | 7 | 2 | 0.5 | Pup | 1 | |
| 4 | RS1 | 7 | 2 | 0.5 | Pup | 1 | |
| 4 | RS1 | 7 | 2 | 1 | Pup | | 1 |
| 4 | RS1 | 7 | 2 | 1 | Pup | 1 | |
| 4 | RS1 | 7 | 2 | 1 | Pup | 1 | |
| 4 | RS1 | 7 | 2 | 0.5 | Pup | 1 | |
| 4 | RS1 | 7 | 2 | 0.5 | Pup | 1 | |
| 4 | RS1 | 7 | 2 | 0.5 | Pup | 1 | |
| 4 | RS1 | 7 | 2 | 0.5 | Pup | 1 | |
| 4 | RS1 | 7 | 2 | 1 | Pup | 1 | |
| 4 | RS1 | 7 | 2 | 1 | Pup | 1 | |
| 4 | RS1 | 7 | 2 | 0.5 | Pup | 1 | |
| 4 | RS1 | 7 | 2 | 0.5 | Pup | 1 | |
| 4 | RS1 | 7 | 2 | 3 | Pup | | 1 |
| 4 | RS1 | 7 | 2 | 3 | Pup | | 1 |
| 4 | RS1 | 7 | 2 | 2 | Pup | 1 | |
| 4 | RS1 | 7 | 2 | 2 | Pup | 1 | |
| 4 | RS1 | 7 | 2 | 2 | Pup | 1 | |
| 4 | RS1 | 7 | 2 | 0.5 | Pup | | 1 |
| 4 | RS1 | 7 | 2 | 0.5 | Pup | | 1 |
| 4 | RS1 | 7 | 2 | 1 | Pup | | 1 |
| 4 | RS1 | 7 | 2 | 1 | Pup | | 1 |
| 4 | RS1 | 7 | 2 | 1 | Pup | 1 | |
| 4 | RS1 | 7 | 2 | 1 | Pup | 1 | |
| 4 | RS1 | 7 | 2 | 1 | Pup | | 1 |
| 4 | RS1 | 7 | 2 | 1 | Pup | | 1 |
| 4 | RS1 | 7 | 2 | 0.5 | Pup | | 1 |
| 4 | RS1 | 7 | 2 | 0.5 | Pup | | 1 |
| 4 | RS1 | 7 | 2 | 0.5 | Pup | | 1 |
| 4 | RS1 | 7 | 2 | 0.5 | Pup | | 1 |
| 4 | RS1 | 7 | 2 | 0.5 | Pup | | 1 |
| 4 | RS1 | 7 | 2 | 0.5 | Pup | | 1 |
| 4 | RS1 | 7 | 2 | 1 | Pup | | 1 |
| 4 | RS1 | 7 | 2 | 1 | Pup | | 1 |
| 4 | RS1 | 7 | 2 | 1 | Pup | 1 | |
| 4 | RS1 | 7 | 2 | 1 | Pup | 1 | |
| 4 | RS1 | 7 | 2 | 0.5 | Pup | | 1 |
| 4 | RS1 | 7 | 2 | 0.5 | Pup | | 1 |
| 4 | RS1 | 7 | 2 | 0.5 | Pup | | 1 |
| 4 | RS1 | 7 | 2 | 0.5 | Pup | | 1 |
| 4 | RS1 | 7 | 2 | 0.5 | Pup | 1 | |
| 4 | RS1 | 7 | 2 | 0.5 | Pup | 1 | |
| 4 | RS1 | 7 | 2 | 0.5 | Pup | | 1 |
| 4 | RS1 | 7 | 2 | 0.5 | Pup | | 1 |
| 4 | RS1 | 7 | 2 | 2 | Pup | | 1 |
| 4 | RS1 | 7 | 2 | 0.5 | Pup | | 1 |
| 4 | RS1 | 7 | 2 | 2 | Pup | | 1 |
| 4 | RS1 | 7 | 2 | 1 | Pup | | 1 |
| 4 | RS1 | 7 | 2 | 1 | Pup | | 1 |
| 4 | RS1 | 7 | 2 | 1 | Pup | | 1 |
| 4 | RS1 | 7 | 2 | 1 | Pup | | 1 |
| 4 | RS1 | 7 | 2 | 1 | Pup | | 1 |
| 4 | RS1 | 7 | 2 | 0.5 | Pup | 1 | |
| 4 | RS1 | 7 | 2 | 1 | Pup | | 1 |
| 4 | RS1 | 7 | 2 | 1 | Pup | | 1 |
| 4 | RS1 | 7 | 2 | 1 | Pup | | 1 |
| 4 | RS1 | 7 | 2 | 7 | Pup | | 1 |
| 4 | RS1 | 7 | 2 | 7 | Pup | | 1 |
| 4 | RS1 | 7 | 2 | 7 | Pup | | 1 |
| 4 | RS1 | 7 | 2 | 3 | Pup | | 1 |
| 4 | RS1 | 7 | 2 | 3 | Pup | | 1 |
| 4 | RS1 | 7 | 2 | 3 | Pup | | 1 |
| 4 | RS2 | 7 | 2 | 1 | Pup | 1 | |
| 4 | RS2 | 7 | 2 | 6 | Pup | 1 | |
| 4 | RS2 | 7 | 2 | 6 | Pup | 1 | |
| 4 | RS2 | 7 | 2 | 1 | Pup | 1 | |
| 4 | RS2 | 7 | 2 | 0.5 | Pup | 1 | |
| 4 | RS2 | 7 | 2 | 1 | Pup | 1 | |
| 4 | RS2 | 7 | 2 | 1 | Pup | 1 | |
| 4 | RS2 | 7 | 2 | 1 | Pup | 1 | |
| 4 | RS2 | 7 | 2 | 1 | Pup | 1 | |
| 4 | RS2 | 7 | 2 | 3 | Pup | 1 | |
| 4 | RS2 | 7 | 2 | 3 | Pup | 1 | |
| 4 | RS2 | 7 | 2 | 3 | Pup | 1 | |
| 4 | RS2 | 7 | 2 | 1 | Pup | | 1 |
| 4 | RS2 | 7 | 2 | 1 | Pup | | 1 |
| 4 | RS2 | 7 | 2 | 1 | Pup | 1 | |
| 4 | RS2 | 7 | 2 | 6 | Pup | 1 | |
| 4 | RS2 | 7 | 2 | 6 | Pup | 1 | |
| 4 | RS2 | 7 | 2 | 5 | Pup | 1 | |
| 4 | RS2 | 7 | 2 | 1 | Pup | 1 | |
| 4 | RS2 | 7 | 2 | 2 | Pup | 1 | |
| 4 | RS2 | 7 | 2 | 2 | Pup | 1 | |
| 4 | RS2 | 7 | 2 | 1 | Pup | 1 | |
| 4 | RS2 | 7 | 2 | 1 | Pup | 1 | |
| 4 | RS2 | 7 | 2 | 1 | Pup | 1 | |
| 4 | RS2 | 7 | 2 | 1 | Pup | 1 | |
| 4 | RS2 | 7 | 2 | 1 | Pup | 1 | |
| 4 | RS2 | 7 | 2 | 1 | Pup | 1 | |
| 4 | RS2 | 7 | 2 | 2 | Pup | 1 | |
| 4 | RS2 | 7 | 2 | 2 | Pup | 1 | |
| 4 | RS2 | 7 | 2 | 2 | Pup | 1 | |
| 4 | RS2 | 7 | 2 | 1 | Pup | 1 | |
| 4 | RS2 | 7 | 2 | 1 | Pup | 1 | |
| 4 | RS2 | 7 | 2 | 2 | Pup | | 1 |
| 4 | RS2 | 7 | 2 | 2 | Pup | | 1 |
| 4 | RS2 | 7 | 2 | 3 | Pup | 1 | |
| 4 | RS2 | 7 | 2 | 3 | Pup | 1 | |
| 4 | RS2 | 7 | 2 | 2 | Pup | 1 | |
| 4 | RS3 | 7 | 2 | 6 | MO | 1 | |
| 4 | RS3 | 7 | 2 | 5 | MO | 1 | |
| 4 | RS3 | 7 | 2 | 0.5 | Pup | 1 | |
| 4 | RS3 | 7 | 2 | 1 | Pup | 1 | |
| 4 | RS3 | 7 | 2 | 1 | Pup | 1 | |
| 4 | RS3 | 7 | 2 | 3 | MO | 1 | |
| 4 | RS3 | 7 | 2 | 3 | MO | 1 | |
| 4 | RS3 | 7 | 2 | 2 | Pup | 1 | |
| 4 | RS3 | 7 | 2 | 2 | Pup | 1 | |
| 4 | RS3 | 7 | 2 | 4 | Pup | 1 | |
| 4 | RS3 | 7 | 2 | 5 | Pup | 1 | |
| 4 | RS3 | 7 | 2 | 3 | Pup | 1 | |
| 4 | RS3 | 7 | 2 | 1 | Pup | 1 | |
| 4 | RS3 | 7 | 2 | 1 | Pup | 1 | |
| 5 | BRN | 7 | 2 | 1 | Pup | 1 | |
| 5 | BRN | 7 | 2 | 0.5 | Pup | 1 | |
| 5 | BRN | 7 | 2 | 0.5 | Pup | 1 | |
| 5 | BRN | 7 | 2 | 1 | Pup | 1 | |
| 5 | BRN | 7 | 2 | 1 | Pup | 1 | |
| 5 | BRN | 7 | 2 | 1 | Pup | 1 | |
| 5 | BRN | 7 | 2 | 0.5 | Pup | 1 | |
| 5 | BRN | 7 | 2 | 0.5 | Pup | 1 | |
| 5 | BRN | 7 | 2 | 0.5 | Pup | 1 | |
| 5 | BRN | 7 | 2 | 1 | Pup | 1 | |
| 5 | BRN | 7 | 2 | 1 | Pup | 1 | |
| 5 | BRN | 7 | 2 | 0.5 | Pup | 1 | |
| 5 | BRN | 7 | 2 | 1 | Pup | 1 | |
| 5 | BRN | 7 | 2 | 1 | Pup | 1 | |
| 5 | BRN | 7 | 2 | 1 | Pup | 1 | |
| 5 | BRN | 7 | 2 | 1 | Pup | 1 | |
| 5 | BRN | 7 | 2 | 1 | Pup | 1 | |
| 5 | BRN | 7 | 2 | 0.5 | Pup | 1 | |
| 5 | BRN | 7 | 2 | 4 | Pup | 1 | |
| 5 | BRN | 7 | 2 | 1 | Pup | 1 | |
| 5 | BRN | 7 | 2 | 5 | Pup | 1 | |
| 5 | BRN | 7 | 2 | 1 | Pup | 1 | |
| 5 | BRN | 7 | 2 | 1 | Pup | 1 | |
| 5 | BRN | 7 | 2 | 1 | Pup | 1 | |
| 5 | BRN | 7 | 2 | 5 | Pup | 1 | |
| 5 | BRN | 7 | 2 | 4 | Pup | 1 | |
| 5 | BRN | 7 | 2 | 8 | Pup | 1 | |
| 5 | BRN | 7 | 2 | 5 | Pup | 1 | |
| 5 | BRN | 7 | 2 | 5 | Pup | 1 | |
| 5 | BRN | 7 | 2 | 1 | Pup | | 1 |
| 5 | BRN | 7 | 2 | 0.5 | Pup | | 1 |
| 5 | BRN | 7 | 2 | 2 | Pup | | 1 |
| 5 | BRN | 7 | 2 | 3 | Pup | 1 | |
| 5 | BRN | 7 | 2 | 2 | Pup | 1 | |
| 5 | BRN | 7 | 2 | 1 | Pup | 1 | |
| 5 | BRN | 7 | 2 | 1 | Pup | 1 | |
| 5 | BRN | 7 | 2 | 0.5 | Pup | 1 | |
| 5 | BRN | 7 | 2 | 3 | Pup | 1 | |
| 5 | BRN | 7 | 2 | 2 | Pup | 1 | |
| 5 | BRN | 7 | 2 | 5 | Pup | 1 | |
| 5 | BRN | 7 | 2 | 3 | Pup | 1 | |
| 5 | BRN | 7 | 2 | 3 | Pup | 1 | |
| 5 | BRN | 7 | 2 | 3 | Pup | 1 | |
| 5 | BRN | 7 | 2 | 0.5 | Pup | 1 | |
| 5 | BRN | 7 | 2 | 1 | Pup | | 1 |
| 5 | BRN | 7 | 2 | 1 | Pup | | 1 |
| 5 | BRN | 7 | 2 | 0.5 | Pup | | 1 |
| 5 | BRN | 7 | 2 | 0.5 | Pup | | 1 |
| 5 | BRN | 7 | 2 | 0.5 | Pup | | 1 |
| 5 | BRN | 7 | 2 | 2 | Pup | 1 | |
| 5 | BRN | 7 | 2 | 1 | Pup | 1 | |
| 5 | BRN | 7 | 2 | 1 | Pup | 1 | |
| 5 | BRN | 7 | 2 | 1 | Pup | 1 | |
| 5 | BRN | 7 | 2 | 1 | Pup | 1 | |
| 5 | BRN | 7 | 2 | 1 | Pup | | 1 |
| 5 | BRN | 7 | 2 | 1 | Pup | | 1 |
| 5 | BRN | 7 | 2 | 1 | Pup | | 1 |
| 5 | KTI | 7 | 2 | 1 | Pup | 1 | |
| 5 | KTI | 7 | 2 | 2 | Pup | | 1 |
| 5 | KTI | 7 | 2 | 2 | Pup | 1 | |
| 5 | KTI | 7 | 2 | 3 | Pup | 1 | |
| 5 | KTI | 7 | 2 | 3 | Pup | 1 | |
| 5 | KTI | 7 | 2 | 1 | Pup | 1 | |
| 5 | KTI | 7 | 2 | 1 | Pup | 1 | |
| 5 | KTI | 7 | 2 | 2 | Pup | | 1 |
| 5 | KTI | 7 | 2 | 2 | Pup | | 1 |
| 5 | KTI | 7 | 2 | 2 | Pup | | 1 |
| 5 | KTI | 7 | 2 | 2 | Pup | | 1 |
| 5 | KTI | 7 | 2 | 1 | Pup | 1 | |
| 5 | KTI | 7 | 2 | 1 | Pup | 1 | |
| 5 | KTI | 7 | 2 | 3 | Pup | | 1 |
| 5 | KTI | 7 | 2 | 3 | Pup | | 1 |
| 5 | KTI | 7 | 2 | 2 | Pup | | 1 |
| 5 | KTI | 7 | 2 | 4 | Pup | | 1 |
| 5 | KTI | 7 | 2 | 4 | Pup | | 1 |
| 5 | KTI | 7 | 2 | 3 | Pup | | 1 |
| 5 | KTI | 7 | 2 | 2 | Pup | 1 | |
| 5 | KTI | 7 | 2 | 2 | Pup | 1 | |
| 5 | KTI | 7 | 2 | 2 | Pup | 1 | |
| 5 | KTI | 7 | 2 | 1 | Pup | 1 | |
| 5 | KTI | 7 | 2 | 1 | Pup | 1 | |
| 5 | BBR | 7 | 4 | 1 | Pup | 1 | |
| 5 | BBR | 7 | 4 | 0.5 | Pup | 1 | |
| 5 | BBR | 7 | 4 | 2 | Pup | 1 | |
| 5 | BBR | 7 | 4 | 2 | Pup | 1 | |
| 5 | BBR | 7 | 4 | 2 | Pup | 1 | |
| 5 | BBR | 7 | 4 | 2 | Pup | 1 | |
| 5 | BBR | 7 | 4 | 2 | Pup | 1 | |
| 5 | BBR | 7 | 4 | 0.5 | Pup | | 1 |
| 5 | BBR | 7 | 4 | 0.5 | Pup | | 1 |
| 5 | BBR | 7 | 4 | 0.5 | Pup | | 1 |
| 5 | BBR | 7 | 4 | 1 | Pup | 1 | |
| 5 | BBR | 7 | 4 | 1 | Pup | | 1 |
| 5 | BBR | 7 | 4 | 1 | Pup | | 1 |
| 5 | BBR | 7 | 4 | 4 | Pup | 1 | |
| 5 | BBR | 7 | 4 | 4 | Pup | 1 | |
| 5 | BBR | 7 | 4 | 4 | Pup | 1 | |
| 5 | BBR | 7 | 4 | 4 | Pup | 1 | |
| 5 | BBR | 7 | 4 | 1 | Pup | | 1 |
| 5 | BBR | 7 | 4 | 1 | Pup | | 1 |
| 5 | BBR | 7 | 4 | 1 | Pup | | 1 |
| 5 | BBR | 7 | 4 | 1 | Pup | 1 | |
| 5 | BBR | 7 | 4 | 1 | Pup | 1 | |
| 5 | BBR | 7 | 4 | 1 | Pup | 1 | |
| 5 | BBR | 7 | 4 | 1 | Pup | 1 | |
| 5 | BBR | 7 | 4 | 1 | Pup | 1 | |
| 5 | BBR | 7 | 4 | 1 | Pup | 1 | |
| 5 | BBR | 7 | 4 | 0.5 | Pup | | 1 |
| 5 | BBR | 7 | 4 | 0.5 | Pup | | 1 |
| 5 | BBR | 7 | 4 | 1 | Pup | | 1 |
| 5 | BBR | 7 | 4 | 1 | Pup | | 1 |
| 5 | BBR | 7 | 4 | 1 | Pup | | 1 |
| 5 | BBR | 7 | 4 | 1 | Pup | | 1 |
| 5 | BBR | 7 | 4 | 1 | Pup | | 1 |
| 5 | BBR | 7 | 4 | 0.5 | Pup | | 1 |
| 5 | BBR | 7 | 4 | 0.5 | Pup | | 1 |
| 5 | BBR | 7 | 4 | 0.5 | Pup | | 1 |
| 5 | BBR | 7 | 4 | 2 | Pup | | 1 |
| 5 | BBR | 7 | 4 | 2 | Pup | | 1 |
| 5 | BBR | 7 | 4 | 2 | Pup | | 1 |
| 5 | BBR | 7 | 4 | 3 | Pup | | 1 |
| 5 | BBR | 7 | 4 | 3 | Pup | | 1 |
| 5 | BBR | 7 | 4 | 1 | Pup | | 1 |
| 5 | BBR | 7 | 4 | 2 | Pup | | 1 |
| 5 | BBR | 7 | 4 | 2 | Pup | | 1 |
| 5 | BBR | 7 | 4 | 2 | Pup | | 1 |
| 5 | BBR | 7 | 4 | 2 | Pup | | 1 |
| 5 | BBR | 7 | 4 | 2 | Pup | | 1 |
| 5 | BBR | 7 | 4 | 1 | Pup | | 1 |
| 5 | BBR | 7 | 4 | 1 | Pup | | 1 |
| 5 | BBR | 7 | 4 | 1 | Pup | | 1 |
| 5 | BBR | 7 | 4 | 1 | Pup | | 1 |
| 5 | BBR | 7 | 4 | 1 | Pup | | 1 |
| 5 | BBR | 7 | 4 | 2 | Pup | 1 | |
| 5 | BBR | 7 | 4 | 1 | Pup | 1 | |
| 5 | BBR | 7 | 4 | 0.5 | Pup | 1 | |
| 5 | BBR | 7 | 4 | 1 | Pup | | 1 |
| 5 | BBR | 7 | 4 | 1 | Pup | | 1 |
| 5 | BBR | 7 | 4 | 1 | Pup | | 1 |
| 5 | BBR | 7 | 4 | 0.5 | Pup | | 1 |
| 5 | BBR | 7 | 4 | 0.5 | Pup | | 1 |
| 5 | BBR | 7 | 4 | 0.5 | Pup | | 1 |
| 5 | BBR | 7 | 4 | 1 | Pup | 1 | |
| 5 | BBR | 7 | 4 | 1 | Pup | 1 | |
| 5 | BBR | 7 | 4 | 1 | Pup | 1 | |
| 5 | BBR | 7 | 4 | 1 | Pup | 1 | |
| 5 | BBR | 7 | 4 | 1 | Pup | 1 | |
| 5 | BBR | 7 | 4 | 2 | Pup | 1 | |
| 5 | BBR | 7 | 4 | 2 | Pup | 1 | |
| 5 | BBR | 7 | 4 | 2 | Pup | 1 | |
| 5 | BBR | 7 | 4 | 2 | Pup | 1 | |
| 5 | BBR | 7 | 4 | 2 | Pup | 1 | |
| 5 | BBR | 7 | 4 | 1 | Pup | | 1 |
| 5 | BBR | 7 | 4 | 1 | Pup | | 1 |
| 5 | BBR | 7 | 4 | 1 | Pup | | 1 |
| 5 | BBR | 7 | 4 | 1 | Pup | | 1 |
| 5 | BBR | 7 | 4 | 0.5 | Pup | 1 | |
| 5 | BBR | 7 | 4 | 0.5 | Pup | 1 | |
| 5 | BBR | 7 | 4 | 0.5 | Pup | 1 | |
| 5 | BBR | 7 | 4 | 1 | Pup | 1 | |
| 5 | BBR | 7 | 4 | 1 | Pup | 1 | |
| 5 | BBR | 7 | 4 | 0.5 | Pup | 1 | |
| 5 | BBR | 7 | 4 | 0.5 | Pup | 1 | |
| 5 | BBR | 7 | 4 | 0.5 | Pup | 1 | |
| 5 | BBR | 7 | 4 | 1 | Pup | 1 | |
| 5 | BBR | 7 | 4 | 1 | Pup | 1 | |
| 5 | BBR | 7 | 4 | 0.5 | Pup | 1 | |
| 5 | BBR | 7 | 4 | 0.5 | Pup | 1 | |
| 5 | BBR | 7 | 4 | 0.5 | Pup | 1 | |
| 5 | BBR | 7 | 4 | 0.5 | Pup | 1 | |
| 5 | BBR | 7 | 4 | 0.5 | Pup | 1 | |
| 5 | BBR | 7 | 4 | 0.5 | Pup | 1 | |
| 5 | BBR | 7 | 4 | 0.5 | Pup | 1 | |
| 5 | WHI | 7 | 2 | 2 | MO | 1 | |
| 5 | WHI | 7 | 2 | 2 | MO | 1 | |
| 5 | WHI | 7 | 2 | 6 | Pup | 1 | |
| 5 | WHI | 7 | 2 | 3 | Pup | 1 | |
| 5 | WHI | 7 | 2 | 1 | Pup | 1 | |
| 5 | WHI | 7 | 2 | 0.5 | Pup | 1 | |
| 5 | WHI | 7 | 2 | 0.5 | Pup | 1 | |
| 5 | WHI | 7 | 2 | 4 | Pup | 1 | |
| 5 | WHI | 7 | 2 | 4 | MO | 1 | |
| 5 | WHI | 7 | 2 | 4 | MO | 1 | |
| 5 | WHI | 7 | 2 | 3 | Pup | 1 | |
| 5 | WHI | 7 | 2 | 2 | Pup | 1 | |
| 5 | WHI | 7 | 2 | 4 | MO | 1 | |
| 5 | WHI | 7 | 2 | 4 | MO | 1 | |
| 5 | WHI | 7 | 2 | 1 | Pup | 1 | |
| 5 | WHI | 7 | 2 | 1 | Pup | 1 | |
| 5 | WHI | 7 | 2 | 4 | Pup | 1 | |
| 5 | WHI | 7 | 2 | 3 | Pup | 1 | |
| 5 | WHI | 7 | 2 | 1 | Pup | 1 | |
| 5 | WHI | 7 | 2 | 2 | Pup | | 1 |
| 5 | WHI | 7 | 2 | 4 | Pup | | 1 |
| 5 | WHI | 7 | 2 | 4 | Pup | | 1 |
| 5 | WHI | 7 | 2 | 2 | Pup | 1 | |
| 5 | WHI | 7 | 2 | 2 | Pup | 1 | |
| 5 | WHI | 7 | 2 | 3 | MO | 1 | |
| 5 | WHI | 7 | 2 | 3 | MO | 1 | |
| 5 | WHI | 7 | 2 | 1 | Pup | 1 | |
| 5 | WHI | 7 | 2 | 1 | Pup | 1 | |
| 5 | RS5 | 7 | 3 | 3 | Pup | 1 | |
| 5 | RS5 | 7 | 3 | 3 | Pup | 1 | |
| 5 | RS5 | 7 | 3 | 3 | Pup | 1 | |
| 5 | RS5 | 7 | 3 | 4 | MO | 1 | |
| 5 | RS5 | 7 | 3 | 4 | MO | 1 | |
| 5 | RS5 | 7 | 3 | 4 | MO | 1 | |
| 5 | RS5 | 7 | 3 | 2 | Pup | 1 | |
| 5 | RS5 | 7 | 3 | 2 | Pup | 1 | |
| 5 | RS5 | 7 | 3 | 2 | Pup | 1 | |
| 5 | RS5 | 7 | 3 | 3 | Pup | | 1 |
| 5 | RS5 | 7 | 3 | 1 | Pup | 1 | |
| 5 | RS5 | 7 | 3 | 1 | Pup | 1 | |
| 5 | RS5 | 7 | 3 | 1 | Pup | 1 | |
| 5 | RS5 | 7 | 3 | 5 | Pup | 1 | |
| 5 | RS5 | 7 | 3 | 5 | Pup | 1 | |
| 5 | RS5 | 7 | 3 | 5 | Pup | 1 | |
| 4 | CAN | 8 | 5 | 0.5 | Pup | 1 | |
| 4 | CAN | 8 | 5 | 0.5 | Pup | 1 | |
| 4 | CAN | 8 | 5 | 0.5 | Pup | 1 | |
| 4 | CAN | 8 | 5 | 0.5 | Pup | 1 | |
| 4 | CAN | 8 | 5 | 0.5 | Pup | 1 | |
| 4 | CAN | 8 | 5 | 0.5 | Pup | 1 | |
| 4 | CAN | 8 | 5 | 0.5 | Pup | 1 | |
| 4 | CAN | 8 | 5 | 0.5 | Pup | 1 | |
| 4 | CAN | 8 | 5 | 7 | Pup | 1 | |
| 4 | CAN | 8 | 5 | 7 | Pup | 1 | |
| 4 | CAN | 8 | 5 | 7 | Pup | 1 | |
| 4 | CAN | 8 | 5 | 7 | Pup | 1 | |
| 4 | CAN | 8 | 5 | 7 | Pup | 1 | |
| 4 | CAN | 8 | 5 | 0.5 | Pup | 1 | |
| 4 | CAN | 8 | 5 | 0.5 | Pup | 1 | |
| 4 | CAN | 8 | 5 | 0.5 | Pup | 1 | |
| 4 | CAN | 8 | 5 | 7 | MO | 1 | |
| 4 | CAN | 8 | 5 | 7 | MO | 1 | |
| 4 | CAN | 8 | 5 | 7 | MO | 1 | |
| 4 | CAN | 8 | 5 | 7 | MO | 1 | |
| 4 | CAN | 8 | 5 | 7 | MO | 1 | |
| 4 | PF1 | 8 | 3 | 0.5 | Pup | 1 | |
| 4 | PF1 | 8 | 3 | 2 | Pup | 1 | |
| 4 | PF1 | 8 | 3 | 2 | Pup | 1 | |
| 4 | PF1 | 8 | 3 | 3 | Pup | 1 | |
| 4 | PF1 | 8 | 3 | 3 | Pup | 1 | |
| 4 | PF1 | 8 | 3 | 2 | Pup | 1 | |
| 4 | RS1 | 8 | 2 | 3 | Pup | | 1 |
| 4 | RS1 | 8 | 2 | 3 | Pup | | 1 |
| 4 | RS1 | 8 | 2 | 1 | Pup | 1 | |
| 4 | RS1 | 8 | 2 | 2 | Pup | | 1 |
| 4 | RS1 | 8 | 2 | 4 | Pup | | 1 |
| 4 | RS1 | 8 | 2 | 4 | Pup | | 1 |
| 4 | RS1 | 8 | 2 | 1 | Pup | | 1 |
| 4 | RS1 | 8 | 2 | 0.5 | Pup | | 1 |
| 4 | RS1 | 8 | 2 | 1 | Pup | 1 | |
| 4 | RS1 | 8 | 2 | 1 | Pup | 1 | |
| 4 | RS1 | 8 | 2 | 2 | Pup | | 1 |
| 4 | RS1 | 8 | 2 | 0.5 | Pup | | 1 |
| 4 | RS1 | 8 | 2 | 1 | Pup | | 1 |
| 4 | RS1 | 8 | 2 | 2 | Pup | 1 | |
| 4 | RS1 | 8 | 2 | 1 | Pup | 1 | |
| 4 | RS1 | 8 | 2 | 2 | Pup | 1 | |
| 4 | RS1 | 8 | 2 | 2 | Pup | 1 | |
| 4 | RS1 | 8 | 2 | 2 | Pup | 1 | |
| 4 | RS1 | 8 | 2 | 5 | Pup | | 1 |
| 4 | RS1 | 8 | 2 | 4 | Pup | | 1 |
| 4 | RS1 | 8 | 2 | 3 | Pup | | 1 |
| 4 | RS1 | 8 | 2 | 3 | Pup | | 1 |
| 4 | RS1 | 8 | 2 | 1 | Pup | 1 | |
| 4 | RS1 | 8 | 2 | 1 | Pup | 1 | |
| 4 | RS1 | 8 | 2 | 0.5 | Pup | | 1 |
| 4 | RS1 | 8 | 2 | 0.5 | Pup | | 1 |
| 4 | RS1 | 8 | 2 | 1 | Pup | 1 | |
| 4 | RS1 | 8 | 2 | 0.5 | Pup | 1 | |
| 4 | RS1 | 8 | 2 | 0.5 | Pup | 1 | |
| 4 | RS1 | 8 | 2 | 1 | Pup | | 1 |
| 4 | RS1 | 8 | 2 | 1 | Pup | | 1 |
| 4 | RS1 | 8 | 2 | 1 | Pup | | 1 |
| 4 | RS1 | 8 | 2 | 0.5 | Pup | | 1 |
| 4 | RS1 | 8 | 2 | 1 | Pup | | 1 |
| 4 | RS1 | 8 | 2 | 0.5 | Pup | | 1 |
| 4 | RS3 | 8 | 2 | 1 | Pup | 1 | |
| 4 | RS3 | 8 | 2 | 4 | Pup | 1 | |
| 4 | RS3 | 8 | 2 | 1 | Pup | 1 | |
| 4 | RS3 | 8 | 2 | 5 | Pup | | 1 |
| 4 | RS3 | 8 | 2 | 5 | Pup | | 1 |
| 4 | RS3 | 8 | 2 | 1 | Pup | | 1 |
| 4 | RS3 | 8 | 2 | 1 | Pup | | 1 |
| 4 | RS3 | 8 | 2 | 1 | Pup | 1 | |
| 4 | RS3 | 8 | 2 | 1 | Pup | 1 | |
| 4 | RS3 | 8 | 2 | 0.5 | Pup | 1 | |
| 4 | RS3 | 8 | 2 | 0.5 | Pup | 1 | |
| 4 | RS3 | 8 | 2 | 1 | Pup | 1 | |
| 4 | RS3 | 8 | 2 | 1 | Pup | 1 | |
| 4 | RS3 | 8 | 2 | 0.5 | Pup | 1 | |
| 4 | RS3 | 8 | 2 | 0.5 | Pup | 1 | |
| 4 | RS3 | 8 | 2 | 4 | Pup | 1 | |
| 4 | RS3 | 8 | 2 | 2 | Pup | 1 | |
| 5 | BRN | 8 | 2 | 2 | Pup | | 1 |
| 5 | BRN | 8 | 2 | 2 | Pup | | 1 |
| 5 | BRN | 8 | 2 | 0.5 | Pup | | 1 |
| 5 | BRN | 8 | 2 | 0.5 | Pup | | 1 |
| 5 | BRN | 8 | 2 | 1 | Pup | | 1 |
| 5 | BRN | 8 | 2 | 1 | Pup | | 1 |
| 5 | BRN | 8 | 2 | 0.5 | Pup | | 1 |
| 5 | BRN | 8 | 2 | 1 | Pup | | 1 |
| 5 | BRN | 8 | 2 | 1 | Pup | | 1 |
| 5 | BRN | 8 | 2 | 1 | Pup | | 1 |
| 5 | BRN | 8 | 2 | 3 | Pup | | 1 |
| 5 | BRN | 8 | 2 | 2 | Pup | | 1 |
| 5 | BRN | 8 | 2 | 1 | Pup | | 1 |
| 5 | BRN | 8 | 2 | 1 | Pup | | 1 |
| 5 | BRN | 8 | 2 | 1 | Pup | 1 | |
| 5 | BRN | 8 | 2 | 1 | Pup | 1 | |
| 5 | BRN | 8 | 2 | 1 | Pup | | 1 |
| 5 | BRN | 8 | 2 | 1 | Pup | | 1 |
| 5 | BRN | 8 | 2 | 1 | Pup | | 1 |
| 5 | BRN | 8 | 2 | 1 | Pup | | 1 |
| 5 | BRN | 8 | 2 | 0.5 | Pup | 1 | |
| 5 | BRN | 8 | 2 | 0.5 | Pup | 1 | |
| 5 | KTI | 8 | 2 | 1 | Pup | | 1 |
| 5 | KTI | 8 | 2 | 0.5 | Pup | | 1 |
| 5 | KTI | 8 | 2 | 1 | Pup | 1 | |
| 5 | KTI | 8 | 2 | 1 | Pup | 1 | |
| 5 | KTI | 8 | 2 | 0.5 | Pup | 1 | |
| 5 | KTI | 8 | 2 | 0.5 | Pup | 1 | |
| 5 | KTI | 8 | 2 | 0.5 | Pup | | 1 |
| 5 | KTI | 8 | 2 | 6 | Pup | 1 | |
| 5 | KTI | 8 | 2 | 6 | Pup | 1 | |
| 5 | KTI | 8 | 2 | 0.5 | Pup | 1 | |
| 5 | KTI | 8 | 2 | 0.5 | Pup | 1 | |
| 5 | KTI | 8 | 2 | 1 | Pup | | 1 |
| 5 | KTI | 8 | 2 | 1 | Pup | | 1 |
| 5 | KTI | 8 | 2 | 1 | Pup | | 1 |
| 5 | KTI | 8 | 2 | 0.5 | Pup | | 1 |
| 5 | BBR | 8 | 4 | 3 | Pup | | 1 |
| 5 | BBR | 8 | 4 | 3 | Pup | | 1 |
| 5 | BBR | 8 | 4 | 2 | Pup | | 1 |
| 5 | BBR | 8 | 4 | 2 | Pup | 1 | |
| 5 | BBR | 8 | 4 | 2 | Pup | 1 | |
| 5 | BBR | 8 | 4 | 2 | Pup | 1 | |
| 5 | BBR | 8 | 4 | 2 | Pup | 1 | |
| 5 | BBR | 8 | 4 | 1 | Pup | 1 | |
| 5 | BBR | 8 | 4 | 4 | Pup | 1 | |
| 5 | BBR | 8 | 4 | 4 | Pup | 1 | |
| 5 | BBR | 8 | 4 | 4 | Pup | 1 | |
| 5 | BBR | 8 | 4 | 4 | Pup | 1 | |
| 5 | BBR | 8 | 4 | 4 | Pup | 1 | |
| 5 | BBR | 8 | 4 | 4 | Pup | | 1 |
| 5 | BBR | 8 | 4 | 4 | Pup | | 1 |
| 5 | BBR | 8 | 4 | 4 | Pup | 1 | |
| 5 | BBR | 8 | 4 | 4 | Pup | 1 | |
| 5 | BBR | 8 | 4 | 4 | Pup | 1 | |
| 5 | BBR | 8 | 4 | 5 | Pup | 1 | |
| 5 | BBR | 8 | 4 | 5 | Pup | 1 | |
| 5 | BBR | 8 | 4 | 5 | Pup | 1 | |
| 5 | BBR | 8 | 4 | 5 | Pup | 1 | |
| 5 | BBR | 8 | 4 | 5 | Pup | 1 | |
| 5 | BBR | 8 | 4 | 4 | Pup | 1 | |
| 5 | BBR | 8 | 4 | 3 | Pup | 1 | |
| 5 | BBR | 8 | 4 | 3 | Pup | 1 | |
| 5 | BBR | 8 | 4 | 1 | Pup | 1 | |
| 5 | BBR | 8 | 4 | 1 | Pup | 1 | |
| 5 | BBR | 8 | 4 | 2 | Pup | 1 | |
| 5 | BBR | 8 | 4 | 2 | Pup | | 1 |
| 5 | BBR | 8 | 4 | 2 | Pup | | 1 |
| 5 | BBR | 8 | 4 | 1 | Pup | 1 | |
| 5 | BBR | 8 | 4 | 1 | Pup | 1 | |
| 5 | BBR | 8 | 4 | 1 | Pup | 1 | |
| 5 | BBR | 8 | 4 | 1 | Pup | 1 | |
| 5 | BBR | 8 | 4 | 0.5 | Pup | 1 | |
| 5 | BBR | 8 | 4 | 2 | Pup | 1 | |
| 5 | BBR | 8 | 4 | 2 | Pup | 1 | |
| 5 | BBR | 8 | 4 | 2 | Pup | 1 | |
| 5 | BBR | 8 | 4 | 2 | Pup | 1 | |
| 5 | BBR | 8 | 4 | 1 | Pup | 1 | |
| 5 | BBR | 8 | 4 | 1 | Pup | 1 | |
| 5 | BBR | 8 | 4 | 1 | Pup | | 1 |
| 5 | BBR | 8 | 4 | 1 | Pup | | 1 |
| 5 | BBR | 8 | 4 | 1 | Pup | 1 | |
| 5 | BBR | 8 | 4 | 1 | Pup | 1 | |
| 5 | BBR | 8 | 4 | 1 | Pup | 1 | |
| 5 | BBR | 8 | 4 | 1 | Pup | 1 | |
| 5 | BBR | 8 | 4 | 2 | Pup | 1 | |
| 5 | BBR | 8 | 4 | 2 | Pup | 1 | |
| 5 | BBR | 8 | 4 | 2 | Pup | 1 | |
| 5 | BBR | 8 | 4 | 2 | Pup | 1 | |
| 5 | BBR | 8 | 4 | 3 | Pup | 1 | |
| 5 | BBR | 8 | 4 | 3 | Pup | 1 | |
| 5 | BBR | 8 | 4 | 3 | Pup | 1 | |
| 5 | BBR | 8 | 4 | 3 | Pup | 1 | |
| 5 | WHI | 8 | 2 | 1 | Pup | | 1 |
| 5 | WHI | 8 | 2 | 1 | Pup | | 1 |
| 5 | WHI | 8 | 2 | 2 | Pup | | 1 |
| 5 | WHI | 8 | 2 | 1 | Pup | 1 | |
| 5 | WHI | 8 | 2 | 2 | Pup | 1 | |
| 5 | WHI | 8 | 2 | 2 | Pup | 1 | |
| 5 | WHI | 8 | 2 | 2 | Pup | 1 | |
| 5 | WHI | 8 | 2 | 2 | Pup | 1 | |
| 5 | WHI | 8 | 2 | 1 | Pup | 1 | |
| 5 | WHI | 8 | 2 | 4 | Pup | | 1 |
| 5 | WHI | 8 | 2 | 2 | Pup | | 1 |
| 5 | WHI | 8 | 2 | 0.5 | Pup | 1 | |
| 5 | WHI | 8 | 2 | 2 | Pup | 1 | |
| 5 | WHI | 8 | 2 | 2 | Pup | 1 | |
| 5 | WHI | 8 | 2 | 2 | Pup | 1 | |
| 5 | WHI | 8 | 2 | 2 | Pup | 1 | |
| 5 | WHI | 8 | 2 | 1 | Pup | 1 | |
| 5 | WHI | 8 | 2 | 1 | Pup | 1 | |
| 5 | WHI | 8 | 2 | 2 | Pup | 1 | |
| 5 | WHI | 8 | 2 | 2 | Pup | 1 | |
| 5 | WHI | 8 | 2 | 3 | Pup | | 1 |
| 5 | WHI | 8 | 2 | 3 | Pup | | 1 |
| 5 | WHI | 8 | 2 | 2 | Pup | | 1 |
| 5 | WHI | 8 | 2 | 2 | Pup | | 1 |
| 5 | WHI | 8 | 2 | 2 | Pup | | 1 |
| 5 | WHI | 8 | 2 | 0.5 | Pup | | 1 |
| 5 | WHI | 8 | 2 | 0.5 | Pup | | 1 |
| 5 | WHI | 8 | 2 | 1 | Pup | | 1 |
| 5 | WHI | 8 | 2 | 1 | Pup | | 1 |
| 5 | WHI | 8 | 2 | 1 | Pup | | 1 |
| 5 | WHI | 8 | 2 | 1 | Pup | | 1 |
| 5 | WHI | 8 | 2 | 4 | Pup | 1 | |
| 5 | WHI | 8 | 2 | 1 | Pup | 1 | |
| 5 | WHI | 8 | 2 | 0.5 | Pup | 1 | |
| 5 | WHI | 8 | 2 | 3 | Pup | 1 | |
| 5 | WHI | 8 | 2 | 3 | Pup | 1 | |
| 5 | WHI | 8 | 2 | 1 | Pup | 1 | |
| 5 | WHI | 8 | 2 | 1 | Pup | 1 | |
| 5 | WHI | 8 | 2 | 2 | Pup | 1 | |
| 5 | WHI | 8 | 2 | 2 | Pup | 1 | |
| 5 | WHI | 8 | 2 | 2 | Pup | 1 | |
| 5 | WHI | 8 | 2 | 0.5 | Pup | 1 | |
| 5 | WHI | 8 | 2 | 0.5 | Pup | 1 | |
| 5 | WHI | 8 | 2 | 2 | Pup | 1 | |
| 5 | WHI | 8 | 2 | 1 | Pup | 1 | |
| 4 | PF1 | 9 | 2 | 1 | Pup | 1 | |
| 4 | PF1 | 9 | 2 | 1 | Pup | | 1 |
| 4 | PF1 | 9 | 2 | 1 | Pup | | 1 |
| 4 | RS1 | 9 | 2 | 5 | Pup | 1 | |
| 4 | RS1 | 9 | 2 | 2 | Pup | 1 | |
| 4 | RS1 | 9 | 2 | 2 | Pup | 1 | |
| 4 | RS1 | 9 | 2 | 2 | Pup | 1 | |
| 4 | RS1 | 9 | 2 | 1 | Pup | | 1 |
| 4 | RS1 | 9 | 2 | 1 | Pup | | 1 |
| 4 | RS1 | 9 | 2 | 1 | Pup | | 1 |
| 4 | RS1 | 9 | 2 | 1 | Pup | | 1 |
| 4 | RS1 | 9 | 2 | 1 | Pup | | 1 |
| 4 | RS1 | 9 | 2 | 0.5 | Pup | | 1 |
| 4 | RS1 | 9 | 2 | 1 | Pup | | 1 |
| 4 | RS1 | 9 | 2 | 1 | Pup | | 1 |
| 4 | RS1 | 9 | 2 | 1 | Pup | | 1 |
| 4 | RS1 | 9 | 2 | 2 | Pup | | 1 |
| 4 | RS1 | 9 | 2 | 1 | Pup | | 1 |
| 4 | RS1 | 9 | 2 | 1 | Pup | | 1 |
| 4 | RS1 | 9 | 2 | 1 | Pup | | 1 |
| 4 | RS1 | 9 | 2 | 3 | Pup | | 1 |
| 4 | RS1 | 9 | 2 | 3 | Pup | | 1 |
| 4 | RS1 | 9 | 2 | 3 | Pup | | 1 |
| 4 | RS1 | 9 | 2 | 1 | Pup | | 1 |
| 4 | RS1 | 9 | 2 | 0.5 | Pup | | 1 |
| 4 | RS1 | 9 | 2 | 0.5 | Pup | | 1 |
| 4 | RS1 | 9 | 2 | 0.5 | Pup | | 1 |
| 4 | RS1 | 9 | 2 | 4 | Pup | | 1 |
| 4 | RS1 | 9 | 2 | 1 | Pup | | 1 |
| 4 | RS1 | 9 | 2 | 0.5 | Pup | | 1 |
| 4 | RS1 | 9 | 2 | 1 | Pup | | 1 |
| 4 | RS1 | 9 | 2 | 1 | Pup | | 1 |
| 4 | RS1 | 9 | 2 | 1 | Pup | | 1 |
| 4 | RS1 | 9 | 2 | 1 | Pup | 1 | |
| 4 | RS1 | 9 | 2 | 2 | Pup | | 1 |
| 4 | RS1 | 9 | 2 | 0.5 | Pup | | 1 |
| 4 | RS2 | 9 | 2 | 0.5 | Pup | 1 | |
| 4 | RS2 | 9 | 2 | 0.5 | Pup | 1 | |
| 4 | RS2 | 9 | 2 | 0.5 | Pup | 1 | |
| 4 | RS2 | 9 | 2 | 1 | Pup | 1 | |
| 4 | RS2 | 9 | 2 | 0.5 | Pup | 1 | |
| 4 | RS2 | 9 | 2 | 1 | Pup | 1 | |
| 4 | RS2 | 9 | 2 | 0.5 | Pup | 1 | |
| 4 | RS2 | 9 | 2 | 0.5 | Pup | 1 | |
| 4 | RS2 | 9 | 2 | 2 | Pup | 1 | |
| 4 | RS2 | 9 | 2 | 0.5 | Pup | 1 | |
| 4 | RS2 | 9 | 2 | 1 | Pup | 1 | |
| 4 | RS2 | 9 | 2 | 1 | Pup | 1 | |
| 4 | RS2 | 9 | 2 | 4 | Pup | 1 | |
| 4 | RS2 | 9 | 2 | 1 | Pup | 1 | |
| 4 | RS2 | 9 | 2 | 2 | Pup | 1 | |
| 4 | RS2 | 9 | 2 | 0.5 | Pup | 1 | |
| 4 | RS2 | 9 | 2 | 1 | Pup | 1 | |
| 4 | RS2 | 9 | 2 | 2 | Pup | 1 | |
| 4 | RS2 | 9 | 2 | 1 | Pup | 1 | |
| 4 | RS2 | 9 | 2 | 3 | Pup | 1 | |
| 4 | RS2 | 9 | 2 | 1 | Pup | 1 | |
| 4 | RS2 | 9 | 2 | 0.5 | Pup | 1 | |
| 4 | RS2 | 9 | 2 | 1 | Pup | 1 | |
| 4 | RS2 | 9 | 2 | 3 | Pup | 1 | |
| 4 | RS2 | 9 | 2 | 3 | Pup | 1 | |
| 4 | RS2 | 9 | 2 | 3 | Pup | 1 | |
| 4 | RS2 | 9 | 2 | 0.5 | Pup | 1 | |
| 4 | RS2 | 9 | 2 | 1 | Pup | | 1 |
| 4 | RS2 | 9 | 2 | 2 | Pup | 1 | |
| 4 | RS2 | 9 | 2 | 2 | Pup | 1 | |
| 4 | RS2 | 9 | 2 | 1 | Pup | 1 | |
| 4 | RS2 | 9 | 2 | 2 | Pup | 1 | |
| 4 | RS2 | 9 | 2 | 2 | Pup | 1 | |
| 4 | RS2 | 9 | 2 | 0.5 | Pup | 1 | |
| 4 | RS2 | 9 | 2 | 0.5 | Pup | 1 | |
| 4 | RS2 | 9 | 2 | 3 | Pup | 1 | |
| 4 | RS2 | 9 | 2 | 3 | Pup | 1 | |
| 4 | RS2 | 9 | 2 | 1 | Pup | 1 | |
| 4 | RS2 | 9 | 2 | 0.5 | Pup | 1 | |
| 4 | RS2 | 9 | 2 | 2 | Pup | 1 | |
| 4 | RS2 | 9 | 2 | 2 | Pup | 1 | |
| 4 | RS2 | 9 | 2 | 5 | Pup | 1 | |
| 4 | RS2 | 9 | 2 | 5 | Pup | 1 | |
| 4 | RS2 | 9 | 2 | 1 | Pup | 1 | |
| 4 | RS2 | 9 | 2 | 1 | Pup | 1 | |
| 4 | RS2 | 9 | 2 | 1 | Pup | 1 | |
| 4 | RS2 | 9 | 2 | 1 | Pup | 1 | |
| 4 | RS2 | 9 | 2 | 0.5 | Pup | 1 | |
| 4 | RS2 | 9 | 2 | 4 | Pup | 1 | |
| 4 | RS2 | 9 | 2 | 4 | Pup | 1 | |
| 4 | RS2 | 9 | 2 | 1 | Pup | 1 | |
| 4 | RS2 | 9 | 2 | 1 | Pup | 1 | |
| 4 | RS2 | 9 | 2 | 0.5 | Pup | 1 | |
| 4 | RS2 | 9 | 2 | 0.5 | Pup | 1 | |
| 4 | RS2 | 9 | 2 | 2 | Pup | 1 | |
| 4 | RS2 | 9 | 2 | 0.5 | Pup | 1 | |
| 4 | RS2 | 9 | 2 | 1 | Pup | 1 | |
| 4 | RS2 | 9 | 2 | 1 | Pup | 1 | |
| 4 | RS2 | 9 | 2 | 0.5 | Pup | 1 | |
| 4 | RS2 | 9 | 2 | 0.5 | Pup | 1 | |
| 4 | RS2 | 9 | 2 | 1 | Pup | 1 | |
| 4 | RS3 | 9 | 1 | 2 | Pup | 1 | |
| 4 | RS3 | 9 | 1 | 2 | Pup | 1 | |
| 4 | RS3 | 9 | 1 | 1 | Pup | | 1 |
| 5 | BRN | 9 | 2 | 2 | Pup | 1 | |
| 5 | BRN | 9 | 2 | 1 | Pup | 1 | |
| 5 | BRN | 9 | 2 | 1 | Pup | 1 | |
| 5 | BRN | 9 | 2 | 1 | Pup | 1 | |
| 5 | BRN | 9 | 2 | 1 | Pup | 1 | |
| 5 | BRN | 9 | 2 | 1 | Pup | | 1 |
| 5 | BRN | 9 | 2 | 1 | Pup | | 1 |
| 5 | BRN | 9 | 2 | 2 | Pup | | 1 |
| 5 | BRN | 9 | 2 | 2 | Pup | | 1 |
| 5 | BRN | 9 | 2 | 1 | Pup | | 1 |
| 5 | BRN | 9 | 2 | 1 | Pup | | 1 |
| 5 | BRN | 9 | 2 | 1 | Pup | | 1 |
| 5 | BRN | 9 | 2 | 0.5 | Pup | | 1 |
| 5 | BRN | 9 | 2 | 1 | Pup | | 1 |
| 5 | BRN | 9 | 2 | 1 | Pup | | 1 |
| 5 | BRN | 9 | 2 | 1 | Pup | | 1 |
| 5 | BRN | 9 | 2 | 0.5 | Pup | | 1 |
| 5 | BRN | 9 | 2 | 0.5 | Pup | | 1 |
| 5 | BRN | 9 | 2 | 3 | Pup | | 1 |
| 5 | BRN | 9 | 2 | 3 | Pup | | 1 |
| 5 | BRN | 9 | 2 | 3 | Pup | | 1 |
| 5 | BRN | 9 | 2 | 2 | Pup | | 1 |
| 5 | BRN | 9 | 2 | 2 | Pup | | 1 |
| 5 | BRN | 9 | 2 | 1 | Pup | | 1 |
| 5 | BRN | 9 | 2 | 1 | Pup | | 1 |
| 5 | BRN | 9 | 2 | 0.5 | Pup | | 1 |
| 5 | BRN | 9 | 2 | 0.5 | Pup | | 1 |
| 5 | BRN | 9 | 2 | 0.5 | Pup | | 1 |
| 5 | BRN | 9 | 2 | 2 | Pup | | 1 |
| 5 | BRN | 9 | 2 | 2 | Pup | | 1 |
| 5 | BRN | 9 | 2 | 2 | Pup | | 1 |
| 5 | BRN | 9 | 2 | 2 | Pup | 1 | |
| 5 | BRN | 9 | 2 | 1 | Pup | 1 | |
| 5 | BRN | 9 | 2 | 3 | Pup | | 1 |
| 5 | BRN | 9 | 2 | 3 | Pup | | 1 |
| 5 | BRN | 9 | 2 | 3 | Pup | | 1 |
| 5 | BRN | 9 | 2 | 1 | Pup | 1 | |
| 5 | BRN | 9 | 2 | 0.5 | Pup | | 1 |
| 5 | BRN | 9 | 2 | 0.5 | Pup | | 1 |
| 5 | BRN | 9 | 2 | 2 | Pup | | 1 |
| 5 | BRN | 9 | 2 | 2 | Pup | | 1 |
| 5 | BRN | 9 | 2 | 2 | Pup | | 1 |
| 5 | BRN | 9 | 2 | 2 | Pup | | 1 |
| 5 | BRN | 9 | 2 | 2 | Pup | | 1 |
| 5 | BRN | 9 | 2 | 2 | Pup | | 1 |
| 5 | BRN | 9 | 2 | 1 | Pup | | 1 |
| 5 | BRN | 9 | 2 | 1 | Pup | | 1 |
| 5 | BRN | 9 | 2 | 1 | Pup | | 1 |
| 5 | BRN | 9 | 2 | 1 | Pup | | 1 |
| 5 | BRN | 9 | 2 | 1 | Pup | | 1 |
| 5 | BRN | 9 | 2 | 1 | Pup | | 1 |
| 5 | BRN | 9 | 2 | 1 | Pup | | 1 |
| 5 | BRN | 9 | 2 | 1 | Pup | | 1 |
| 5 | BRN | 9 | 2 | 3 | Pup | | 1 |
| 5 | BRN | 9 | 2 | 3 | Pup | | 1 |
| 5 | BRN | 9 | 2 | 2 | Pup | | 1 |
| 5 | BRN | 9 | 2 | 2 | Pup | | 1 |
| 5 | BRN | 9 | 2 | 2 | Pup | 1 | |
| 5 | BRN | 9 | 2 | 2 | Pup | 1 | |
| 5 | BRN | 9 | 2 | 1 | Pup | | 1 |
| 5 | BRN | 9 | 2 | 1 | Pup | | 1 |
| 5 | BRN | 9 | 2 | 1 | Pup | 1 | |
| 5 | BRN | 9 | 2 | 1 | Pup | | 1 |
| 5 | BRN | 9 | 2 | 1 | Pup | | 1 |
| 5 | BRN | 9 | 2 | 1 | Pup | | 1 |
| 5 | BRN | 9 | 2 | 1 | Pup | 1 | |
| 5 | BRN | 9 | 2 | 1 | Pup | 1 | |
| 5 | BRN | 9 | 2 | 2 | Pup | 1 | |
| 5 | BRN | 9 | 2 | 1 | Pup | 1 | |
| 5 | KTI | 9 | 2 | 1 | Pup | | 1 |
| 5 | KTI | 9 | 2 | 1 | Pup | | 1 |
| 5 | KTI | 9 | 2 | 1 | Pup | | 1 |
| 5 | KTI | 9 | 2 | 1 | Pup | | 1 |
| 5 | KTI | 9 | 2 | 0.5 | Pup | | 1 |
| 5 | KTI | 9 | 2 | 0.5 | Pup | | 1 |
| 5 | KTI | 9 | 2 | 0.5 | Pup | | 1 |
| 5 | KTI | 9 | 2 | 1 | Pup | | 1 |
| 5 | KTI | 9 | 2 | 1 | Pup | | 1 |
| 5 | KTI | 9 | 2 | 1 | Pup | | 1 |
| 5 | KTI | 9 | 2 | 0.5 | Pup | | 1 |
| 5 | KTI | 9 | 2 | 3 | Pup | | 1 |
| 5 | KTI | 9 | 2 | 2 | Pup | | 1 |
| 5 | KTI | 9 | 2 | 2 | Pup | | 1 |
| 5 | KTI | 9 | 2 | 2 | Pup | | 1 |
| 5 | KTI | 9 | 2 | 2 | Pup | | 1 |
| 5 | KTI | 9 | 2 | 2 | Pup | | 1 |
| 5 | KTI | 9 | 2 | 1 | Pup | | 1 |
| 5 | KTI | 9 | 2 | 3 | Pup | | 1 |
| 5 | KTI | 9 | 2 | 3 | Pup | | 1 |
| 5 | KTI | 9 | 2 | 3 | Pup | | 1 |
| 5 | KTI | 9 | 2 | 1 | Pup | | 1 |
| 5 | KTI | 9 | 2 | 1 | Pup | | 1 |
| 5 | KTI | 9 | 2 | 0.5 | Pup | | 1 |
| 5 | KTI | 9 | 2 | 0.5 | Pup | | 1 |
| 5 | KTI | 9 | 2 | 0.5 | Pup | | 1 |
| 5 | KTI | 9 | 2 | 0.5 | Pup | 1 | |
| 5 | KTI | 9 | 2 | 2 | Pup | | 1 |
| 5 | KTI | 9 | 2 | 1 | Pup | | 1 |
| 5 | KTI | 9 | 2 | 1 | Pup | | 1 |
| 5 | KTI | 9 | 2 | 2 | Pup | | 1 |
| 5 | KTI | 9 | 2 | 2 | Pup | | 1 |
| 5 | KTI | 9 | 2 | 2 | Pup | | 1 |
| 5 | KTI | 9 | 2 | 2 | Pup | | 1 |
| 5 | KTI | 9 | 2 | 2 | Pup | | 1 |
| 5 | KTI | 9 | 2 | 2 | Pup | | 1 |
| 5 | KTI | 9 | 2 | 2 | Pup | | 1 |
| 5 | KTI | 9 | 2 | 1 | Pup | | 1 |
| 5 | KTI | 9 | 2 | 1 | Pup | | 1 |
| 5 | BBR | 9 | 4 | 2 | Pup | | 1 |
| 5 | BBR | 9 | 4 | 2 | Pup | | 1 |
| 5 | BBR | 9 | 4 | 2 | Pup | | 1 |
| 5 | BBR | 9 | 4 | 2 | Pup | | 1 |
| 5 | BBR | 9 | 4 | 1 | Pup | | 1 |
| 5 | BBR | 9 | 4 | 1 | Pup | | 1 |
| 5 | BBR | 9 | 4 | 1 | Pup | | 1 |
| 5 | BBR | 9 | 4 | 1 | Pup | 1 | |
| 5 | BBR | 9 | 4 | 1 | Pup | 1 | |
| 5 | BBR | 9 | 4 | 0.5 | Pup | 1 | |
| 5 | BBR | 9 | 4 | 1 | Pup | 1 | |
| 5 | BBR | 9 | 4 | 1 | Pup | 1 | |
| 5 | BBR | 9 | 4 | 1 | Pup | 1 | |
| 5 | BBR | 9 | 4 | 1 | Pup | 1 | |
| 5 | BBR | 9 | 4 | 1 | Pup | 1 | |
| 5 | BBR | 9 | 4 | 1 | Pup | 1 | |
| 5 | BBR | 9 | 4 | 2 | Pup | 1 | |
| 5 | BBR | 9 | 4 | 2 | Pup | 1 | |
| 5 | BBR | 9 | 4 | 2 | Pup | 1 | |
| 5 | BBR | 9 | 4 | 1 | Pup | 1 | |
| 5 | BBR | 9 | 4 | 1 | Pup | 1 | |
| 5 | BBR | 9 | 4 | 2 | Pup | 1 | |
| 5 | BBR | 9 | 4 | 4 | Pup | 1 | |
| 5 | BBR | 9 | 4 | 4 | Pup | 1 | |
| 5 | BBR | 9 | 4 | 4 | Pup | 1 | |
| 5 | BBR | 9 | 4 | 4 | Pup | 1 | |
| 5 | BBR | 9 | 4 | 4 | Pup | 1 | |
| 5 | BBR | 9 | 4 | 1 | Pup | | 1 |
| 5 | BBR | 9 | 4 | 1 | Pup | | 1 |
| 5 | BBR | 9 | 4 | 1 | Pup | | 1 |
| 5 | BBR | 9 | 4 | 1 | Pup | | 1 |
| 5 | BBR | 9 | 4 | 1 | Pup | 1 | |
| 5 | BBR | 9 | 4 | 1 | Pup | 1 | |
| 5 | BBR | 9 | 4 | 1 | Pup | 1 | |
| 5 | BBR | 9 | 4 | 1 | Pup | 1 | |

| | | | | | | | | |
|---|---|---|---|---|---|---|---|---|
| 5 | BBR | 9 | 4 | 1 | Pup | 1 | | |
| 5 | BBR | 9 | 4 | 1 | Pup | 1 | | |
| 5 | BBR | 9 | 4 | 2 | Pup | 1 | | |
| 5 | BBR | 9 | 4 | 2 | Pup | 1 | | |
| 5 | BBR | 9 | 4 | 2 | Pup | 1 | | |
| 5 | BBR | 9 | 4 | 2 | Pup | 1 | | |
| 5 | BBR | 9 | 4 | 1 | Pup | 1 | | |
| 5 | BBR | 9 | 4 | 1 | Pup | 1 | | |
| 5 | BBR | 9 | 4 | 1 | Pup | 1 | | |
| 5 | WHI | 9 | 2 | 1 | Pup | | | 1 |
| 5 | WHI | 9 | 2 | 1 | Pup | | | 1 |
| 5 | WHI | 9 | 2 | 0.5 | Pup | 1 | | |
| 5 | WHI | 9 | 2 | 1 | Pup | 1 | | |
| 5 | WHI | 9 | 2 | 1 | Pup | 1 | | |
| 5 | WHI | 9 | 2 | 1 | Pup | 1 | | |
| 5 | WHI | 9 | 2 | 1 | Pup | 1 | | |
| 5 | WHI | 9 | 2 | 1 | Pup | 1 | | |
| 5 | WHI | 9 | 2 | 5 | Pup | | | 1 |
| 5 | WHI | 9 | 2 | 4 | Pup | | | 1 |
| 5 | WHI | 9 | 2 | 4 | Pup | | | 1 |
| 5 | WHI | 9 | 2 | 5 | Pup | | | 1 |
| 5 | WHI | 9 | 2 | 5 | Pup | | | 1 |
| 5 | WHI | 9 | 2 | 5 | Pup | | | 1 |
| 5 | WHI | 9 | 2 | 1 | Pup | | | 1 |
| 5 | WHI | 9 | 2 | 3 | Pup | | | 1 |
| 5 | WHI | 9 | 2 | 2 | Pup | | | 1 |
| 5 | WHI | 9 | 2 | 1 | Pup | | | 1 |
| 5 | WHI | 9 | 2 | 1 | Pup | | | 1 |
| 5 | WHI | 9 | 2 | 1 | Pup | | | 1 |
| 5 | WHI | 9 | 2 | 1 | Pup | | | 1 |
| 5 | WHI | 9 | 2 | 1 | Pup | | | 1 |
| 5 | WHI | 9 | 2 | 2 | Pup | | | 1 |
| 5 | WHI | 9 | 2 | 2 | Pup | | | 1 |
| 5 | WHI | 9 | 2 | 2 | Pup | | | 1 |
| 5 | WHI | 9 | 2 | 0.5 | Pup | | | 1 |

| | | | | | | | |
|---|---|---|---|---|---|---|---|
| 5 | WHI | 9 | 2 | 0.5 | Pup | | 1 |
| 5 | WHI | 9 | 2 | 0.5 | Pup | | 1 |
| 5 | WHI | 9 | 2 | 1 | Pup | | 1 |
| 5 | WHI | 9 | 2 | 1 | Pup | | 1 |
| 5 | WHI | 9 | 2 | 1 | Pup | | 1 |
| 5 | WHI | 9 | 2 | 1 | Pup | | 1 |
| 5 | WHI | 9 | 2 | 1 | Pup | | 1 |
| 5 | WHI | 9 | 2 | 1 | Pup | | 1 |
| 5 | WHI | 9 | 2 | 1 | Pup | | 1 |
| 5 | WHI | 9 | 2 | 1 | Pup | | 1 |
| 5 | WHI | 9 | 2 | 1 | Pup | | 1 |
| 5 | WHI | 9 | 2 | 1 | Pup | | 1 |
| 5 | WHI | 9 | 2 | 1 | Pup | | 1 |
| 5 | WHI | 9 | 2 | 2 | Pup | | 1 |
| 5 | WHI | 9 | 2 | 2 | Pup | | 1 |
| 5 | WHI | 9 | 2 | 2 | Pup | | 1 |
| 5 | WHI | 9 | 2 | 2 | Pup | 1 | |
| 5 | WHI | 9 | 2 | 1 | Pup | 1 | |
| 5 | WHI | 9 | 2 | 2 | Pup | 1 | |
| 5 | WHI | 9 | 2 | 0.5 | Pup | | 1 |
| 5 | WHI | 9 | 2 | 1 | Pup | | 1 |
| 5 | WHI | 9 | 2 | 1 | Pup | | 1 |
| 5 | WHI | 9 | 2 | 3 | Pup | | 1 |
| 5 | WHI | 9 | 2 | 3 | Pup | | 1 |
| 5 | WHI | 9 | 2 | 3 | Pup | | 1 |
| 5 | WHI | 9 | 2 | 1 | Pup | | 1 |
| 5 | WHI | 9 | 2 | 3 | Pup | | 1 |
| 5 | WHI | 9 | 2 | 3 | Pup | | 1 |
| 5 | WHI | 9 | 2 | 1 | Pup | 1 | |
| 5 | WHI | 9 | 2 | 1 | Pup | 1 | |
| 5 | WHI | 9 | 2 | 1 | Pup | 1 | |
| 5 | WHI | 9 | 2 | 0.5 | Pup | 1 | |
| 5 | WHI | 9 | 2 | 0.5 | Pup | 1 | |
| 5 | WHI | 9 | 2 | 0.5 | Pup | 1 | |
| 5 | WHI | 9 | 2 | 1 | Pup | | 1 |
| 5 | WHI | 9 | 2 | 0.5 | Pup | | 1 |
| 5 | WHI | 9 | 2 | 0.5 | Pup | | 1 |
| 5 | WHI | 9 | 2 | 3 | Pup | 1 | |
| 5 | WHI | 9 | 2 | 2 | Pup | 1 | |
| 5 | WHI | 9 | 2 | 0.5 | Pup | | 1 |
| 5 | WHI | 9 | 2 | 0.5 | Pup | | 1 |
| 5 | WHI | 9 | 2 | 2 | Pup | | 1 |
| 5 | WHI | 9 | 2 | 2 | Pup | | 1 |
| 4 | CAN | 10 | 5 | 2 | MO | 1 | |
| 4 | CAN | 10 | 5 | 2 | MO | 1 | |
| 4 | CAN | 10 | 5 | 2 | MO | 1 | |
| 4 | CAN | 10 | 5 | 2 | MO | 1 | |
| 4 | CAN | 10 | 5 | 2 | MO | 1 | |
| 4 | CAN | 10 | 5 | 2 | Pup | 1 | |
| 4 | CAN | 10 | 5 | 2 | Pup | 1 | |
| 4 | CAN | 10 | 5 | 2 | Pup | 1 | |
| 4 | CAN | 10 | 5 | 2 | Pup | 1 | |
| 4 | CAN | 10 | 5 | 2 | Pup | 1 | |
| 4 | CAN | 10 | 5 | 0.5 | Pup | 1 | |
| 4 | CAN | 10 | 5 | 0.5 | Pup | 1 | |
| 4 | CAN | 10 | 5 | 0.5 | Pup | 1 | |
| 4 | CAN | 10 | 5 | 0.5 | Pup | 1 | |
| 4 | CAN | 10 | 5 | 2 | Pup | 1 | |
| 4 | CAN | 10 | 5 | 2 | Pup | 1 | |
| 4 | CAN | 10 | 5 | 2 | Pup | 1 | |
| 4 | CAN | 10 | 5 | 2 | Pup | 1 | |
| 4 | CAN | 10 | 5 | 3 | Pup | 1 | |
| 4 | CAN | 10 | 5 | 3 | Pup | 1 | |
| 4 | CAN | 10 | 5 | 3 | Pup | 1 | |
| 4 | CAN | 10 | 5 | 3 | Pup | 1 | |
| 4 | CAN | 10 | 5 | 0.5 | Pup | 1 | |
| 4 | CAN | 10 | 5 | 0.5 | Pup | 1 | |
| 4 | CAN | 10 | 5 | 0.5 | Pup | 1 | |
| 4 | CAN | 10 | 5 | 0.5 | Pup | 1 | |
| 4 | CAN | 10 | 5 | 1 | Pup | 1 | |
| 4 | CAN | 10 | 5 | 1 | Pup | 1 | |
| 4 | CAN | 10 | 5 | 1 | Pup | 1 | |
| 4 | CAN | 10 | 5 | 1 | Pup | 1 | |
| 4 | CAN | 10 | 5 | 0.5 | Pup | 1 | |
| 4 | CAN | 10 | 5 | 0.5 | Pup | 1 | |
| 4 | CAN | 10 | 5 | 0.5 | Pup | 1 | |
| 4 | CAN | 10 | 5 | 0.5 | Pup | 1 | |
| 4 | CAN | 10 | 5 | 4 | Pup | 1 | |
| 4 | CAN | 10 | 5 | 4 | Pup | 1 | |
| 4 | CAN | 10 | 5 | 4 | Pup | 1 | |
| 4 | CAN | 10 | 5 | 3 | Pup | 1 | |
| 4 | CAN | 10 | 5 | 2 | Pup | 1 | |
| 4 | CAN | 10 | 5 | 2 | Pup | 1 | |
| 4 | CAN | 10 | 5 | 2 | Pup | 1 | |
| 4 | PF1 | 10 | 2 | 1 | Pup | 1 | |
| 4 | PF1 | 10 | 2 | 1 | Pup | 1 | |
| 4 | PF1 | 10 | 2 | 1 | Pup | 1 | |
| 4 | PF1 | 10 | 2 | 1 | Pup | 1 | |
| 4 | RS2 | 10 | 2 | 2 | Pup | 1 | |
| 4 | RS2 | 10 | 2 | 0.5 | Pup | 1 | |
| 4 | RS2 | 10 | 2 | 0.5 | Pup | 1 | |
| 4 | RS3 | 10 | 1 | 1 | Pup | 1 | |
| 4 | RS3 | 10 | 1 | 1 | Pup | 1 | |
| 4 | RS3 | 10 | 1 | 0.5 | Pup | 1 | |
| 5 | BRN | 10 | 2 | 1 | Pup | 1 | |
| 5 | BRN | 10 | 2 | 1 | Pup | | 1 |
| 5 | BRN | 10 | 2 | 1 | Pup | | 1 |
| 5 | BRN | 10 | 2 | 1 | Pup | | 1 |
| 5 | BRN | 10 | 2 | 1 | Pup | | 1 |
| 5 | BRN | 10 | 2 | 0.5 | Pup | | 1 |
| 5 | BRN | 10 | 2 | 2 | Pup | | 1 |
| 5 | BRN | 10 | 2 | 2 | Pup | | 1 |
| 5 | BRN | 10 | 2 | 2 | Pup | | 1 |
| 5 | BRN | 10 | 2 | 2 | Pup | | 1 |
| 5 | BRN | 10 | 2 | 1 | Pup | | 1 |
| 5 | BRN | 10 | 2 | 1 | Pup | | 1 |
| 5 | BRN | 10 | 2 | 0.5 | Pup | | 1 |
| 5 | BRN | 10 | 2 | 0.5 | Pup | | 1 |
| 5 | BRN | 10 | 2 | 1 | Pup | | 1 |
| 5 | BRN | 10 | 2 | 1 | Pup | | 1 |
| 5 | BRN | 10 | 2 | 1 | Pup | | 1 |
| 5 | BRN | 10 | 2 | 1 | Pup | | 1 |
| 5 | BRN | 10 | 2 | 1 | Pup | 1 | |
| 5 | BRN | 10 | 2 | 2 | Pup | 1 | |
| 5 | BRN | 10 | 2 | 1 | Pup | 1 | |
| 5 | BRN | 10 | 2 | 1 | Pup | | 1 |
| 5 | BRN | 10 | 2 | 1 | Pup | | 1 |
| 5 | BRN | 10 | 2 | 1 | Pup | | 1 |
| 5 | BRN | 10 | 2 | 1 | Pup | | 1 |
| 5 | BRN | 10 | 2 | 1 | Pup | | 1 |
| 5 | BRN | 10 | 2 | 1 | Pup | | 1 |
| 5 | BRN | 10 | 2 | 0.5 | Pup | | 1 |
| 5 | BRN | 10 | 2 | 1 | Pup | | 1 |
| 5 | BRN | 10 | 2 | 1 | Pup | | 1 |
| 5 | KTI | 10 | 2 | 2 | Pup | | 1 |
| 5 | KTI | 10 | 2 | 1 | Pup | | 1 |
| 5 | KTI | 10 | 2 | 1 | Pup | | 1 |
| 5 | KTI | 10 | 2 | 1 | Pup | | 1 |
| 5 | KTI | 10 | 2 | 1 | Pup | | 1 |
| 5 | KTI | 10 | 2 | 1 | Pup | | 1 |
| 5 | BBR | 10 | 4 | 1 | Pup | 1 | |
| 5 | BBR | 10 | 4 | 1 | Pup | 1 | |
| 5 | BBR | 10 | 4 | 1 | Pup | 1 | |
| 5 | BBR | 10 | 4 | 2 | Pup | 1 | |
| 5 | BBR | 10 | 4 | 2 | Pup | 1 | |
| 5 | BBR | 10 | 4 | 0.5 | Pup | 1 | |
| 5 | BBR | 10 | 4 | 0.5 | Pup | 1 | |
| 5 | BBR | 10 | 4 | 1 | Pup | 1 | |
| 5 | BBR | 10 | 4 | 1 | Pup | 1 | |
| 5 | BBR | 10 | 4 | 0.5 | Pup | 1 | |
| 5 | BBR | 10 | 4 | 0.5 | Pup | 1 | |
| 5 | BBR | 10 | 4 | 1 | Pup | 1 | |
| 5 | BBR | 10 | 4 | 1 | Pup | 1 | |
| 5 | BBR | 10 | 4 | 1 | Pup | 1 | |
| 5 | BBR | 10 | 4 | 1 | Pup | 1 | |
| 5 | BBR | 10 | 4 | 0.5 | Pup | 1 | |
| 5 | BBR | 10 | 4 | 0.5 | Pup | 1 | |
| 5 | WHI | 10 | 2 | 1 | Pup | 1 | |
| 5 | WHI | 10 | 2 | 4 | Pup | 1 | |
| 5 | WHI | 10 | 2 | 3 | Pup | 1 | |
| 5 | WHI | 10 | 2 | 3 | Pup | 1 | |
| 5 | WHI | 10 | 2 | 2 | Pup | | 1 |
| 5 | WHI | 10 | 2 | 2 | Pup | | 1 |
| 5 | WHI | 10 | 2 | 0.5 | Pup | | 1 |
| 5 | WHI | 10 | 2 | 0.5 | Pup | | 1 |
| 5 | WHI | 10 | 2 | 0.5 | Pup | | 1 |
| 5 | WHI | 10 | 2 | 0.5 | Pup | | 1 |
| 5 | WHI | 10 | 2 | 1 | Pup | 1 | |
| 5 | WHI | 10 | 2 | 1 | Pup | | 1 |
| 5 | WHI | 10 | 2 | 1 | Pup | | 1 |
| 5 | WHI | 10 | 2 | 1 | Pup | | 1 |
| 5 | WHI | 10 | 2 | 1 | Pup | | 1 |
| 5 | WHI | 10 | 2 | 1 | Pup | | 1 |
| 5 | WHI | 10 | 2 | 1 | Pup | | 1 |
| 5 | WHI | 10 | 2 | 5 | Pup | 1 | |
| 5 | WHI | 10 | 2 | 1 | Pup | | 1 |
| 5 | WHI | 10 | 2 | 1 | Pup | | 1 |
| 5 | WHI | 10 | 2 | 1 | Pup | | 1 |
| 5 | WHI | 10 | 2 | 1 | Pup | | 1 |
| 5 | WHI | 10 | 2 | 1 | Pup | | 1 |
| 5 | WHI | 10 | 2 | 1 | Pup | | 1 |
| 5 | WHI | 10 | 2 | 1 | Pup | | 1 |
| 5 | WHI | 10 | 2 | 1 | Pup | | 1 |
| 5 | WHI | 10 | 2 | 0.5 | Pup | | 1 |
| 5 | WHI | 10 | 2 | 0.5 | Pup | | 1 |
| 5 | WHI | 10 | 2 | 1 | Pup | | 1 |
| 5 | WHI | 10 | 2 | 1 | Pup | | 1 |
| 5 | WHI | 10 | 2 | 0.5 | Pup | | 1 |
| 5 | WHI | 10 | 2 | 0.5 | Pup | | 1 |
| 5 | WHI | 10 | 2 | 1 | Pup | | 1 |
| 5 | WHI | 10 | 2 | 1 | Pup | | 1 |
| 5 | WHI | 10 | 2 | 1 | Pup | | 1 |
| 5 | WHI | 10 | 2 | 1 | Pup | | 1 |
| 5 | WHI | 10 | 2 | 1 | Pup | | 1 |
| 5 | WHI | 10 | 2 | 1 | Pup | | 1 |
| 5 | WHI | 10 | 2 | 0.5 | Pup | | 1 |
| 5 | WHI | 10 | 2 | 0.5 | Pup | | 1 |
| 5 | WHI | 10 | 2 | 0.5 | Pup | | 1 |
| 5 | WHI | 10 | 2 | 0.5 | Pup | | 1 |
| 5 | WHI | 10 | 2 | 0.5 | Pup | | 1 |
| 5 | WHI | 10 | 2 | 0.5 | Pup | | 1 |
| 5 | WHI | 10 | 2 | 1 | Pup | 1 | |
| 5 | WHI | 10 | 2 | 1 | Pup | 1 | |
| 5 | WHI | 10 | 2 | 1 | Pup | 1 | |
| 5 | WHI | 10 | 2 | 1 | Pup | 1 | |
| 5 | WHI | 10 | 2 | 1 | Pup | 1 | |
| 5 | WHI | 10 | 2 | 1 | Pup | 1 | |
| 5 | WHI | 10 | 2 | 1 | Pup | 1 | |
| 5 | WHI | 10 | 2 | 1 | Pup | 1 | |
| 5 | WHI | 10 | 2 | 1 | Pup | 1 | |
| 5 | WHI | 10 | 2 | 1 | Pup | 1 | |
| 5 | WHI | 10 | 2 | 1 | Pup | | 1 |
| 5 | WHI | 10 | 2 | 1 | Pup | | 1 |
| 4 | CAN | 11 | 5 | 0.5 | Pup | 1 | |
| 4 | CAN | 11 | 5 | 2 | Pup | 1 | |
| 4 | CAN | 11 | 5 | 2 | Pup | 1 | |
| 4 | CAN | 11 | 5 | 2 | Pup | 1 | |
| 4 | CAN | 11 | 5 | 2 | Pup | 1 | |
| 4 | CAN | 11 | 5 | 2 | Pup | 1 | |
| 4 | CAN | 11 | 5 | 3 | Pup | 1 | |
| 4 | CAN | 11 | 5 | 2 | Pup | 1 | |
| 4 | CAN | 11 | 5 | 0.5 | Pup | 1 | |
| 4 | RS2 | 11 | 2 | 1 | Pup | 1 | |
| 4 | RS2 | 11 | 2 | 0.5 | Pup | 1 | |
| 4 | RS2 | 11 | 2 | 2 | Pup | 1 | |
| 4 | RS2 | 11 | 2 | 1 | Pup | 1 | |
| 4 | RS2 | 11 | 2 | 2 | Pup | 1 | |
| 4 | RS3 | 11 | 1 | 4 | Pup | 1 | |
| 4 | RS3 | 11 | 1 | 5 | Pup | 1 | |
| 5 | BRN | 11 | 1 | 3 | Pup | | 1 |
| 5 | BRN | 11 | 1 | 3 | Pup | | 1 |
| 5 | BRN | 11 | 1 | 0.5 | Pup | | 1 |
| 5 | BRN | 11 | 1 | 0.5 | Pup | | 1 |
| 5 | BRN | 11 | 1 | 2 | Pup | | 1 |
| 5 | BRN | 11 | 1 | 2 | Pup | | 1 |
| 5 | KTI | 11 | 2 | 1 | Pup | | 1 |
| 5 | KTI | 11 | 2 | 1 | Pup | | 1 |
| 5 | KTI | 11 | 2 | 2 | Pup | | 1 |
| 5 | KTI | 11 | 2 | 2 | Pup | | 1 |
| 5 | KTI | 11 | 2 | 2 | Pup | | 1 |
| 5 | KTI | 11 | 2 | 1 | Pup | | 1 |
| 5 | KTI | 11 | 2 | 1 | Pup | | 1 |
| 5 | WHI | 11 | 2 | 1 | Pup | | 1 |
| 5 | WHI | 11 | 2 | 1 | Pup | | 1 |
| 5 | WHI | 11 | 2 | 1 | Pup | | 1 |
| 5 | WHI | 11 | 2 | 2 | Pup | | 1 |
| 5 | WHI | 11 | 2 | 2 | Pup | | 1 |
| 5 | WHI | 11 | 2 | 2 | Pup | | 1 |
| 5 | WHI | 11 | 2 | 0.5 | Pup | | 1 |
| 5 | WHI | 11 | 2 | 0.5 | Pup | | 1 |
| 5 | WHI | 11 | 2 | 0.5 | Pup | | 1 |
| 5 | WHI | 11 | 2 | 0.5 | Pup | | 1 |
| 5 | WHI | 11 | 2 | 2 | Pup | | 1 |
| 5 | WHI | 11 | 2 | 2 | Pup | | 1 |
| 5 | WHI | 11 | 2 | 1 | Pup | | 1 |
| 5 | WHI | 11 | 2 | 2 | Pup | | 1 |
| 5 | WHI | 11 | 2 | 2 | Pup | | 1 |
| 5 | WHI | 11 | 2 | 2 | Pup | | 1 |
| 5 | WHI | 11 | 2 | 2 | Pup | | 1 |
| 5 | WHI | 11 | 2 | 2 | Pup | | 1 |
| 5 | WHI | 11 | 2 | 0.5 | Pup | | 1 |
| 5 | WHI | 11 | 2 | 0.5 | Pup | | 1 |
| 5 | WHI | 11 | 2 | 0.5 | Pup | | 1 |
| 5 | WHI | 11 | 2 | 0.5 | Pup | | 1 |
| 5 | WHI | 11 | 2 | 0.5 | Pup | | 1 |
| 4 | CAN | 12 | 5 | 2 | Pup | 1 | |
| 4 | CAN | 12 | 5 | 2 | Pup | 1 | |
| 4 | CAN | 12 | 5 | 2 | Pup | 1 | |
| 4 | RS2 | 12 | 2 | 0.5 | Pup | 1 | |
| 4 | RS2 | 12 | 2 | 0.5 | Pup | 1 | |
| 4 | RS3 | 12 | 1 | 1 | Pup | 1 | |
| 4 | RS3 | 12 | 1 | 3 | Pup | 1 | |
| 5 | KTI | 12 | 2 | 1 | Pup | | 1 |
| 5 | KTI | 12 | 2 | 0.5 | Pup | | 1 |
| 5 | WHI | 12 | 2 | 2 | Pup | | 1 |
| 5 | WHI | 12 | 2 | 2 | Pup | | 1 |
| 5 | WHI | 12 | 2 | 1 | Pup | | 1 |
| 5 | WHI | 12 | 2 | 1 | Pup | | 1 |
| 5 | WHI | 12 | 2 | 1 | Pup | | 1 |
| 5 | WHI | 12 | 2 | 1 | Pup | | 1 |
| 5 | WHI | 12 | 2 | 1 | Pup | | 1 |
| 5 | WHI | 12 | 2 | 1 | Pup | | 1 |
| 5 | WHI | 12 | 2 | 1 | Pup | | 1 |
| 5 | WHI | 12 | 2 | 2 | Pup | | 1 |
| 5 | WHI | 12 | 2 | 1 | Pup | | 1 |
| 5 | WHI | 12 | 2 | 1 | Pup | | 1 |
| 5 | WHI | 12 | 2 | 1 | Pup | | 1 |
| 5 | WHI | 12 | 2 | 0.5 | Pup | | 1 |
| 5 | WHI | 12 | 2 | 1 | Pup | | 1 |
| 5 | WHI | 12 | 2 | 1 | Pup | | 1 |
| 5 | WHI | 12 | 2 | 1 | Pup | | 1 |
| 5 | WHI | 12 | 2 | 1 | Pup | | 1 |
| 5 | WHI | 12 | 2 | 0.5 | Pup | | 1 |
| 5 | WHI | 12 | 2 | 0.5 | Pup | | 1 |
| 5 | WHI | 12 | 2 | 0.5 | Pup | | 1 |
| 5 | WHI | 12 | 2 | 1 | Pup | | 1 |
| 5 | WHI | 12 | 2 | 1 | Pup | | 1 |
| 5 | WHI | 12 | 2 | 2 | Pup | | 1 |
| 5 | WHI | 12 | 2 | 2 | Pup | | 1 |
| 4 | CAN | 13 | 5 | 1 | Pup | 1 | |
| 4 | CAN | 13 | 5 | 1 | Pup | 1 | |
| 4 | CAN | 13 | 5 | 0.5 | Pup | 1 | |
| 4 | CAN | 13 | 5 | 0.5 | Pup | 1 | |
| 4 | CAN | 14 | | 0 | | | |
| 4 | CAN | 15 | | 0 | | | |
| 4 | CAN | 16 | | 0 | | | |
| 4 | CAN | 17 | | 0 | | | |
| 4 | CAN | 14 | | 0 | | | |
| 4 | CAN | 15 | | 0 | | | |
| 4 | CAN | 16 | | 0 | | | |
| 4 | CAN | 17 | | 0 | | | |

**Raw data for Model 2**

| Group | Current litter size | Age | Initiated by | |
|---|---|---|---|---|
| | | | mother | pup |
| CAN | 6 | 3 | 18 | 2 |
| CAN | 6 | 4 | 18 | 0 |
| CAN | 6 | 5 | 24 | 16 |
| CAN | 6 | 6 | 7 | 4 |
| CAN | 5 | 7 | 10 | 5 |
| CAN | 5 | 8 | 8 | 13 |
| CAN | 5 | 9 | | |
| CAN | 5 | 10 | 5 | 36 |
| CAN | 5 | 11 | 0 | 9 |
| CAN | 5 | 12 | 0 | 3 |
| CAN | 5 | 13 | 0 | 4 |
| CAN | 5 | 14 | | |
| CAN | 5 | 15 | | |
| MDB | 5 | 3 | 15 | 0 |
| MDB | 5 | 4 | 22 | 8 |
| MDB | 5 | 5 | | |
| MDB | 5 | 6 | 10 | 29 |
| MDB | 5 | 7 | | |
| PF1 | 5 | 3 | 30 | 12 |
| PF1 | 5 | 4 | 0 | 39 |
| PF1 | 4 | 5 | 0 | 29 |
| PF1 | 3 | 6 | 0 | 16 |
| PF1 | 3 | 7 | 0 | 7 |
| PF1 | 3 | 8 | 0 | 6 |
| PF1 | 2 | 9 | 0 | 3 |
| PF1 | 2 | 10 | 0 | 4 |
| PF1 | 2 | 11 | | |
| PF1 | 2 | 12 | | |
| PF1 | 2 | 13 | | |
| RS1 | 2 | 3 | 10 | 18 |
| RS1 | 2 | 4 | 10 | 8 |
| RS1 | 2 | 5 | 2 | 3 |
| RS1 | 2 | 6 | 2 | 5 |
| RS1 | 2 | 7 | 0 | 96 |
| RS1 | 2 | 8 | 0 | 32 |
| RS1 | 2 | 9 | 0 | 30 |
| RS2 | 4 | 3 | 37 | 13 |
| RS2 | 4 | 4 | 12 | 22 |
| RS2 | 4 | 5 | 28 | 1 |
| RS2 | 4 | 6 | 8 | 73 |
| RS2 | 2 | 7 | 0 | 15 |
| RS2 | 2 | 8 | | |
| RS2 | 2 | 9 | 0 | 61 |
| RS2 | 2 | 10 | 0 | 3 |
| RS2 | 2 | 11 | 0 | 5 |
| RS2 | 2 | 12 | 0 | 2 |
| RS2 | 2 | 13 | | |
| RS2 | 2 | 14 | | |
| RS3 | 2 | 3 | 5 | 20 |
| RS3 | 2 | 4 | 4 | 10 |
| RS3 | 2 | 5 | 8 | 11 |
| RS3 | 2 | 6 | 2 | 8 |
| RS3 | 2 | 7 | 4 | 7 |
| RS3 | 1 | 8 | 0 | 11 |
| RS3 | 1 | 9 | 0 | 3 |
| RS3 | 1 | 10 | 0 | 3 |
| RS3 | 1 | 11 | 0 | 2 |
| RS3 | 1 | 12 | 0 | 2 |
| RS3 | 1 | 13 | | |
| BRN | 2 | 3 | 5 | 22 |
| BRN | 2 | 4 | 2 | 12 |
| BRN | 2 | 5 | 0 | 12 |
| BRN | 1 | 6 | 0 | 12 |
| BRN | 2 | 7 | 0 | 41 |
| BRN | 2 | 8 | 0 | 13 |
| BRN | 2 | 9 | 0 | 44 |
| BRN | 2 | 10 | 0 | 17 |
| BRN | 1 | 11 | 0 | 3 |
| KTI | 2 | 3 | 6 | 10 |
| KTI | 2 | 4 | 14 | 26 |
| KTI | 2 | 5 | 14 | 25 |
| KTI | 2 | 6 | 0 | 16 |
| KTI | 2 | 7 | 0 | 22 |
| KTI | 2 | 8 | 0 | 15 |
| KTI | 2 | 9 | 0 | 25 |
| KTI | 2 | 10 | 0 | 3 |
| KTI | 2 | 11 | 0 | 4 |
| KTI | 2 | 12 | 0 | 2 |
| BBR | 4 | 3 | 30 | 19 |
| BBR | 4 | 4 | 24 | 6 |
| BBR | 4 | 5 | 24 | 14 |
| BBR | 4 | 6 | 12 | 29 |
| BBR | 4 | 7 | 0 | 62 |
| BBR | 4 | 8 | 0 | 37 |
| BBR | 4 | 9 | 0 | 30 |
| BBR | 4 | 10 | 0 | 11 |
| WHI | 2 | 3 | 9 | 8 |
| WHI | 2 | 4 | 10 | 43 |
| WHI | 2 | 5 | 12 | 12 |
| WHI | 2 | 6 | 10 | 11 |
| WHI | 2 | 7 | 8 | 17 |
| WHI | 2 | 8 | 0 | 35 |
| WHI | 2 | 9 | 0 | 50 |
| WHI | 2 | 10 | 0 | 37 |
| WHI | 2 | 11 | 0 | 15 |
| WHI | 2 | 12 | 0 | 16 |
| RS5 | 3 | 3 | 15 | 22 |
| RS5 | 3 | 4 | 11 | 7 |
| RS5 | 3 | 5 | 12 | 15 |
| RS5 | 3 | 6 | 3 | 22 |
| RS5 | 3 | 7 | 3 | 13 |

**Raw data for Model 3**

| Group | Current litter size | Age | Termination of suckling bouts through refusal | |
|---|---|---|---|---|
| | | | Mother mediated | Allomother mediated |
| CAN | 6 | 3 | 0.1 | 0 |
| CAN | 6 | 4 | 0.333333333 | |
| CAN | 6 | 5 | 0.1 | |
| CAN | 6 | 6 | 0.181818182 | |
| CAN | 5 | 7 | 1 | |
| CAN | 5 | 8 | 1 | |
| CAN | 5 | 9 | | |
| CAN | 5 | 10 | 1 | |
| CAN | 5 | 11 | 1 | |
| CAN | 5 | 12 | 1 | |
| CAN | 5 | 13 | 1 | |
| CAN | 5 | 14 | | |
| CAN | 5 | 15 | | |
| MDB | 5 | 3 | 0.666666667 | |
| MDB | 5 | 4 | 0.533333333 | |
| MDB | 5 | 5 | | |
| MDB | 5 | 6 | 1 | |
| MDB | 5 | 7 | | |
| PF1 | 5 | 3 | 0.380952381 | |
| PF1 | 5 | 4 | 1 | 0 |
| PF1 | 4 | 5 | 1 | |
| PF1 | 3 | 6 | 0.5 | 0 |
| PF1 | 3 | 7 | 1 | 1 |
| PF1 | 3 | 8 | 0.333333333 | |
| PF1 | 2 | 9 | 1 | 1 |
| PF1 | 2 | 10 | 1 | |
| PF1 | 2 | 11 | | |
| PF1 | 2 | 12 | | |
| PF1 | 2 | 13 | | |
| RS1 | 2 | 3 | 0.357142857 | |
| RS1 | 2 | 4 | 0.277777778 | |
| RS1 | 2 | 5 | 0.6 | |
| RS1 | 2 | 6 | 1 | |
| RS1 | 2 | 7 | 1 | 0.890410959 |
| RS1 | 2 | 8 | 1 | 0.954545455 |
| RS1 | 2 | 9 | 0.8 | 0.892857143 |
| RS2 | 4 | 3 | 0.8 | |
| RS2 | 4 | 4 | 0.828571429 | |
| RS2 | 4 | 5 | 0.651515152 | |
| RS2 | 4 | 6 | 0.879310345 | 1 |
| RS2 | 2 | 7 | 0.96969697 | 1 |
| RS2 | 2 | 8 | | |
| RS2 | 2 | 9 | 0.833333333 | 1 |
| RS2 | 2 | 10 | 0.666666667 | |
| RS2 | 2 | 11 | 0.6 | |
| RS2 | 2 | 12 | 1 | |
| RS2 | 2 | 13 | | |
| RS2 | 2 | 14 | | |
| RS3 | 2 | 3 | 0 | |
| RS3 | 2 | 4 | 0.428571429 | |
| RS3 | 2 | 5 | 0.578947368 | |
| RS3 | 2 | 6 | 0.666666667 | 1 |
| RS3 | 2 | 7 | 0.642857143 | |
| RS3 | 1 | 8 | 0.923076923 | 1 |
| RS3 | 1 | 9 | 1 | 1 |
| RS3 | 1 | 10 | 1 | |
| RS3 | 1 | 11 | 0.5 | |
| RS3 | 1 | 12 | 0.5 | |
| RS3 | 1 | 13 | | |
| BRN | 2 | 3 | 0.296296296 | |
| BRN | 2 | 4 | 1 | |
| BRN | 2 | 5 | 0.818181818 | 1 |
| BRN | 1 | 6 | 0.705882353 | |
| BRN | 2 | 7 | 0.847826087 | 1 |
| BRN | 2 | 8 | 1 | 1 |
| BRN | 2 | 9 | 0.866666667 | 0.962962963 |
| BRN | 2 | 10 | 1 | 1 |
| BRN | 1 | 11 | | 1 |
| KTI | 2 | 3 | 0.4375 | |
| KTI | 2 | 4 | 0.1 | |
| KTI | 2 | 5 | 0.230769231 | |
| KTI | 2 | 6 | 1 | 1 |
| KTI | 2 | 7 | 1 | 1 |
| KTI | 2 | 8 | 1 | 1 |
| KTI | 2 | 9 | 1 | 1 |
| KTI | 2 | 10 | | 1 |
| KTI | 2 | 11 | | 1 |
| KTI | 2 | 12 | | 1 |
| BBR | 4 | 3 | 0.489795918 | |
| BBR | 4 | 4 | 0.575757576 | |
| BBR | 4 | 5 | 0.714285714 | |
| BBR | 4 | 6 | 0.852941176 | 1 |
| BBR | 4 | 7 | 1 | 1 |
| BBR | 4 | 8 | 1 | 1 |
| BBR | 4 | 9 | 1 | 1 |
| BBR | 4 | 10 | 1 | |
| WHI | 2 | 3 | 0.352941176 | |
| WHI | 2 | 4 | 0.132075472 | |
| WHI | 2 | 5 | 0.166666667 | |
| WHI | 2 | 6 | 0.238095238 | |
| WHI | 2 | 7 | 0.913043478 | 1 |
| WHI | 2 | 8 | 1 | 1 |
| WHI | 2 | 9 | 1 | 0.980769231 |
| WHI | 2 | 10 | 0.875 | 1 |
| WHI | 2 | 11 | | 1 |
| WHI | 2 | 12 | | 1 |
| RS5 | 3 | 3 | 0.081081081 | |
| RS5 | 3 | 4 | 0.166666667 | |
| RS5 | 3 | 5 | 0 | |
| RS5 | 3 | 6 | 0.56 | |
| RS5 | 3 | 7 | 1 | 1 |
| RS5 | 3 | 8 | | |
| RS5 | 3 | 9 | | |
| RS5 | 3 | 10 | | |
| RS5 | 3 | 11 | | |
| RS5 | 3 | 12 | | |
| RS5 | 3 | 13 | | |
| RS5 | 3 | 14 | | |
| RS5 | 3 | 15 | | |
| RS5 | 3 | 16 | | |
| RS5 | 3 | 17 | | |

| erial no. | Year | Group name | Mother-litter group id | Litter size | Location | Habitat type | Latitude and Longitude | Suckling details noted | Allonursi present |
|---|---|---|---|---|---|---|---|---|---|
| 1 | 2010-11 | CAN1 | CAN1 | 5 | IISER-K Campus | Suburban | 22.9638° N, 88.5246° E | No | No |
| 2 | 2010-11 | BUD | BUD1 | 4 | IISER-K Campus | Suburban | 22.9638° N, 88.5246° E | No | Yes |
| 3 | 2010-11 | LEL1 | LEL1 | 2 | IISER-K Campus | Suburban | 22.9638° N, 88.5246° E | No | No |
| 4 | 2010-11 | S1 | S1 | 2 | Saltlake, Kolkata | Urban | 22.5800° N, 88.4200° E | No | No |
| 5 | 2011-12 | BSF1 | RS4 | 5 | Kalyani | Suburban | 22.9750° N, 88.4344° E | No | No |
| 6 | 2011-12 | PLT1 | JCB | 2 | IISER-K Campus | Suburban | 22.9638° N, 88.5246° E | No | Yes |
| 7 | 2011-12 | PLT1 | MDB1 | 5 | IISER-K Campus | Suburban | 22.9638° N, 88.5246° E | No | No |
| 8 | 2011-12 | CAN2 | CAN2 | 5 | IISER-K Campus | Suburban | 22.9638° N, 88.5246° E | No | No |
| 9 | 2011-12 | GH | GH2 | 6 | IISER-K Campus | Suburban | 22.9638° N, 88.5246° E | No | No |
| 10 | 2011-12 | LEL2 | LEL2 | 2 | IISER-K Campus | Suburban | 22.9638° N, 88.5246° E | No | No |
| 11 | 2011-12 | S2 | S2 | 3 | Saltlake, Kolkata | Urban | 22.5800° N, 88.4200° E | No | No |
| 12 | 2013-14 | BSF2 | **RS1** | 2 | Kalyani | Suburban | 22.9750° N, 88.4344° E | **Yes** | Yes |
| 13 | 2013-14 | BSF2 | **RS2** | 4 | Kalyani | Suburban | 22.9750° N, 88.4344° E | **Yes** | Yes |
| 14 | 2013-14 | BSF2 | **RS3** | 2 | Kalyani | Suburban | 22.9750° N, 88.4344° E | **Yes** | Yes |
| 15 | 2013-14 | PF | **PF1** | 5 | IISER-K Campus | Suburban | 22.9638° N, 88.5246° E | **Yes** | Yes |
| 16 | 2013-14 | CAN3 | **CAN3** | 6 | IISER-K Campus | Suburban | 22.9638° N, 88.5246° E | **Yes** | No |
| 17 | 2013-14 | PLT2 | **MDB2** | 5 | IISER-K Campus | Suburban | 22.9638° N, 88.5246° E | **Yes** | No |
| 18 | 2014-15 | BSF3 | **BBR** | 4 | Kalyani | Suburban | 22.9750° N, 88.4344° E | **Yes** | Yes |
| 19 | 2014-15 | BSF3 | **KTI** | 2 | Kalyani | Suburban | 22.9750° N, 88.4344° E | **Yes** | Yes |
| 20 | 2014-15 | BSF3 | **WHI** | 2 | Kalyani | Suburban | 22.9750° N, 88.4344° E | **Yes** | Yes |
| 21 | 2014-15 | BSF3 | **BRN** | 2 | Kalyani | Suburban | 22.9750° N, 88.4344° E | **Yes** | Yes |
| 22 | 2014-15 | BSF3 | **RS5** | 3 | Kalyani | Suburban | 22.9750° N, 88.4344° E | **Yes** | Yes |

| **Serial no.** | **Year** | **Group name** | **Mother-litter group id** | **Litter size** | **Location** | **Habitat type** | **Latitude and Longitude** | **Suckling details noted** | **Allonursing present** |
|---|---|---|---|---|---|---|---|---|---|
| 1 | 2010-11 | CAN1 | CAN1 | 5 | IISER-K Campus | Suburban | 22.9638° N, 88.5246° E | No | No |
| 2 | 2010-11 | BUD | BUD1 | 4 | IISER-K Campus | Suburban | 22.9638° N, 88.5246° E | No | Yes |
| 3 | 2010-11 | LEL1 | LEL1 | 2 | IISER-K Campus | Suburban | 22.9638° N, 88.5246° E | No | No |
| 4 | 2010-11 | S1 | S1 | 2 | Saltlake, Kolkata | Urban | 22.5800° N, 88.4200° E | No | No |
| 5 | 2011-12 | BSF1 | RS4 | 5 | Kalyani | Suburban | 22.9750° N, 88.4344° E | No | No |
| 6 | 2011-12 | PLT1 | JCB | 2 | IISER-K Campus | Suburban | 22.9638° N, 88.5246° E | No | Yes |
| 7 | 2011-12 | PLT1 | MDB1 | 5 | IISER-K Campus | Suburban | 22.9638° N, 88.5246° E | No | No |
| 8 | 2011-12 | CAN2 | CAN2 | 5 | IISER-K Campus | Suburban | 22.9638° N, 88.5246° E | No | No |
| 9 | 2011-12 | GH | GH2 | 6 | IISER-K Campus | Suburban | 22.9638° N, 88.5246° E | No | No |
| 10 | 2011-12 | LEL2 | LEL2 | 2 | IISER-K Campus | Suburban | 22.9638° N, 88.5246° E | No | No |
| 11 | 2011-12 | S2 | S2 | 3 | Saltlake, Kolkata | Urban | 22.5800° N, 88.4200° E | No | No |
| 12 | 2013-14 | BSF2 | **RS1** | 2 | Kalyani | Suburban | 22.9750° N, 88.4344° E | **Yes** | Yes |
| 13 | 2013-14 | BSF2 | **RS2** | 4 | Kalyani | Suburban | 22.9750° N, 88.4344° E | **Yes** | Yes |
| 14 | 2013-14 | BSF2 | **RS3** | 2 | Kalyani | Suburban | 22.9750° N, 88.4344° E | **Yes** | Yes |
| 15 | 2013-14 | PF | **PF1** | 5 | IISER-K Campus | Suburban | 22.9638° N, 88.5246° E | **Yes** | Yes |
| 16 | 2013-14 | CAN3 | **CAN3** | 6 | IISER-K Campus | Suburban | 22.9638° N, 88.5246° E | **Yes** | No |
| 17 | 2013-14 | PLT2 | **MDB2** | 5 | IISER-K Campus | Suburban | 22.9638° N, 88.5246° E | **Yes** | No |
| 18 | 2014-15 | BSF3 | **BBR** | 4 | Kalyani | Suburban | 22.9750° N, 88.4344° E | **Yes** | Yes |
| 19 | 2014-15 | BSF3 | **KTI** | 2 | Kalyani | Suburban | 22.9750° N, 88.4344° E | **Yes** | Yes |
| 20 | 2014-15 | BSF3 | **WHI** | 2 | Kalyani | Suburban | 22.9750° N, 88.4344° E | **Yes** | Yes |
| 21 | 2014-15 | BSF3 | **BRN** | 2 | Kalyani | Suburban | 22.9750° N, 88.4344° E | **Yes** | Yes |
| 22 | 2014-15 | BSF3 | **RS5** | 3 | Kalyani | Suburban | 22.9750° N, 88.4344° E | **Yes** | Yes |

18